\documentclass[12pt,preprint]{aastex}

\def\d{{\rm d}}

\shorttitle{Electron and Proton Acceleration}
\shortauthors{Petrosian \& Liu}

\begin{document}


\title{Stochastic Acceleration of Electrons and Protons.
I. Acceleration by Parallel Propagating Waves}


\author{Vah\'{e} Petrosian\altaffilmark{1} and Siming Liu\altaffilmark{2}}
\affil{Center for Space Science and Astrophysics, Department of Physics, Stanford
University, Stanford, CA 94305}


\altaffiltext{1}{Department of Applied Physics, Stanford University, Stanford, CA, 
94305 email: vahe@astronomy.stanford.edu}
\altaffiltext{2}{email: liusm@stanford.edu}


\begin{abstract} 

Stochastic acceleration of electrons and protons by waves propagating parallel to the large
scale magnetic fields of magnetized plasmas is studied with emphasis on the feasibility of
accelerating particles from a thermal background to relativistic energies and with the aim
of determining the relative acceleration of the two species in one source.  In general, the
stochastic acceleration by these waves results in two distinct components in the particle
distributions, a quasi-thermal and a hard nonthermal, with the nonthermal one
being more prominent in hotter plasmas and/or with higher level turbulence.  This can
explain many of the observed features of solar flares.  Regarding the proton to electron
ratio, we find that in a pure hydrogen plasma, the dominance of the wave-proton
interaction by the resonant Alfv\'{e}n mode reduces the acceleration rate of protons in the
intermediate energy range significantly, while the electron-cyclotron and Whistler waves
are very efficient in accelerating electrons from a few keV to MeV energies.  The presence
of such an acceleration barrier prohibits the proton acceleration under solar flare
conditions.  This difficulty is alleviated when we include the effects of $^4$He in the
dispersion relation and the damping of the turbulent waves by the thermal background
plasma.  The additional $^4$He cyclotron branch of the turbulent plasma waves suppresses
the proton acceleration barrier significantly and the steep turbulence spectrum in the
dissipation range makes the nonthermal component have a near power law shape.  The relative
acceleration of protons and electrons is very sensitive to a plasma parameter
$\alpha=\omega_{\rm pe}/\Omega_{\rm e}$, where $\omega_{\rm pe}$ and $\Omega_{\rm e}$ are
the electron plasma frequency and gyro-frequency, respectively.  Protons are preferentially
accelerated in weakly magnetized plasmas (large $\alpha$).  The formalism developed here is
applicable to the acceleration of other ion species and to other astrophysical systems.

\end{abstract}





\section{INTRODUCTION}

One of the important questions in acceleration of cosmic particles is the fractions of
energy that go into acceleration of electrons and protons (and other ions).  In this 
paper, we investigate this question for acceleration by plasma wave turbulence, a second 
order Fermi acceleration process, which we call stochastic acceleration (SA).  The theory 
of SA has received little attention in high energy astrophysics except in solar flares 
where it has achieved significant successes during the past few years.  The turbulence or 
plasma waves required for this model are presumably generated during the magnetic 
reconnection which energizes the flares.  The first application of SA was to the 
acceleration of protons and other ions to explain the observed nuclear gamma ray lines from 
solar flares (see e.g Ramaty 1979; Miller \& Roberts 1995).  Combining with the nuclear 
reaction rates (Ramaty, Kozlovsky \& Lingenfelter 1975; 1979; see also Kozlovsky, Murphy \& 
Ramaty 2002) and a magnetic loop model, Hua, Lingenfelter and Ramaty (1987a; 1987b; 1989) 
showed that the SA model can provide natural explanations for the many observed features in 
the 1 to 7 MeV range.  Later this model was also investigated in the acceleration of 
electrons in several studies (Miller \& Ramaty 1987; Bech, Schlickeiser \& Steinacker 1989; 
Miller, LaRosa \& Moore 1996; Park \& Petrosian 1995 and 1996), and the first quantitative 
comparison of predictions of this model with the observed hard X-ray (10 to 200 keV) 
spectra in some solar flares was carried out by Hamilton \& Petrosian (1992).  With a more 
detailed modeling, Park, Petrosian \& Schwartz (1997) showed that the SA of electrons by 
some generic turbulent plasma waves can reproduce the many spectral breaks observed over a 
broad energy range, from tens of keV to $\sim 100$ MeV, in the so-called electron dominated 
flares via the bremsstrahlung process (Rieger, Gan \& Marschh\"{a}user 1998; Petrosian, 
McTiernan \& Marschh\"{a}user 1994).

The strongest evidence supporting the SA model comes from the {\it YOHKOH} discovery of
impulsive hard X-ray emission from the top of a flaring loop, in addition to previously
known emission from loop foot-points (FPs for short. Masuda et al.  1994; Masuda 1994).  
The presence of the loop-top (LT for short) emission requires temporary trapping of the
accelerated electrons at the top of the loop where the reconnection is taking place.  The
turbulence required for SA will naturally accomplish this by repeated scatterings of the
particles (Petrosian \& Donaghy 1999).  More importantly, an analysis of a larger sample of
{\it YOHKOH} flares (Petrosian, Donaghy \& McTiernan 2002) has shown that the LT emission 
is a common property of all flares, and a preliminary investigation of {\it RHESSI} 
data appears to confirm this picture (Liu et al. 2003).  Finally, a third and equally 
important evidence in support of the SA model comes from the spectra and relative 
abundances of the flare accelerated protons and other ions observed at 1 AU from the Sun 
(Mazur et al. 1992; Reames et al. 1994; Miller 2003).  These several independent lines of 
arguments have established the SA as the leading model for solar flares.  This may be also 
true in many other astrophysical nonthermal sources.  Thus, a more detailed investigation 
of the SA model and its comparison with observations are now fully warranted.

In particular, the SA of electrons on the one hand and protons and other ions on
the other are investigated separately, a unified treatment and comparison with the total
nonthermal radiative signatures of all species have not been carried out yet.  The purpose
of this investigation is to obtain the relative acceleration of electrons and protons from 
the thermal backgrounds of solar flare plasmas with the same spectrum of turbulence.  We
will present some general results of the model and qualitative comparisons with
observations.  More detailed comparisons with observations and the acceleration of other
ions, such as the anomalous overabundance of the flare accelerated $^3$He, will be
addressed in subsequent papers.  Specifically, we will address the energy partition between
the flare accelerated electrons and protons.  Observationally, in some flares, or during
the earlier impulsive phase of most flares, there is little evidence for gamma-ray lines
and therefore proton acceleration.  These are called electron dominated cases.  In the
majority of solar flares the energy partition favors electrons but there are a significant
fraction of flares where more energy resides in protons than in electrons in their
respective observed energy bands.  The ratio of energy of the observed electrons (with
$>20$ keV range) to that of protons (with $>1$ MeV range) in solar flares varies
approximately from 0.03 to 100 (see e.g. a compilation by Miller et al.  1997).  In what
follows, we will use solar flare plasma conditions but the formalism described here will be
applicable to other astrophysical sources.

In \S\ \ref{SA} we describe the general theory of SA and argue that in most cases, the
Fokker-Planck (F-P) equation can be reduced to the diffusion-convection equation with the 
corresponding coefficients given by pitch angle averaged combinations of the F-P 
coefficients.  In \S\ \ref{cep}, we study the resonant interactions of electrons and 
protons with parallel propagating waves in a pure hydrogen plasma, and calculate the 
resultant F-P coefficients and acceleration parameters for interactions with a power law 
turbulence spectrum of the wavenumber.  The new and surprising result here is that the 
proton acceleration is suppressed by a barrier in its acceleration rate in the intermediate 
energy range.  This is what is required by observations of electron dominated cases, but as 
we will show this barrier is too strong and makes the acceleration of protons unacceptably 
inefficient relative to the electron acceleration, except for in very weakly magnetized 
plasmas.  In \S\ \ref{ephe4} we point out that this difficulty can be alleviated by a more 
complete description of the dispersion relation which includes the effects of helium ions 
and by an inclusion of the effects of the thermal damping of the turbulence at high 
wavenumbers.  The presence of an appropriate amount of fully ionized helium introduces an 
extra wave branch which lowers the barrier and the thermal damping steepens the turbulence 
spectrum toward high wavenumbers, making the acceleration of electrons and protons more in 
agreement with observations.  The results presented here are summarized in \S\ \ref{discs} 
and their applications to solar flare observations are discussed qualitatively.  Some 
useful approximate analytical expressions for the interaction rates are presented in the 
appendix.

\section{GENERAL THEORY OF STOCHASTIC ACCELERATION}
\label{SA}

In this section, we present the general theory of SA and show that in most astrophysical 
situations, the diffusion-convection equation is adequate to address the particle 
acceleration processes.

\subsection{Fokker-Planck Equation}

The study of SA in a magnetized plasma starts from the collisionless Boltzmann-Vlasov 
equation and the Lorentz force (Schlickeiser 1989).  In the quasilinear approximation, it 
can be treated by the F-P equation (e.g. Kennel \& Engelmann 1966): 
\begin{equation}
{\partial f\over \partial t} +v{\partial f\over \partial s}=
{1\over p^2}{\partial\over\partial p}p^2\left[D_{pp}{\partial f\over\partial p} 
+
D_{p\mu}{\partial f\over\partial \mu}\right]+{\partial \over\partial 
\mu}\left[D_{\mu\mu}{\partial f\over\partial \mu} + 
D_{\mu p}{\partial f\over\partial p}\right]
-{1\over p^2}{\partial\over \partial p}(p^2\dot{p}_Lf)
+S,
\label{FPeq}
\end{equation}
where the wave-particle interaction is parameterized by the F-P coefficients 
$D_{ij}[i,j\in(\mu, p)]$.  Here $f(t, s, p, \mu)$ is the gyro-phase averaged particle 
distribution, $s$, $v$, $\mu$ and $p$ are the spatial coordinate along the field lines, 
the velocity, the pitch angle cosine, and the momentum of the particle, respectively. 
The energy loss (minus systematic energy gains, if any) processes are accounted for by 
$\dot{p}_L$ and $S$ is the source function. 

For weak turbulence ($\delta B\ll B$), as is the case for solar flares, the F-P
coefficients can be evaluated by assuming that the particles and waves are coupled via a
resonant process.  The acceleration of particles at a given energy is then dominated by
interactions with certain specific wave modes, e.g. the Alfv\'{e}n or Whistler waves.  For 
a study of acceleration in a narrow energy band it is usually sufficient to consider
waves in a narrow frequency range (Miller \& Ramaty 1987).  In order to address the energy 
partition between electrons and ions, however, one has to calculate the particle 
acceleration over the whole energy range.  For example, the Alfv\'en waves can efficiently 
accelerate ions but not nonrelativistic electrons.  Models with pure Alfv\'{e}nic 
turbulence are not adequate to address the energy partition of accelerated particles in 
solar flares and many other astrophysical plasmas, especially for the acceleration of 
particles from a thermal background.  For the acceleration of such low energy particles 
interactions with turbulent plasma waves extending over a broad range of wavenumbers (and 
frequencies) must be considered.  We consider here interactions with a broad spectrum of 
plasma waves propagating along a static background magnetic field.  The interactions of 
parallel propagating waves with electrons are described fully by Dung \& Petrosian (1994)  
(DP94) and Pryadko \& Petrosian (1997) (PP97) (see also Steinacker \& Miller 1992).  We 
will use their formalism and evaluate the relative rates of interaction and acceleration of 
electrons and protons in cold but fully ionized plasmas.

\subsection{Dispersion Relation and Resonance Condition} 
\label{dis}

Waves propagating parallel or anti-parallel to the large scale magnetic field in a uniform
plasma have two normal modes that are polarized circularly (Sturrock 1994).  Because their
electric fields are perpendicular to their corresponding wave vectors, the waves are also 
referred to as transverse waves.  The dispersion relation for these waves in a cold plasma 
is (see DP94)
\begin{equation}
(kc)^2=\omega^2\left[1-\sum_{i}{\omega_{{\rm 
p}i}^2\over\omega(\omega+\Omega_{i})}\right]\,,
\label{dispg}
\end{equation} 
where $\omega_{{\rm p}i}=\sqrt{4\pi n_{i} q_{\rm i}^2/m_{i}}$ and
$\Omega_{i}=(q_{i}B_{0})/(m_{i}c)$ are respectively the plasma frequencies and the 
nonrelativistic gyro-frequencies of the background particles (with charges $q_{i}$, masses 
$m_{i}$ and number densities $n_i$).  $B_{0}$ stands for the large 
scale magnetic field, $c$ is the speed of light, $\omega$ and $k$ are the wave 
frequency and wavenumber, respectively. 

One of the important parameters characterizing a 
magnetized plasma is the ratio of the electron plasma frequency to the 
electron nonrelativistic gyro-frequency:
\begin{equation}
\alpha = \omega_{\rm pe}/\Omega_{\rm e}
= 3.2 (n_{\rm e}/10^{10}{\rm cm}^{-3})^{1/2}(B_{0}/100\,{\rm G})^{-1}\,,
\label{alpha}
\end{equation}
where $\Omega_{\rm e} = (eB_0)/(m_{\rm e}c)$ and $e$ and $m_{\rm e}$ are the elemental 
charge unit and the electron mass, respectively. The value of $\alpha$ is small for a
strongly magnetized plasma.

A particle with a velocity $\beta c$ (Lorentz factor $\gamma$) and a pitch angle cosine 
$\mu$ interacts most strongly with waves satisfying the resonance condition:  
\begin{equation} 
\omega - k_{||}\beta \mu = {n\omega_{i}\over \gamma}\,, 
\label{reson} 
\end{equation} 
where $n$ is for the harmonics of the gyro-frequency (not to be confused with the background 
particle number densities $n_i$), $\omega$ and $k_{||}$ are the wave frequency and the 
parallel component of the wave vector in units of $\Omega_{\rm e}$ and $\Omega_{\rm e}/c$, 
respectively (we will use these units in the following discussion unless specified 
otherwise and in our case $k_{||}=k$ and $n=-1$), $\omega_{i} =q_{i}m_{\rm e}/em_{i}$ 
is the particle gyro-frequencies in units of $\Omega_{\rm e}$; $\omega_{\rm e} = -1$ for 
electrons and $\omega_{\rm p} = \delta\equiv m_{\rm e}/m_{\rm p}$ for protons, where 
$m_{\rm p}$ is the proton mass (for more details see also DP94).  
One notes that low energy particles mostly resonate with waves with high wavenumbers 
and only relativistic particles interact with large scale waves with low frequencies.  The 
resonant wave-particle interaction can transfer energy between the turbulence and particles 
with the details depending on the particle distribution and the spectrum and 
polarization of the turbulence.

\subsection{Fokker-Planck Coefficients}
\label{fp}

The evaluation of the F-P coefficients requires a knowledge of the spectrum of the
turbulence.  Following previous studies (DP94; PP97) we will first assume a power law
distribution of unpolarized turbulent plasma waves.  For unpolarized turbulence, the
amplitudes of the waves only depend on $k$.  Then we have ${\cal E}(k) = (q-1){\cal E}_{\rm
tot} k_{\rm min}^{q-1}k^{-q}$ for $k > k_{\rm min}$ ({\it i.e.} a large scale cutoff),
where the turbulence spectral index $q>1$.  For a given turbulence energy density ${\cal 
E}_{\rm tot}$, $k_{\rm min}$, presumably larger than the inverse of the size of the 
acceleration region, determines the maximum energy that the accelerated particles can reach 
and the characteristic time scale of the interaction.  The general features of this 
situation have been explored in the papers cited above.  For the sake of completeness, we 
briefly summarize the key results here.

The F-P coefficients can be written as 
\begin{equation}
D_{ab}={(\mu^{-2}-1)\over\tau_{{\rm p}i}\gamma^2}\sum_{j=1}^N\chi(k_j)
\cases{\mu\mu (1-x_j)^2, &  for $ab=\mu\mu$; \cr
\mu p x_j(1-x_j), & for $ab=\mu p$; \cr
p^2 x_j^2, & for $ab=pp$, \cr} 
\label{coeff}
\end{equation}
where
\begin{equation}
\chi(k_j) = {|k_j|^{-q}\over |\beta \mu-\beta_{\rm g}(k_j)|}\,\,\,\,\,\,\,\,\,\, 
{\rm and}\,\,\,\,\,\,\,\, x_j=\mu\omega_j/\beta k_j\,.
\label{chi}
\end{equation}
The sum over $j$ is for the resonant interactions discussed in the previous section.  The 
characteristic interaction time scale for each of the charged particle species is 
$\tau_{{\rm p}i}=\tau_{\rm p}/\omega_{i}^2$ with that for electrons given by (see DP94):
\begin{equation}
\tau_{\rm p}^{-1} = {\pi\over 2}\Omega_{\rm e}\left[{{\cal E}_{\rm tot}\over 
B_0^2/8\pi}\right](q-1)k_{\rm min}^{q-1}\,.
\label{taup}
\end{equation}

In general, the F-P coefficients have complicated dependence on the turbulence spectral
index $q$, the plasma parameter $\alpha$ and the energy and pitch angle of the particles.  
The exact solution of the full F-P equation is a difficult task.  Fortunately under certain
conditions considerable simplifications are possible.  These conditions are defined by the
relative values of the three F-P coefficients.  The pitch angle change rate of the
particles is proportional to $D_{\mu\mu}$, while the momentum or energy change rate is
proportional to $D_{pp}/p^2$.  As evident from equation (\ref{coeff}) the behavior of
$D_{\mu p}/p$ is intermediate between the two.

\subsection{Diffusion-Convection Equation}
\label{dc}

The relative values of the F-P coefficients determine the type of approximations that 
can be used for solving the F-P equation.  We now show that for most conditions reasonable 
approximations lead to the well known transport equation (eq. [\ref{dceq}]).  In order to 
justify these approximations it is convenient to define two ratios of the coefficients: 
\begin{eqnarray} 
R_1(\mu, p)& = & {D_{pp}\over p^2D_{\mu\mu}}\,,\label{r1}\\ 
R_2(\mu, p)& = & {D_{p\mu}\over pD_{\mu\mu}}\,.
\label{r2} 
\end{eqnarray} 
We will show in the following sections, for most energies and pitch angles both $R_1 \, 
{\rm and}\,  |R_2|\ll 1$, which means that $D_{\mu\mu} \gg D_{pp}/p^2$.  Under these 
conditions the particles are scattered frequently before being significantly accelerated 
and the accelerated particle distribution is nearly isotropic.  Then the 
pitch angle averaged particle distribution function $F(s,t,p)= 0.5\int_{-1}^1\d\mu f(\mu, 
s, t, p)$ satisfies the well known diffusion-convection equation (see e.g. Kirk,
Schneider \& Schlickeiser 1988; DP94; PP97). 

In this study we are interested in the relative acceleration of electrons and protons which 
is not sensitive to the detailed geometry or the inhomogeneities of the source.  
Therefore we can assume a homogeneous and finite (size $L$) source, or alternatively 
confine our discussion to spatially integrated spectra. In this case we can treat the 
spatial diffusion or advection of the particles by an energy dependent escape term.  Then 
the above mentioned equation is reduced to
\begin{equation}
{\partial N\over\partial t}= {\partial^2\over \partial E^2}(D_{EE} N) + 
{\partial\over 
\partial E}[({\dot E}_{\rm L}-A) N] -{N\over T_{\rm esc}} + Q\,,
\label{dceq}
\end{equation}
where $E=(\gamma-1)m_i c^2$ is the particle kinetic energy, $N(t,E)\d E=4\pi p^2\d 
p\int_0^LF(s,t,p)\d s$, ${\dot E}_{\rm L}$ describes the net systematic energy loss, and 
$Q(t,E)=0.5\int_{-1}^{1}\d\mu \int_0^L S(s,\mu,t,E)\d s$ is the total injection flux of 
particles into the acceleration region.  $D_{EE}$ describing the diffusion in energy is 
related to $D_{pp}$ and defines the acceleration time, and $T_{\rm esc}$ is related 
to the scattering time $\tau_{\rm sc}$:
\begin{eqnarray}
T_{\rm esc}&=& (L^2/v^2)/\tau_{\rm sc}\,,\,\,\,\,\,\,\, \tau_{\rm sc} = 
{1\over 2}\int_{-1}^1\d\mu{(1-\mu^2)^2\over D_{\mu\mu}} \ll 
L/v\,,\label{scat}\\
\tau_{\rm ac}&=& E^2/D_{EE}\,,\,\,\,\,\,\,\,\, D_{EE}={E^2\over 2} 
\int_{-1}^1\d\mu 
D_{\mu\mu}(R_1-R_2^2)\,.
\label{accel} 
\end{eqnarray}
Note that equation (\ref{dceq}) describes the energy diffusion with two terms, $D_{EE}$ 
and the direct acceleration rate:
\begin{equation}
A(E) ={1\over \beta\gamma^2}{\d \beta\gamma^2 D_{EE}\over \d E}= {\d D_{EE}\over \d E}+ 
{D_{EE}\over E} {2-\gamma^{-2}\over1+\gamma^{-1}}\,.
\label{ae}
\end{equation}

There are several important features in the diffusion coefficients which we emphasize here.

1) The first is that in the {\it extremely relativistic limit} the diffusion coefficients 
(and their ratios) for protons and electrons are identical and assume asymptotic values 
such that both of the ratios are much less than one.  Therefore equations  (\ref{dceq}), 
(\ref{scat}) and (\ref{accel}) are valid.  (Strictly speaking, this is not true for very 
strongly magnetized plasmas $\alpha\le \delta^{1/2}$ where one gets 
$R_1\sim|R_2|\sim 1$. See eq. [\ref{coeff}])

2) The second is that {\it at low energies}, as pointed out by PP97, $R_1$ and $R_2^2$ are 
not necessary less than one, especially for plasmas with low values of $\alpha$. In the 
extreme case of $R_1\gg |R_2|\gg 1$, three of the four diffusion terms in equation 
(\ref{FPeq}) can be ignored.  Again, if we assume a finite homogeneous region, or integrate 
over a finite inhomogeneous source, the resultant equation becomes similar to equation 
(\ref{dceq}).  Now because of the lower rate of pitch angle scatterings, the escape time 
may be equal to the transit time $T_{\rm esc}\sim L/(v\mu)$, the other transport 
coefficients $D_{EE}$ and ${\dot E}_{\rm L}$ (and consequently the accelerated particle 
spectra) may depend on the pitch angle, and the assumption of isotropy may not be 
valid.  However, as can be seen in the next section (Figures \ref{fig5.ps} and 
\ref{fig6.ps}), these coefficients change slowly with $\mu$, except for some negligibly 
small ranges of $\mu$, so that the expected anisotropy is small. In addition, at lower 
energies Coulomb scatterings become increasingly important and can make the
particle distribution isotropic.  In many cases, especially for plasmas not completely 
dominated by the magnetic field ({\it i.e.} for $\alpha\geq 1$) one can neglect the small 
expected anisotropy and integrate the equation over the pitch angle, in which case the 
transport equation becomes identical to equation (\ref{dceq}) except now
\begin{eqnarray} 
T_{\rm esc}&=& L/\sqrt{2}v \ll\tau_{\rm sc} \sim 
<1/D_{\mu\mu}>\,,\label{esc}\\ 
\tau_{\rm ac}&=& {2 p^2\over \int_{-1}^1\d\mu D_{pp}(\mu)}\,, \label{accel2} 
\end{eqnarray} 
where ``$<>$'' denotes averaging over the pitch angle.

3) It is easy to see that one can combine the above two sets of expressions (eq. 
[\ref{scat}]-[\ref{accel2}]) for the acceleration rates (or time scales) and the escape 
times at the nonrelativistic and extremely relativistic cases as
\begin{equation} 
T_{\rm esc} = {L\over \sqrt{2}v}\left(1 + {\sqrt{2} L \over v\tau_{\rm 
sc}}\right), \,\,\,\,\,\, \tau_{\rm ac} = E^2/D_{EE}, 
\label{escall} 
\end{equation} 
and
\begin{equation} 
D_{EE}={E^2\over 2}\int_{-1}^1\d\mu D_{\mu\mu} 
\cases{R_1\,, &if $R_1\gg |R_2| \gg 1$;\cr 
R_1-R_2^2\,, &if $R_1, \, |R_2| \ll 1$\,.\cr} 
\label{tacclall} 
\end{equation} 
The first expression in equation (\ref{tacclall}) is valid at low values of $E$ and 
$\alpha$ and the second at higher energies and in weakly magnetized plasmas.  However, it 
turns out that at extremely relativistic energies and in weakly magnetized plasmas 
($\alpha>1$), independent 
of other conditions, $R_2^2\ll R_1$ and the first expression can be used.  These 
expressions and equation (\ref{dceq}) then describe the problem adequately for most 
purposes in high energy astrophysics, in particular for solar flares, the focus of 
this paper.

4) Finally, in certain cases, especially in the intermediate energy range the quantity 
$R_1-R_2^2$ appearing in equations (\ref{accel}) and (\ref{tacclall}) can be 
small.  The acceleration rate can be reduced dramatically when both $R_1$ and $|R_2|$ are 
much less than one and $R_1\simeq R_2^2$. From the definitions of these ratios and 
expressions for the F-P coefficients (eqs. [\ref{r1}], [\ref{r2}] and [\ref{coeff}]) 
it is clear that if there were only one resonant interaction one would have $R_1=R_2^2$ and 
there would be no acceleration. Thus, strictly speaking the use of equation (\ref{dceq}) 
with interactions involving only one wave mode (say the Alfv\'en) is incorrect.  However, 
as we will show in \S\ \ref{dispep} there are always at least two resonant interactions in 
unpolarized turbulence, in which case $R_1\not= R_2^2$ so that the acceleration rate is 
finite.  But if one of the interactions is much stronger than the others, $R_1-R_2^2$ can 
be small.  In the next section, we will show some examples where this is true (Figure 
\ref{fig5.ps}) and that this happens at the intermediate values of energy (Figure 
\ref{fig6.ps}).  The acceleration rate is then reduced greatly.  The much lower 
acceleration rate at the intermediate energies compared to the higher rates in the 
nonrelativistic and extremely relativistic limits introduces an acceleration barrier. As we 
shall see that in the intermediate energy range the behaviors of protons and electrons are 
quite different and a much stronger acceleration barrier appears for protons.

\subsection{Loss Rate}

To determine the distributions of the accelerated protons and electrons by solving equation 
(\ref{dceq}) with the above formalism, in addition to the transport 
coefficients $D_{EE}$, $A$ and $T_{\rm esc}$, we need to specify the loss term 
$\dot{E}_{\rm L}$.  For electrons the loss processes are dominated by Coulomb collisions at 
low energies and by synchrotron losses at high energies:
\begin{equation}
\dot{E}_{\rm Le} = 4r_0^2 m_{\rm e} c^3[\pi n_{\rm e}\ln{\Lambda}/\beta + 
B_0^2\beta^2\gamma^2/9m_{\rm e}c^2]\,,
\end{equation}
where $r_0=2.8\times10^{-13}$ cm is the classical electron radius and $\ln{\Lambda} = 20$ is 
a reasonable value in our case (See Leach 1984).  The ion losses in a fully ionized plasma 
are mainly due to Coulomb collisions with the background electrons and protons (Post 1956; 
Ginzburg \& Syrovatskii 1964).  For electron-ion collisions, we have
\begin{equation}
\dot{E}_{{\rm L}i} =  2\pi r_0^2 m_{\rm e}c^3 n_{\rm e}(q_i/e)^2\cases{
2\sqrt{6/\pi} \beta^2\beta_{\rm Te}^{-3} \ln\Lambda
& for $\beta<\beta_{\rm Te}$\,; \cr
\beta^{-1}\ln{\left({m_{\rm e}^2c^2\beta^4\over \pi r_0 n_{\rm e} \hbar^2}\right)}
& for $1\gg \beta>\beta_{\rm Te}$ \,;\cr
\ln{\left({m_{\rm e}^2c^2\gamma^2\over 2\pi r_0 n_{\rm e} \hbar^2}\right)}
& for $m_i/m_{\rm e}\gg\gamma\gg 1$ 
\,;\cr
\ln{\left({m_{\rm e}m_ic^2\gamma\over 4\pi r_0 n_{\rm e} \hbar^2}\right)}             
& for $\gamma\gg m_i/m_{\rm e}$
\,,
}
\end{equation}
where $\beta_{\rm Te}=(3k_{\rm b}T_{\rm e}/m_{\rm e}c^2)^{1/2}$ is the mean thermal 
velocity of the background electrons in units of $c$ and $k_{\rm b}$ is the Boltzmann 
constant.  For proton-ion collisions, which are important for ions with even lower 
energies, we have (Spitzer 1956)
\begin{equation}
\dot{E}_{{\rm L}i} =4\pi r_0^2 m_{\rm e} c^3n_{\rm 
p}(q_i/e)^2(m_{\rm e}/m_p)\beta^{-1}\ln{\Lambda}\,\ \ \ {\rm for}\ \ \ 
\beta>\beta_{\rm Tp}.
\end{equation}
where $\beta_{\rm Tp}=(3k_{\rm b}T_{\rm p}/m_{\rm p}c^2)^{1/2}$ is the mean thermal
velocity of the background protons in units of $c$.
These loss processes dominate at different energies and we can define a loss time 
$\tau_{\rm loss}=E/\dot{E}_{\rm L}$.

\subsection{Steady State Solution and Normalization}

We will use the impulsive phase conditions of solar flares for our demonstration.  In this 
case, we can assume that the system is in a steady state because the relevant time scales 
are shorter than the dynamical time (the flare duration).  We also assume the presence of a 
constant spectrum of turbulence.  We are interested in the acceleration from a thermal 
background plasma, therefore, a thermal distribution is assumed for the source term $Q$.  
As described above, equation (\ref{dceq}) may not be valid at low (keV) energies where 
$R_1\gg1$.  However, for solar flare conditions and in the keV energy range, 
Coulomb scatterings become important ($D_{\mu\mu}^{\rm Coul}\gg D_{\mu\mu}^{\rm wave}$, see 
Hamilton \& Petrosian 1992).  In this case $R_1\ll 1$ and the particle distribution will be 
nearly isotropic at all energies.  We therefore calculate the acceleration rate with the 
second expression of equation (\ref{tacclall}) and solve equation (\ref{dceq}) to get the 
distributions of the accelerated particles over all energies.

To appreciate the relevant physical processes, one can compare the acceleration time with
the escape and the loss time.  We are mostly interested in the energy range above the
energy of the injected particles.  So the source term is not as important in shaping the
spectrum as the other terms.  In the energy band where the escape and loss terms are
negligible, from the flux conservation in the energy space, one can show that $AN-\d
(D_{EE}N)/\d E = constant$.  On the other hand, when the acceleration terms are negligible,
no acceleration occurs.  When the escape time becomes much shorter than the acceleration
time and both of them are much shorter than the loss time, particles escape before being
accelerated.  This results in a sharp cutoff in the particle distribution at the energy
where $T_{\rm esc}\simeq E/A(E)\sim E^2/D_{EE}$.  When the escape time is long and the loss
time is much shorter than the acceleration time, one would then expect a quasi-thermal
distribution for the Coulomb collisional losses (Hamilton \& Petrosian 1992) and a sharp
high energy cutoff for the synchrotron losses (Park, Petrosian \& Schwartz 1997).  
Power-law distributions can be produced only in energy ranges where the loss term is small
and the acceleration and escape times have similar energy dependence.

The normalization of the steady state particle distributions is determined by their rates
of acceleration, escape and injection. The injection rates depend on the geometries of the
reconnection and the turbulent acceleration site and on possible contributions of the
charged particles to reverse currents which must exist when a net charge current leaves
the acceleration site. A more detailed time dependent treatment is required to
determine the relative normalization. This is beyond the scope of the paper and will be
dealt with in the future. Here we concentrate on the relative shapes of the electron and
proton spectra in the LT and FP sources. We will assume that the injection flux $\int Q\d
E= 1$ s$^{-1}$ cm$^{-2}$ for both electrons and protons (see also \S\ \ref{discs}).  In the
steady state this is equal to the flux of the escaping particles $N_{\rm esc}^{\rm
tot}=\int_0^\infty N_{\rm LT}(E)/T_{\rm esc}(E)\d E$.  Since the escaping particles lose
most of their energy at the FPs, instead of $N_{\rm esc}(E)=N_{\rm LT}(E)/T_{\rm esc}(E)$
we will show the effective particle distribution for a thick target (complete cooling) FP
source, which is related to the corresponding LT distribution $N_{\rm LT}$ via (Petrosian 
\& Donaghy 1999): 
\begin{equation} 
N_{\rm FP}(E)={1\over \dot{E}_{\rm L}}\int_E^\infty {N_{\rm
LT}(E')\over T_{\rm esc}(E')}\d E'\,.  
\label{Nfp} 
\end{equation}

\section{APPLICATION IN COLD HYDROGEN PLASMAS}
\label{cep}

In this section we describe the relative acceleration of electrons and protons in cold,
fully ionized, pure hydrogen plasmas.  This is an approximation because all astrophysical
plasmas contain some helium and traces of heavy elements.  Ignoring the effects of helium
(trace elements will, in general, have no influence on the following discussion) and
adopting a turbulence spectrum of a single power law of the wavenumber simplify the
mathematics and allow us to demonstrate the differences between the acceleration rates
of electrons and protons more clearly.  Moreover, in some low temperature plasmas, most of
the helium may be neutral and not be involved in the SA processes.  The results presented
here will be a good approximation.  Pure hydrogen plasmas can also be realized in
terrestrial experiments to test the theory.  The formalism can also be easily generalized
to the case of electron-positron plasmas and to more complicated situations.  In the next
section we will present our results for plasmas including about 8\% by number of helium and
for turbulence with a more realistic spectrum.

\subsection{Dispersion Relation and Resonant Interactions}
\label{dispep}

In a pure hydrogen plasma, equation (\ref{dispg}) reduces to (PP97)
\begin{equation} {k^2\over \omega^2} = 1 - 
{\alpha^2(1+\delta)\over(\omega-1)(\omega+\delta)}\,,
\label{disp}
\end{equation}
and the Alfv\'{e}n velocity in units of c is given by $\beta_{\rm A}=\delta^{1/2}/\alpha$. 
(For e$^{\pm}$ pair dominated plasmas $\delta = 1$).

\begin{figure}[htb]
\begin{center}
\includegraphics[height=8.4cm]{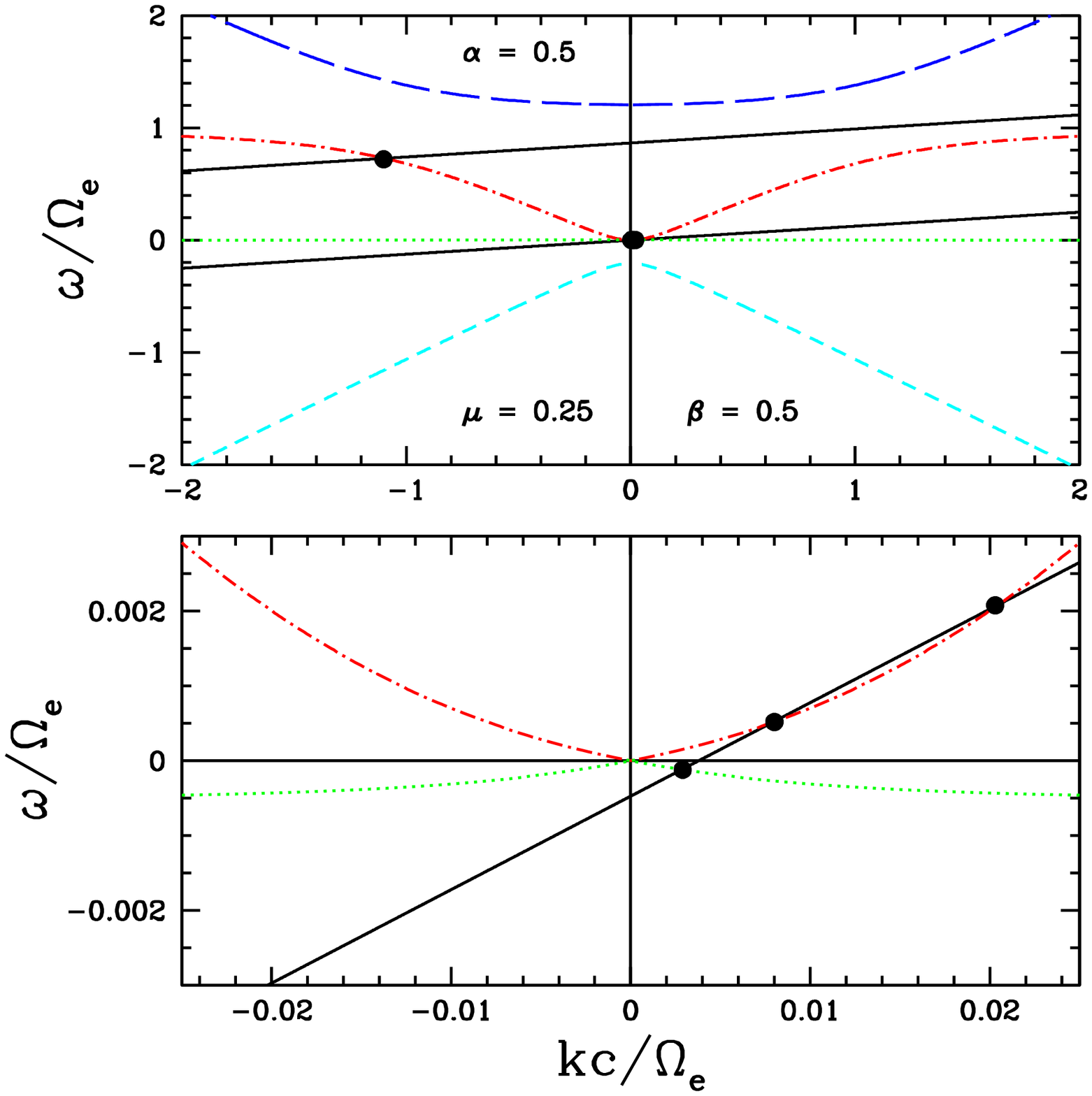}
\hspace{-0.6cm}
\includegraphics[height=8.4cm]{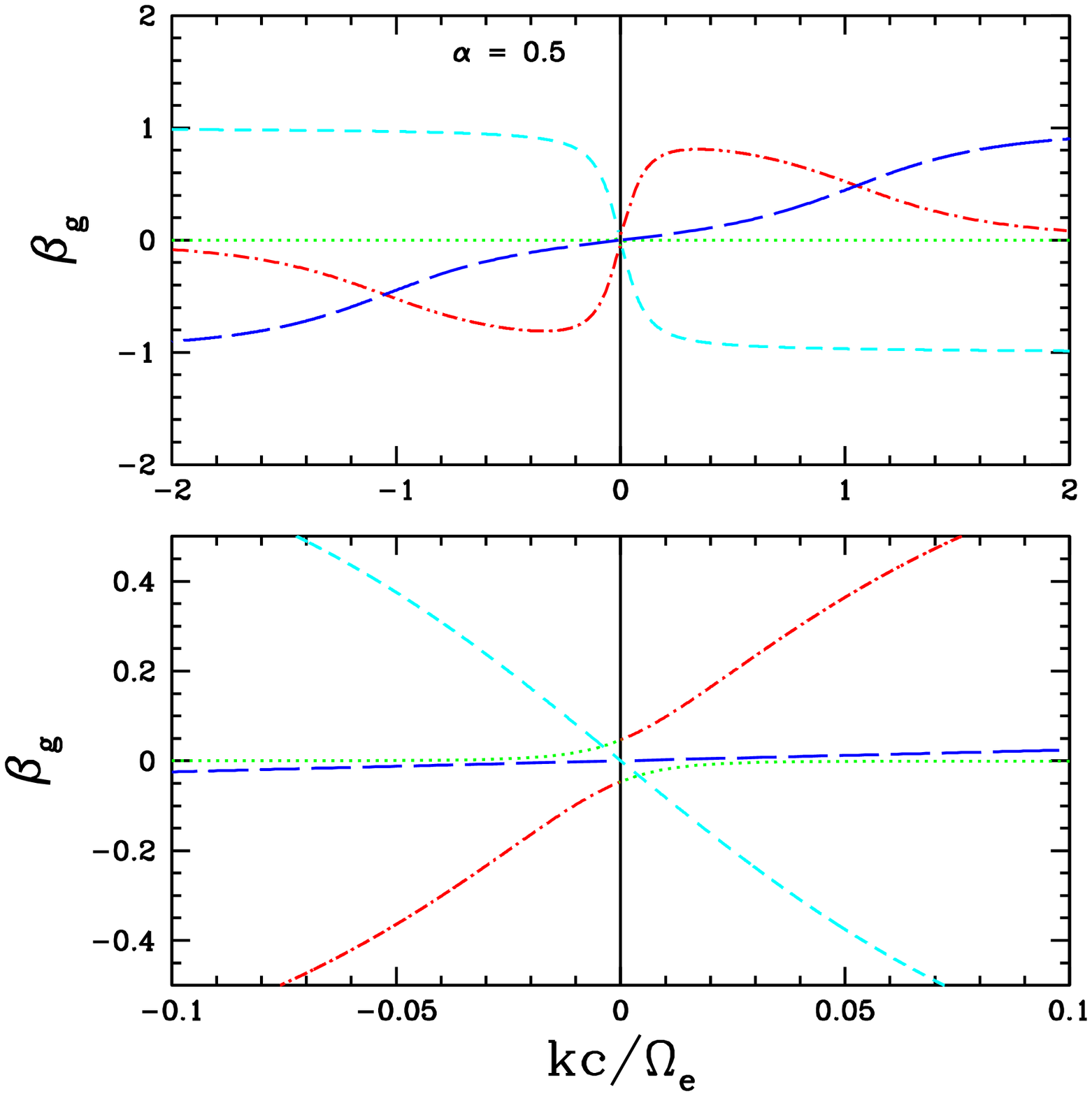}
\end{center}
\caption{ \small
{\bf a) Left panel:} Dispersion relation of parallel propagating waves in a cold pure 
hydrogen plasma with $\alpha=0.5$. The bottom panel is an enlargement of the region around 
the origin. The curves, from top to bottom, describe the EM (long-dashed), EC (dot-dashed), 
PC (dotted) and EM' (short-dashed)  waves. The upper and lower solid lines give 
respectively the resonance conditions for electrons and protons with $v = 0.5c$ 
($\beta = v/c$) and $\mu = 0.25$. Resonant interactions occur at the points where these 
lines cross the curves which depict the waves.   
{\bf b) Right panel:} Same as the left panel but for the group velocity $\beta_{\rm 
g}=\d\omega/\d k$ versus the wavenumber $k$. The line type remains the same for each wave
branch. Negative group velocities mean that the energy fluxes of the waves are in the
direction anti-parallel to the large scale magnetic field.
}
\label{fig1.ps}
\end{figure}

Figure \ref{fig1.ps}a depicts the normal modes of these waves, which compose four distinct 
branches.  From top to bottom, we have the electromagnetic wave branch (EM; long dashed), 
electron-cyclotron branch (EC; dot-dashed), proton-cyclotron branch (PC; dotted), and a 
second electromagnetic wave branch (EM'; short dashed), respectively.  The lower panel is 
an enlargement of the region near the origin.  The positive and negative frequencies mean 
that the waves are right- and left-handed polarized, respectively, where the polarization 
is defined relative to the large scale magnetic field (Schlickeiser 2002).  Figure 
\ref{fig1.ps}b depicts the group velocities $\beta_{\rm g} = \d\omega/\d k$ of these waves.  
One may note that the signs of the phase velocity $\beta_{\rm ph}=\omega/k$ and the group 
velocity of a specific wave mode are always the same.

In Figure \ref{fig1.ps}a, the two solid straight lines depict equation (\ref{reson}) for an 
electron (upper) and a proton (lower) with $\beta=0.5$ and $\mu=0.25$.  The 
intersections of these lines with the wave branches satisfy the resonance condition.  
The electron interacts resonantly at the indicated point with the EC branch and at another
point with the PC branch at a high negative wavenumber that lies outside the figure.  The 
proton, on the other hand, not only resonates with one PC wave, but also with 
\underline{three} EC waves (only two of which are seen in the lower panel of the figure).  
As we shall show below, the fact that certain protons can resonate with more than one EC 
wave has significant implications for the overall proton acceleration process.

\clearpage

\clearpage

\subsection{Critical Angles and Critical Velocities} 
\label{cri}

\begin{figure}[h]
\begin{center}
\includegraphics[height=8.4cm]{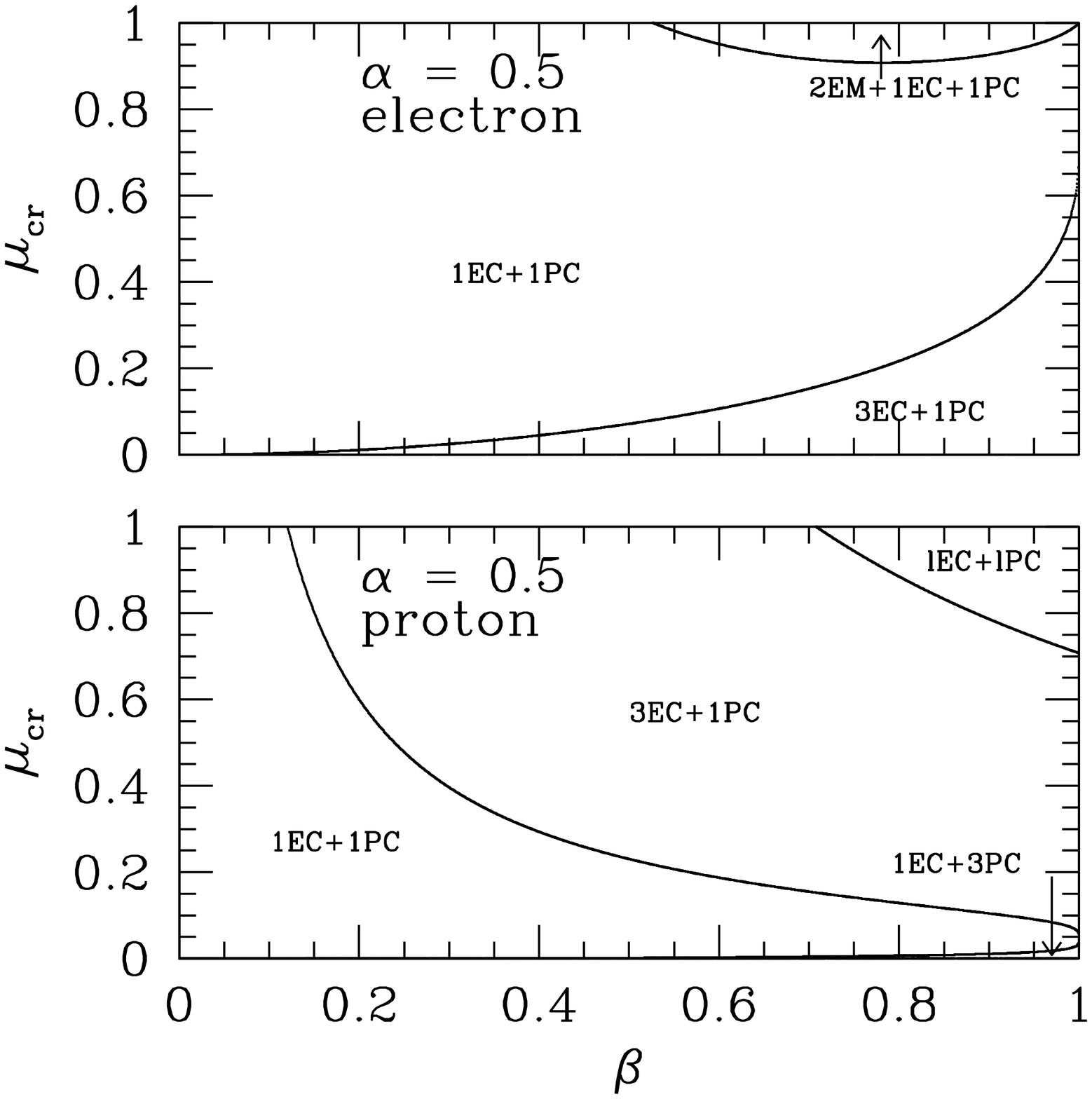}
\hspace{-0.6cm}
\includegraphics[height=8.4cm]{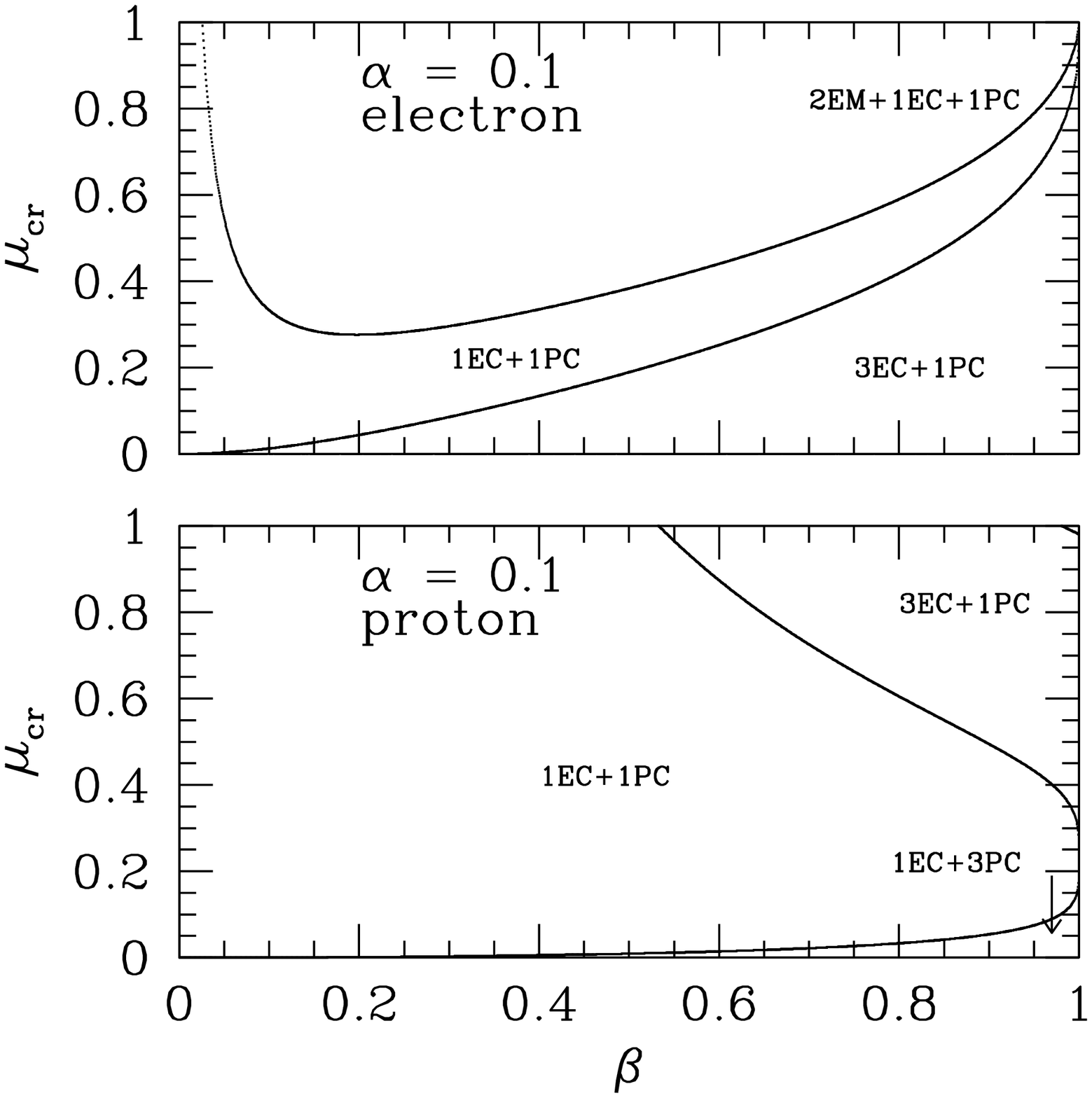}
\end{center}  
\caption{ \small
{\bf Left Panel:} Velocity dependence of the critical angles in a plasma with $\alpha=0.5$ 
for electrons (top) and protons (bottom). At small pitch angles, i.e. $\mu\rightarrow 
1$, electrons can resonate with high frequency electromagnetic waves of the EM branch 
(region labeled with``2EM+1EC+1PC''), while energetic protons mainly interact with 
the Whistler and Alfv\'{e}n waves (region labeled with``3EC+1PC''). {\bf Right Panel:} Same 
as above but for $\alpha = 0.1$ where interactions of protons with the Whistler waves start 
at a higher energy.  See text for details.
}
\label{fig2.ps}
\end{figure}

In general one expects four resonant points.  However, for a given particle
velocity or energy, at {\it critical angles}, where the group velocities of the
waves are equal to the parallel component of the particle velocity, the number of
resonant points can change from four to two or vice verse.  Figure
\ref{fig2.ps} shows the velocity dependence of the critical angles for electrons
(upper panels) and protons (lower panels) in plasmas with $\alpha=0.5$ (left
panels) and $\alpha=0.1$ (right panels).  (The results for electrons are the
same as those given by PP97.)  Both particles have at least two resonant
interactions (one with the PC and one with the EC branch except for $\mu=0$
where electrons interact with two EC waves and protons interact with two PC
waves).

\begin{figure}[htb]
\begin{center}
\includegraphics[height=8.4cm]{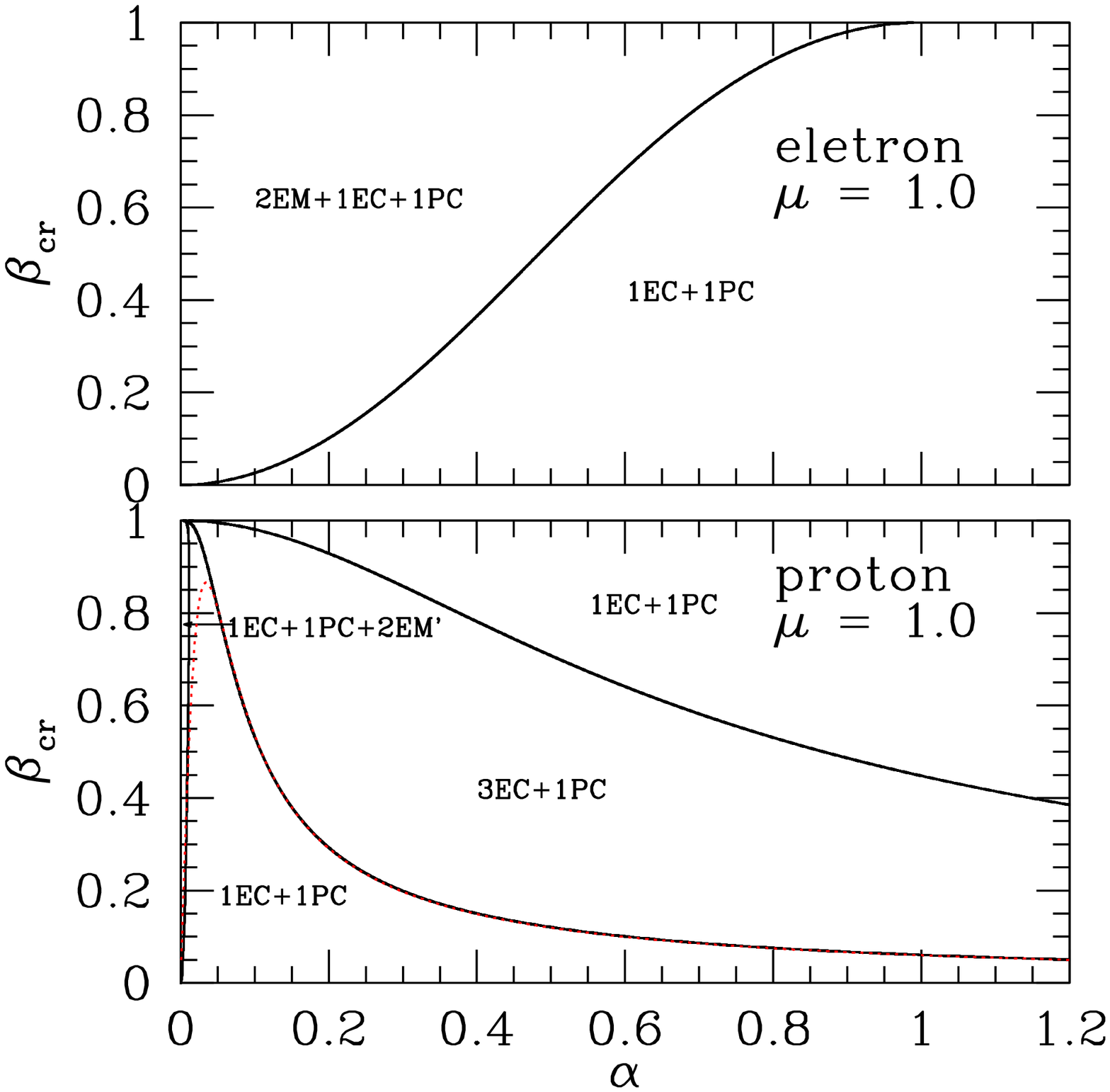}
\hspace{-0.6cm}
\includegraphics[height=8.4cm]{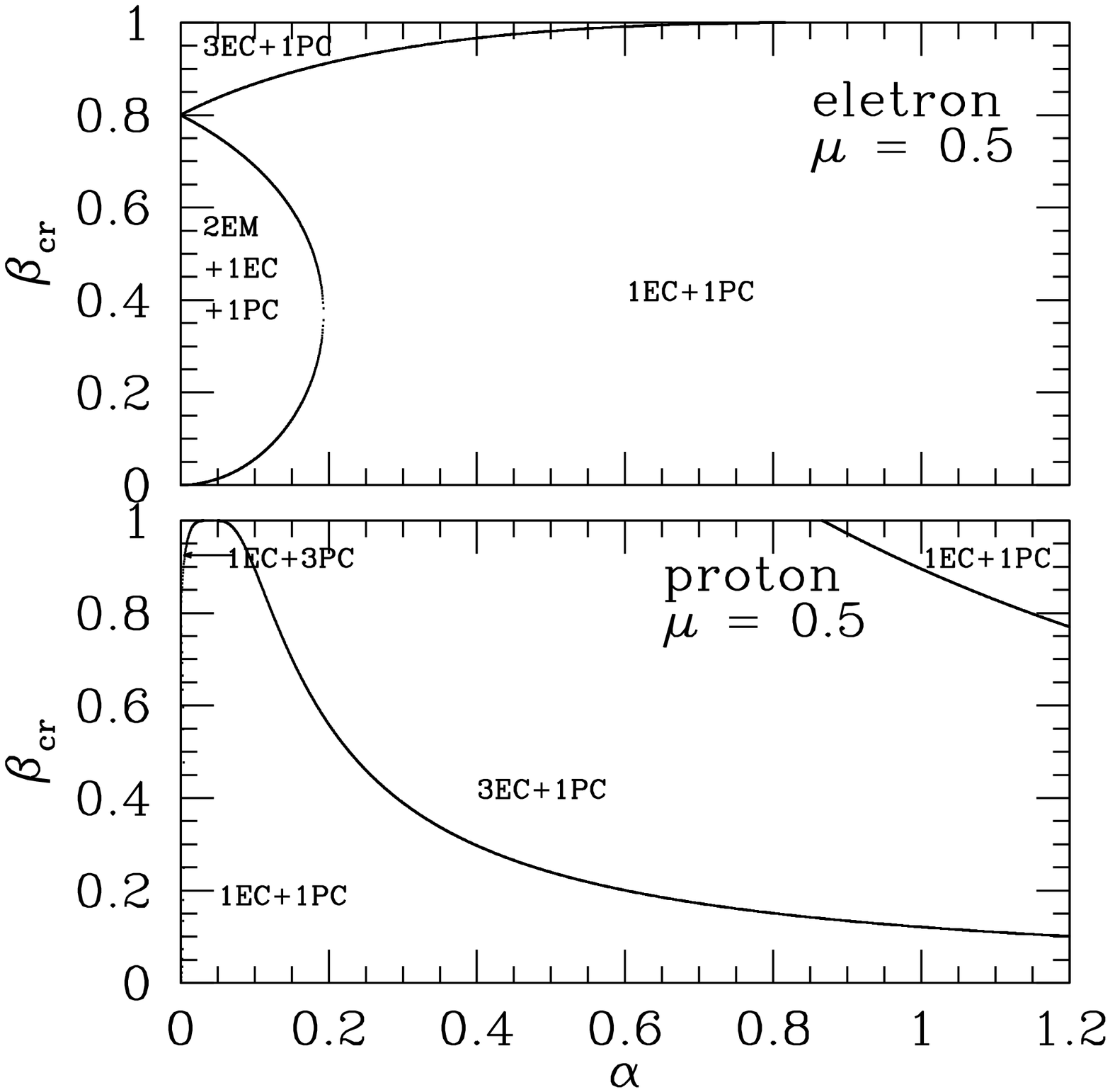}
\end{center}
\caption{ \small
Same as Figure \ref{fig2.ps} but depicting the dependence of the critical 
velocities on the plasma parameter $\alpha$. Combining this with Figure \ref{fig2.ps}, one 
can tell the wave branches responsible for the critical velocities.
}
\label{fig3.ps}
\end{figure}

{\bf Electrons} with a large $\mu$ can have two additional resonances with the EM
branch and those with a small $\mu$ have two additional resonances with the EC
branch.  The two regions with four resonances grow with decreasing $\alpha$ and
shrink as $\alpha$ increases.  For larger values of $\alpha$ the interaction is weaker
because for large ranges of velocities and pitch angles electrons interact with
only two waves (e.g.  the interactions with the EM branch disappear for
$\alpha>1$. See Figures \ref{fig2.ps} and \ref{fig3.ps} ). But as $\alpha$ approaches zero 
the region with two wave interactions diminishes and the two curves for the critical angles  
merge into one, satisfying the relation
\begin{equation}
\mu_{\rm cr}=(\gamma-1)/\beta\gamma\,, \,\,\,\,\, {\rm for}\ \ \ \ \ \alpha\rightarrow 0.
\label{mucr1}
\end{equation}
In the case there are always four resonances and the total interaction is strong at 
all energies.

{\bf Protons} have a similar, but slightly more complicated, behavior.  As $\mu$
increases one obtains interactions with 1EC+3PC, 1EC+1PC, 3EC+1PC and back to
1EC+1PC waves. With the decrease of $\alpha$, the upper two regions diminish,
while the lower portions increase in size. Protons can also be accelerated by the EM' waves
but this only occurs in more highly magnetized plasmas ($\alpha<\delta^{1/2}/2\sim 0.012$) 
as compared with the interactions of electrons with the EM branch.  At such low values 
of $\alpha$ a region with four interactions (1EC+1PC+2EM') appears in the upper portion of 
the $\mu - \beta$ plane and its lower boundary eventually merges with the lower curve for 
$\mu_{\rm cr}$ as $\alpha$ approaches zero. Just like electrons the critical angle is 
given by equation (\ref{mucr1}).  In this limit, particles are basically exchanging energy 
with the Poynting fluxes of the electromagnetic waves.

\clearpage

\clearpage

These behaviors can also be seen in Figure \ref{fig3.ps}, where instead of $\mu_{\rm cr}$
we plot what one may call the {\it critical velocities} as a function of $\alpha$ for two
values of $\mu$. Note that in the proton panel with $\mu=1.0$ there is a small region with
$\alpha<0.012$ where there are four resonances including two with the forward-moving
left-handed polarized electromagnetic waves from the EM' branch. Protons will not resonate
with the electromagnetic waves for larger values of $\alpha$. In general, we have similar
patterns of transition between different regions caused by the electromagnetic branches in
the $\mu-\beta$ space, except that the transitions for protons occur at a value of $\alpha$
which is lower than that for electrons by a factor of $\delta^{1/2}$. The main difference
in the behaviors of electrons and protons resides in their four resonant interactions with
the PC and EC branches. Protons have two such regions where they resonate with ``1EC+3PC''
or ``3EC+1PC'', while electrons only have one with ``3EC+1PC''; electrons never interact
with more than one PC wave. This is where the above scaling symmetry of $\alpha$ between
protons and electrons is broken.

\clearpage

\clearpage

{\it Low Energy Approximations:} Because the acceleration of particles at low energies is
of particular interest we present here some approximate analytic relations, which are 
derived in the appendix. 

The first is for the proton critical velocity curve dividing the region with two and four 
resonances (i.e. the middle curve in the lower left panel of Figure \ref{fig3.ps}). As we 
will see in the following sections, at a given $\alpha$ the acceleration rate (eq. 
[\ref{tacclall}]) increases dramatically once protons attain the critical velocity 
or energy and enter the region with four resonant interactions.  The pitch angle averaged 
acceleration rate also increases sharply above this energy.  It will be useful to have a 
formula to estimate this critical velocity.  We find the following approximate expression 
for this transition
\begin{equation}
\beta_{\rm cr} = {0.06\alpha\over 0.0012+\alpha^2}\ \ \ \ {\rm or} \ \ \ \ E_{\rm 
cr}={1\over 2}m_{\rm p}c^2\beta_{\rm cr}^2 = 1.7{\rm 
MeV}\left({\alpha\over 0.0012+\alpha^2}\right)^2\,, 
\label{betacr}
\end{equation}
which is shown by the dotted curve in the low left panel of Figure \ref{fig3.ps} and agrees 
within $0.2\%$ with the exact result for $\alpha>0.05$.

\begin{figure}[thb]
\begin{center}
\includegraphics[height=8.4cm]{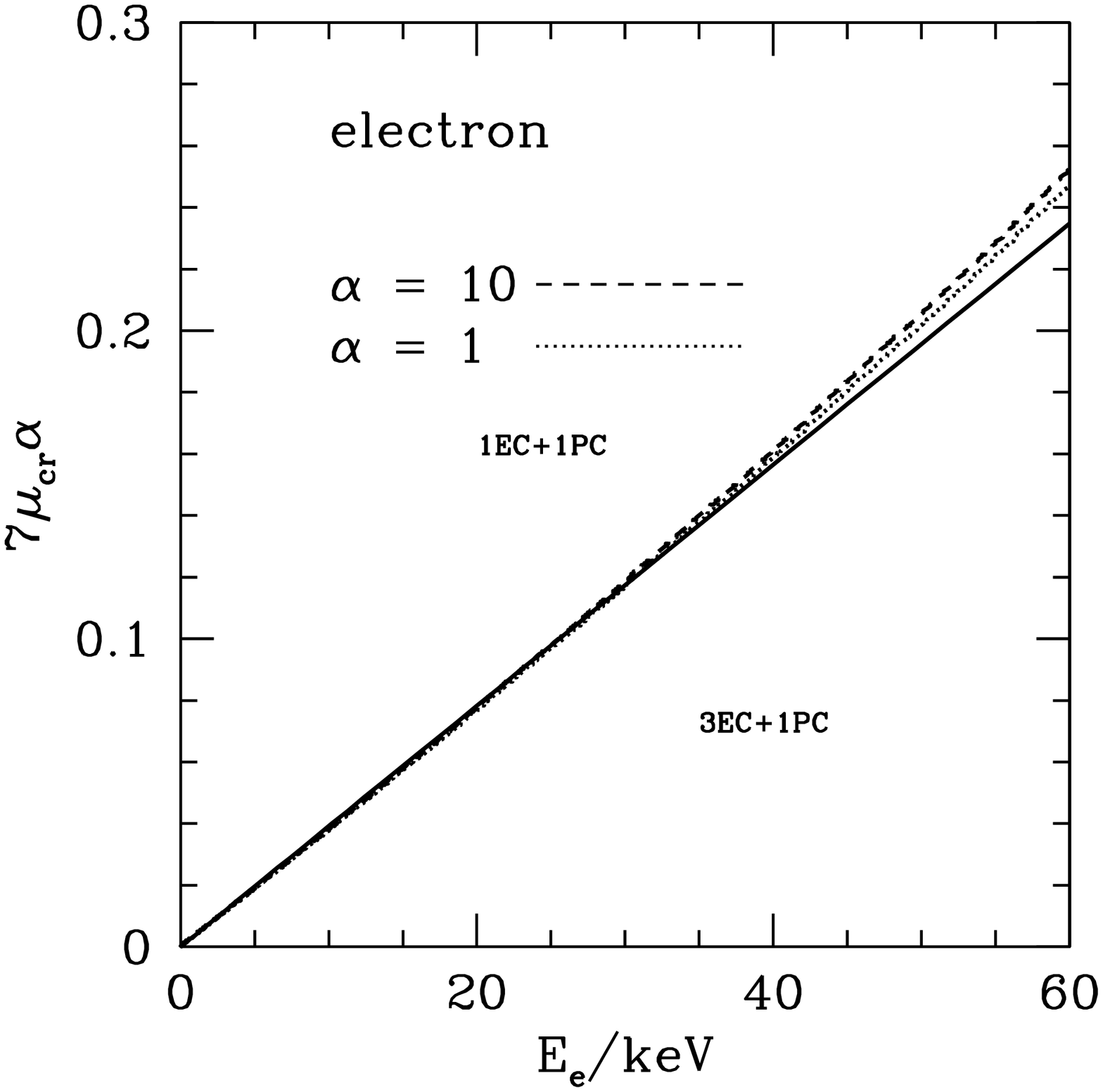}
\hspace{-0.6cm}
\includegraphics[height=8.4cm]{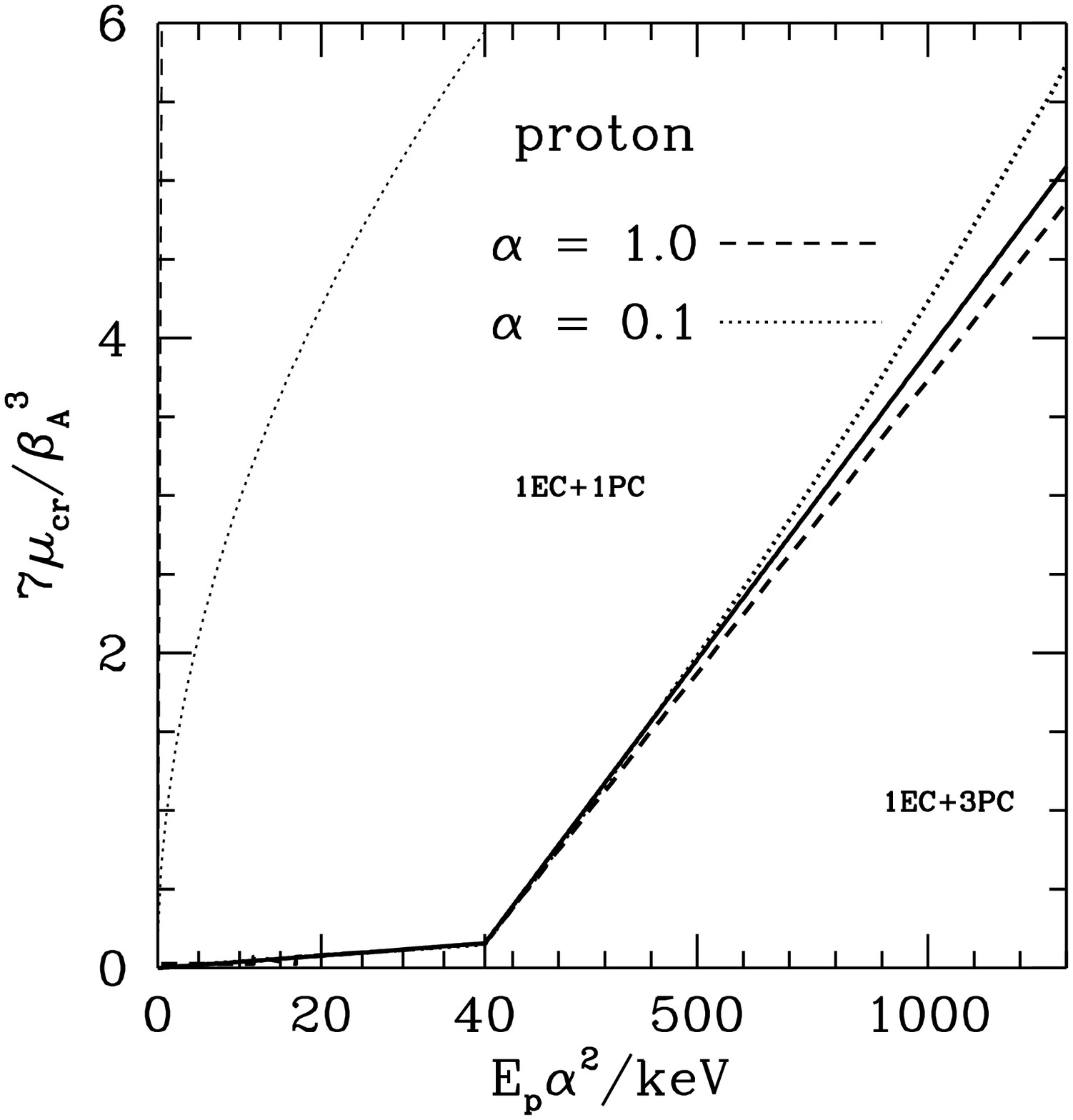}
\end{center}
\caption{ \small
Comparison of the analytical expressions (eq. [\ref{mucr2}] solid curves) with the exact 
values of $\mu_{\rm cr}$ (dotted and dashed curves; in the low energy range) due to 
resonant interactions with the EC (left panel for electrons) or PC (right panel for 
protons. Note that the region to the left of $E_{\rm p}\alpha^2 = 40$ keV has an expanded 
scale) branch. The thin dashed (barely visible near the left axis and for $\alpha=1.0$) and 
dotted (for $\alpha=0.1$) curves in the right panel give the critical angles for waves 
obeying the Alfv\'{e}nic dispersion relation $\omega=-|k|\beta_A$, which clearly give 
incorrect descriptions for the acceleration of low energy protons. Low energy electrons do 
not interact with the Alfv\'en waves.
}
\label{fig4.ps}
\end{figure}

The second approximation is for the critical angles of protons below the critical energy 
(velocity) and low energy electrons, most of which interact only with two waves with one 
dominating over the other. When this happens, the acceleration rate for the particles can 
be very small (see \S\ \ref{fp}). Only particles with very large pitch angles ($\mu\simeq 
0$) have four resonances and significant contributions to the pitch angle averaged 
acceleration rate. The regions for this lie in the small areas below the lowest curves in 
Figure \ref{fig2.ps}, which are barely visible for the proton and $\alpha=0.5$ case. 
As shown in the appendix, using the approximations of equations (\ref{ecapprox}) and 
(\ref{pcapprox}) for the dispersion relations, we can derive analytic expressions for the 
critical angles, which in the nonrelativistic limit give $\mu_{\rm 
cr}\propto\beta^2\propto E/mc^2$. Empirically, we find the following simple approximate 
expressions, as shown in Figure \ref{fig4.ps}, agree with the exact results to better 
than $\sim 10\%$ in the indicated energy ranges:

\begin{equation}
\mu_{\rm cr}={1 \over 3.5\alpha}\cases{\delta^{1/2}E/m_{\rm p}c^2, & for protons;
\ \ \ \ $E<E_{\rm cr}$,\cr
E/m_{\rm e}c^2, & for electrons; \ \ \ \ $E<60$ keV.\cr}
\label{mucr2}
\end{equation}

\clearpage

\clearpage

Here it should be emphasized that the commonly used approximation of accelerating protons
by the Alfv\'{e}n waves with the dispersion relation $\omega=-|k|\beta_A$ for
$|\omega|<\delta$, which is valid at relativistic energies (Barbosa 1979; Schlickeiser
1989; Miller \& Roberts 1995), is invalid at low energies. This can be seen from the lower
left panel of Figure \ref{fig1.ps}, which shows clearly that the dispersion relation of the
waves in resonance with low energy protons deviates far from the simple Alfv\'{e}nic form.
For the simple Alfv\'{e}nic dispersion relation, nonrelativistic protons resonate with
both a forward and a backward moving wave only if $|\mu|<\mu_{cr} =
\beta_A(\gamma-1)/\beta\gamma$.  (Particles only interacting with one wave can not be
accelerated; \S\ \ref{dc}.) This critical angle is indicated by the thin dashed (barely 
visible near the vertical axis) and dotted curves in the right panel of Figure 
\ref{fig4.ps} for $\alpha = 1.0$ and $0.1$ respectively, which clearly overestimate
by several orders of magnitudes the fractions of low energy protons that can be 
accelerated.  The inefficiency of proton acceleration at intermediate energies combined 
with the increase of the interaction rate above $E_{\rm cr}$ gives rise to the acceleration 
barrier to be described in \S \ref{barrier}.

In most of the particle acceleration models, an injection process of high energy particles
is postulated as an input. If the injected particles have an energy above $E_{\rm cr}$, it 
may be appropriate to use the Alfv\'{e}nic dispersion relation to describe the waves. 
If the energy of the injected particles is low, as is the case under studied here, one must 
use the exact dispersion relation to calculate the acceleration of low energy particles. 
Although most of the turbulence energy is carried by waves with low wavenumbers, the 
acceleration of low energy particles is determined by waves with high wavenumbers (see eq. 
[\ref{reson}]), which can constrain the overall acceleration efficiency.  As discussed 
above, the Alfv\'{e}nic dispersion relation for the waves will overestimate the 
acceleration efficiency of low energy protons significantly.

\subsection{Fokker-Planck Coefficients}

\begin{figure}[thb]
\begin{center}
\includegraphics[height=5.4cm, width = 5.4cm]{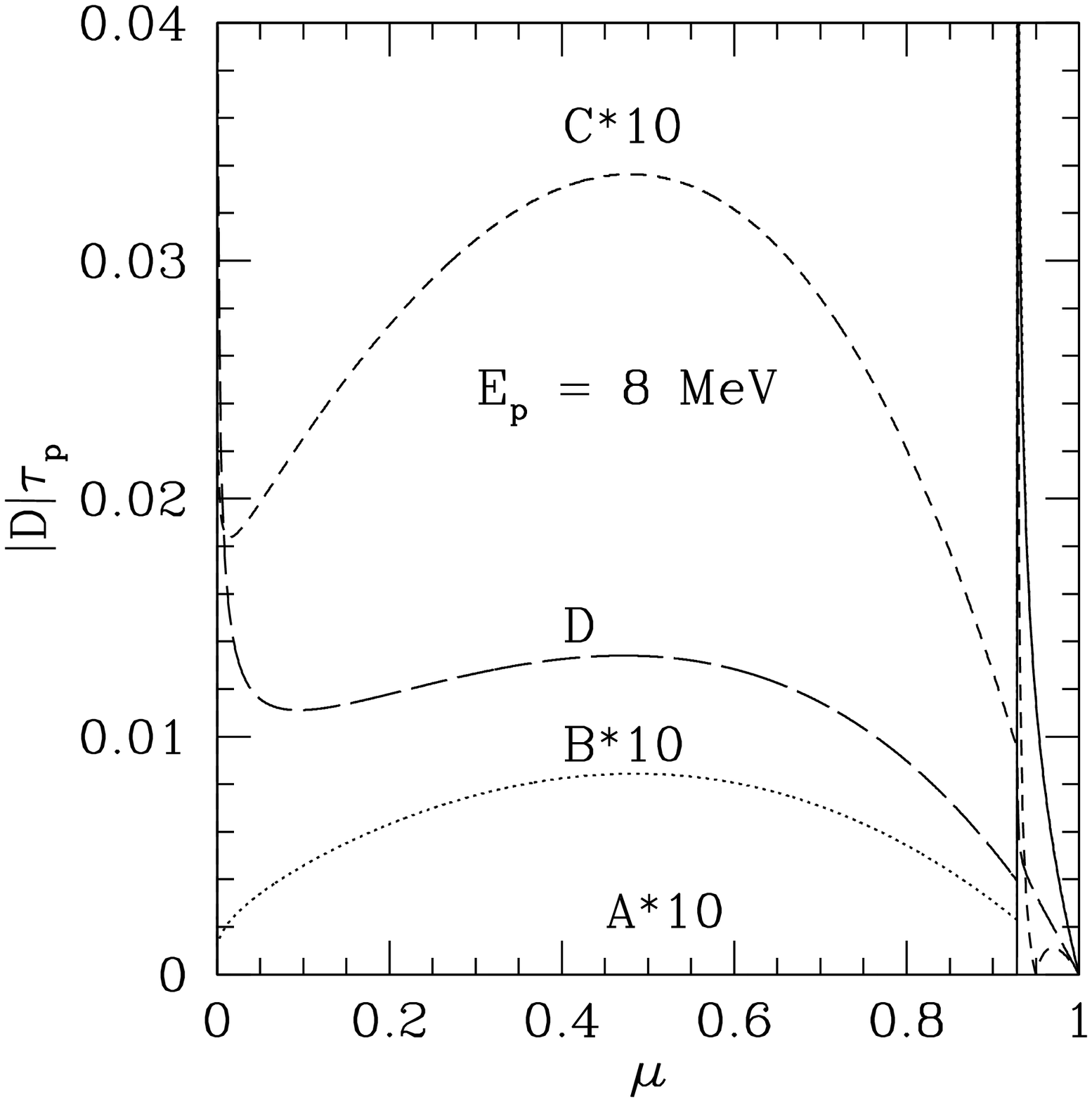}
\includegraphics[height=5.4cm, width = 5.4cm]{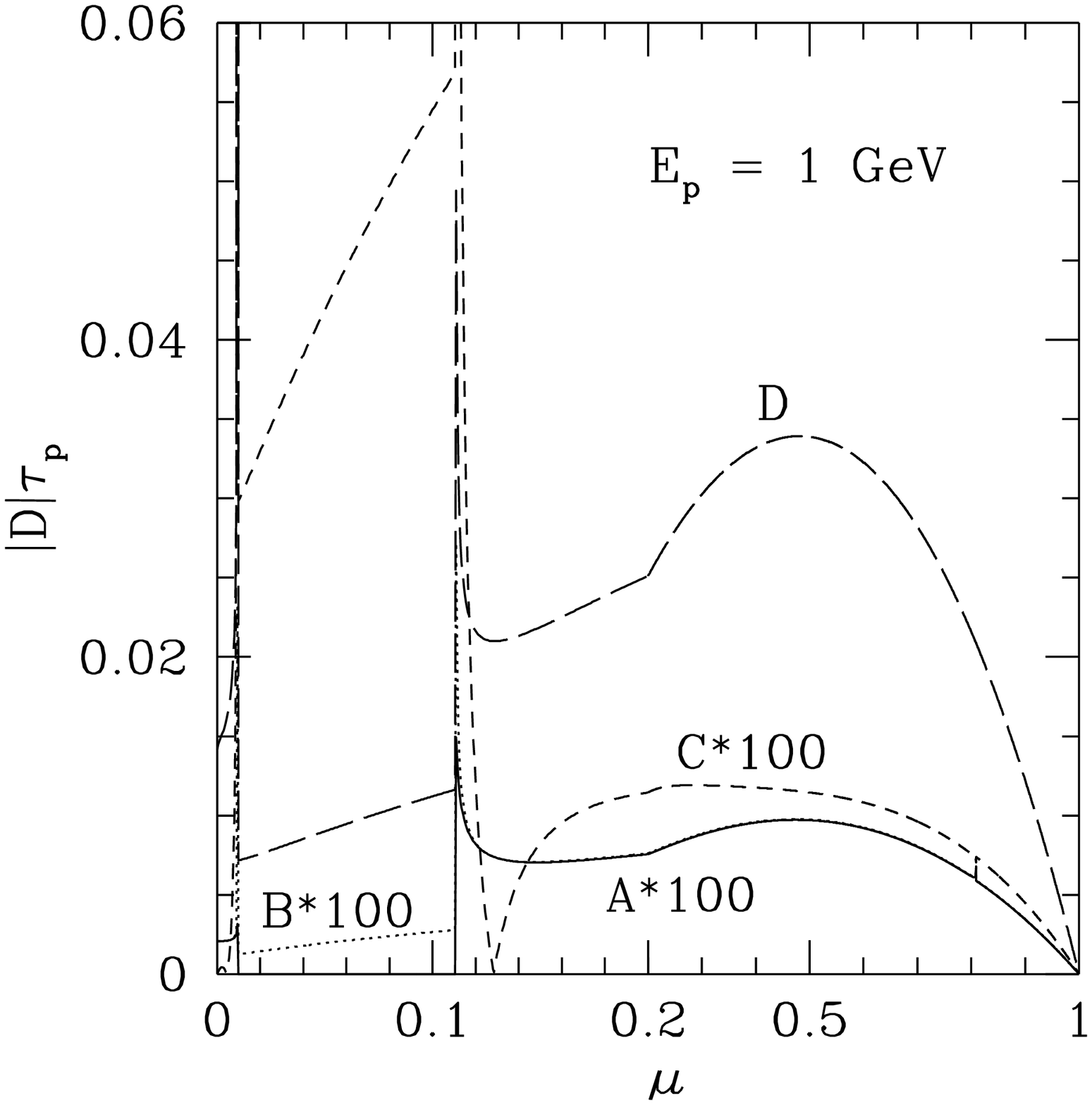}
\includegraphics[height=5.4cm, width = 5.4cm]{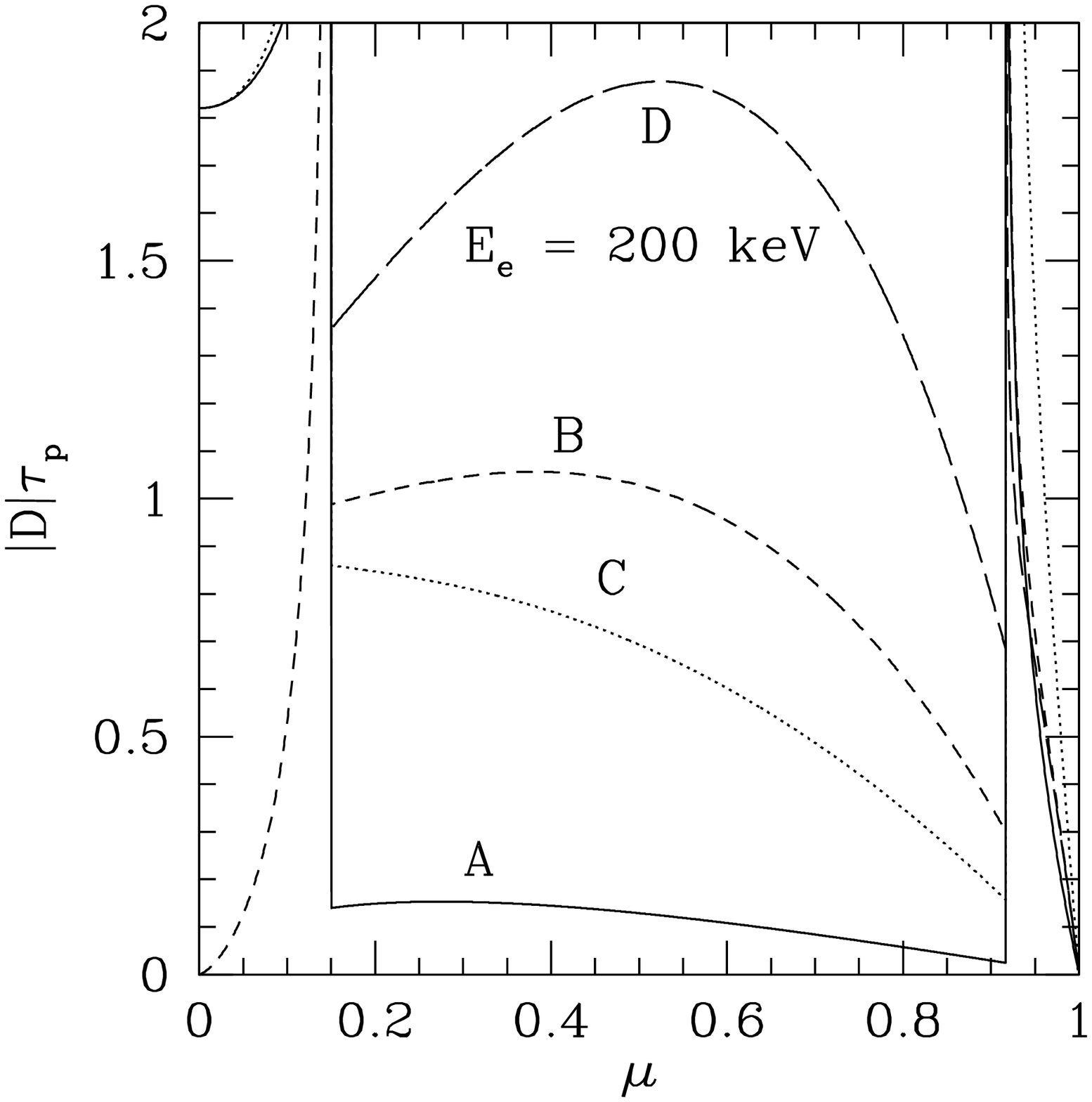}
\end{center}
\caption{ \small
Pitch angle dependence of the F-P coefficients (absolute values) for protons of
two different energies ($8\,$MeV: left panel and $1\,$GeV: middle panel), and $200\,$keV 
electrons (right panel). A, B, C, and D stand for $(D_{pp}-D^2_{p\mu}/D_{\mu\mu})/p^2$, 
$D_{pp}/p^2$, $|D_{p\mu}|/p$, and $D_{\mu\mu}$, respectively. The plasma parameter 
$\alpha=0.5$ and the turbulence spectral index $q=1.6$. Note that different coefficients 
are scaled differently and that for illustration, in the middle panel the region to 
the left of $\mu=0.2$ is expanded. }
\label{fig5.ps}
\end{figure}

\begin{figure}[thb]
\begin{center}
\includegraphics[height=8.4cm]{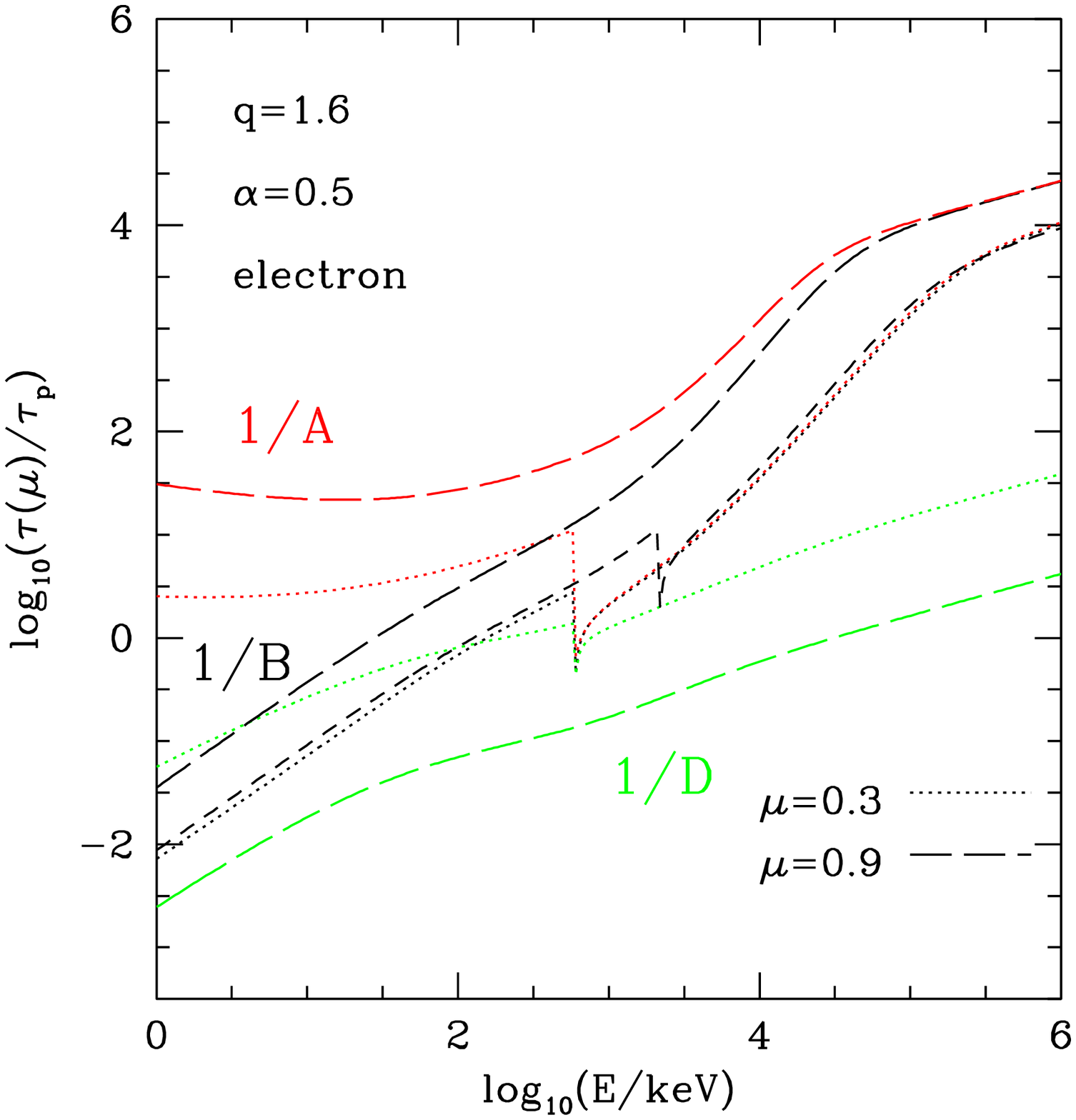}
\hspace{-0.6cm}
\includegraphics[height=8.4cm]{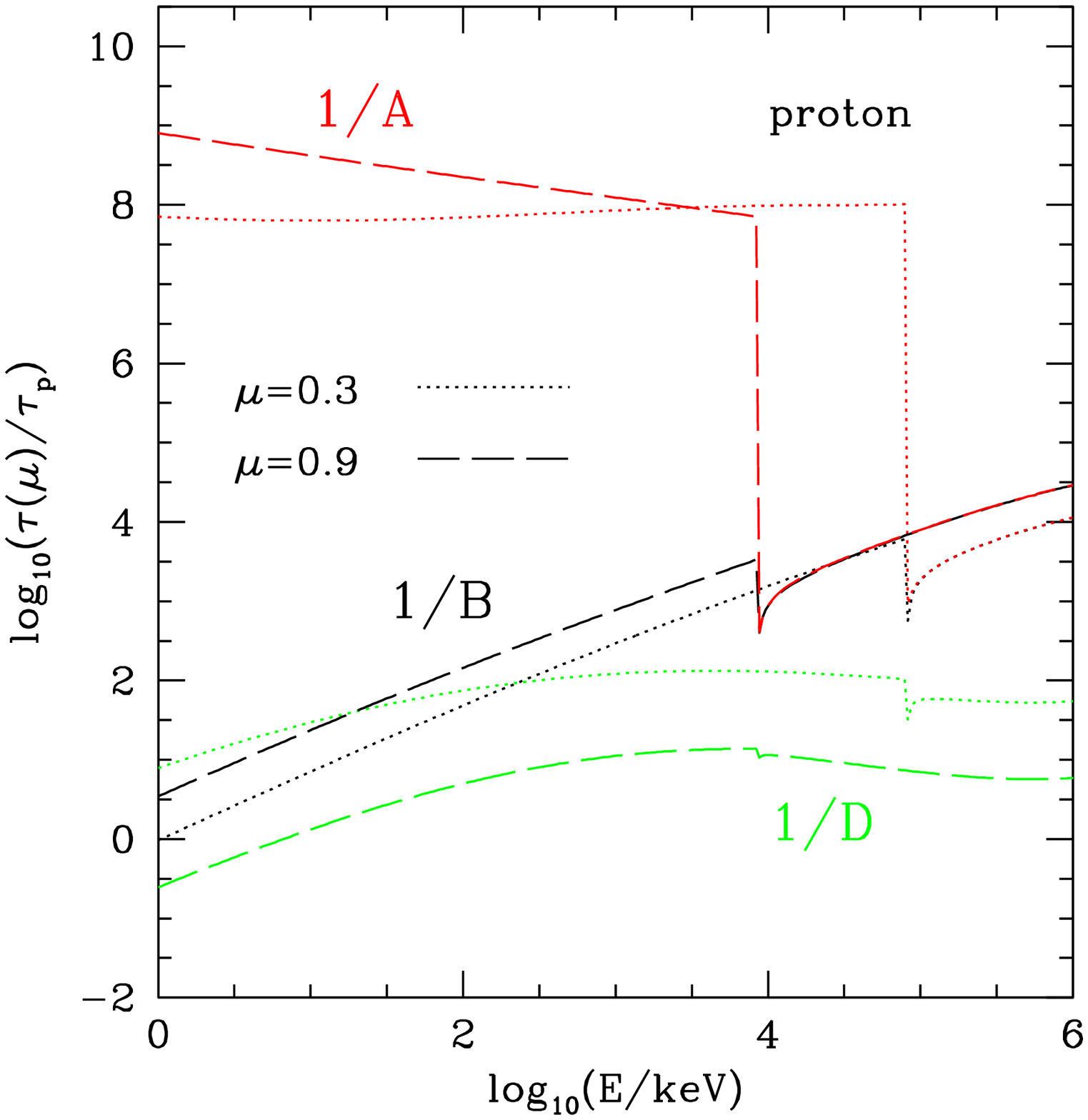}
\end{center}
\caption{ \small
Energy dependence of the scattering, $1/D=D_{\mu\mu}^{-1}$, and acceleration times, 
$1/B = p^2/D_{pp}$ or $1/A = p^2/(D_{pp}-D_{\mu p}^2/D_{\mu\mu})$, in units of $\tau_{\rm 
p}$ at different pitch angles for electrons (left panel) and protons (right panel). The 
plasma parameter $\alpha=0.5$ and the turbulence spectral index $q=1.6$. Note that $A=B$ 
for $\mu=0$. At very low energies $D_{\mu\mu}^{-1}>p^2/D_{pp}$ and the acceleration time 
$\tau_{\rm ac} = p^2/D_{pp}$. At intermediate and high energies $\tau_{\rm 
ac}=p^2/(D_{pp}-D_{\mu 
p}^2/D_{\mu\mu})$.
}
\label{fig6.ps}
\end{figure}

\begin{figure}[thb]
\begin{center}
\includegraphics[height=8.4cm]{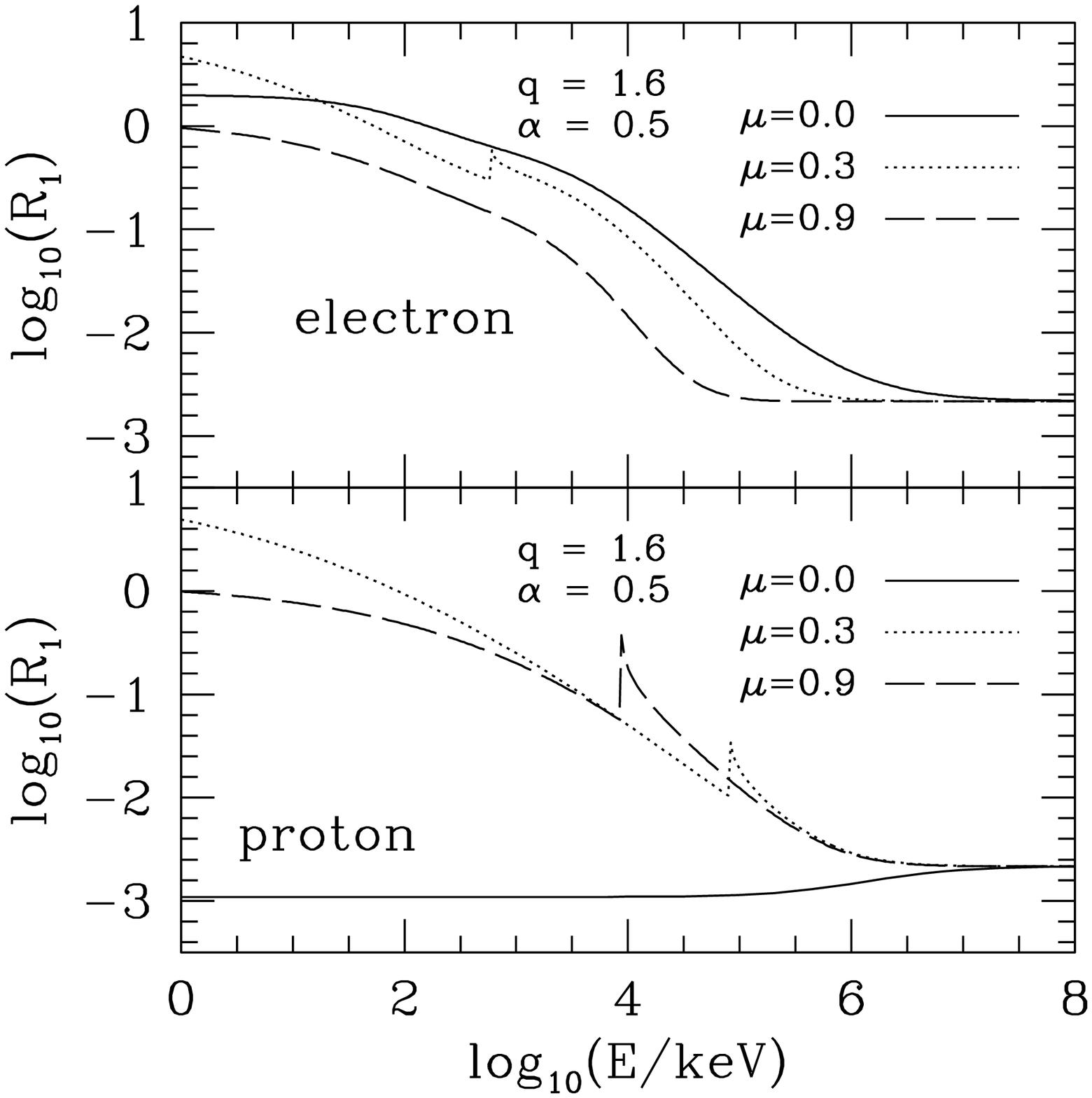}
\hspace{-0.6cm}
\includegraphics[height=8.4cm]{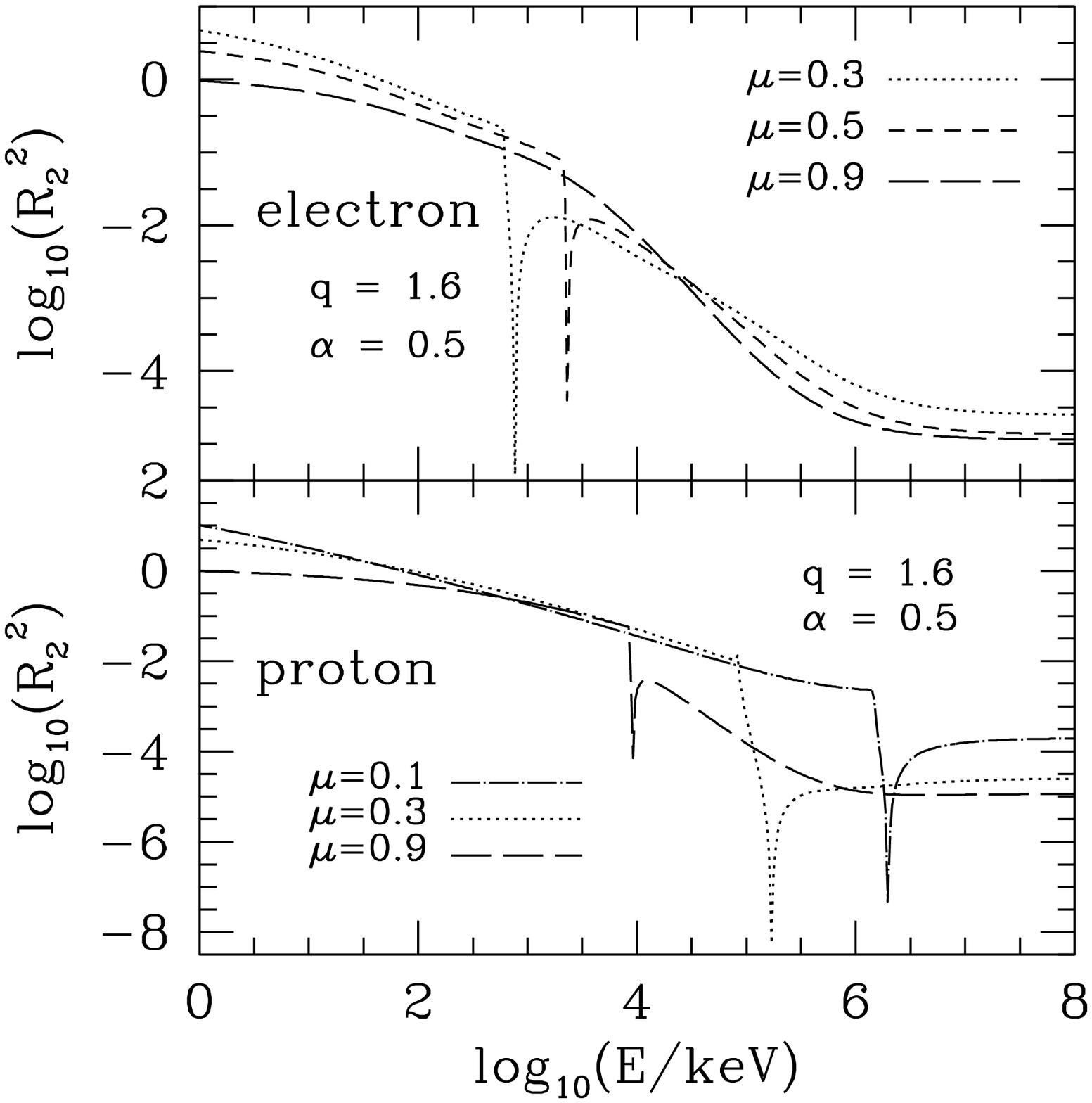}
\end{center}
\caption{ \small
The ratios $R_1$ (left panel) and  $R_2^2$ (right panel) as functions of energy 
for several different pitch angles for electrons (upper panels) and protons (lower panels). 
In general, $R_1$ and $|R_2|$ exceed unit at low energies and $\mu$ (and small values of 
$\alpha$, see PP97). For protons, at certain ranges of $E$ and $\mu$,  $R^2_2\approx R_1 
\ll 1$, and the acceleration rate, $D_{\mu\mu}(R_1-R_2^2)$ can be very small. (see eq. 
[\ref{accel}]).
}
\label{fig7.ps}
\end{figure}

For a given power law spectrum of the turbulence, it is straightforward to calculate the
F-P coefficients with equation (\ref{coeff}).  Figure \ref{fig5.ps} shows variation of
these coefficients with $\mu$ at some representative energies and Figure \ref{fig6.ps}
shows the variation of their inverses (i.e. times) with energy at different values of 
$\mu$, here $\alpha = 0.5$ and $q = 1.6$.  The discontinuous jumps occur at the critical 
values of $\mu_{\rm cr}$ and $\beta_{\rm cr}$ described in \S\ \ref{cri}.  The variations 
of the two ratios defined in equations (\ref{r1}) and (\ref{r2}) with energy at different 
pitch angles are shown in Figure \ref{fig7.ps}.  These results justify the discussion in 
\S\ \ref{dc}.

\clearpage

\clearpage

\clearpage

\clearpage

\subsection{Barrier in the Proton Acceleration}
\label{barrier}

In the previous section, we showed that the pitch angle averaged acceleration rate is one
of the dominating factors in the particle acceleration processes. The relative acceleration
of protons and electrons therefore depends on the contrast of their acceleration times.
Figure \ref{fig2.ps} shows that at low energies particles with $\mu>\mu_{\rm cr}$ only
resonate with one PC and one EC wave: the EC wave dominates the PC wave for electrons while
the reverse is true for protons. These particles have significant contributions to the 
pitch angle averaged acceleration rate for $\mu_{\rm cr}\ll 1$ (eq. [\ref{mucr2}]).  
Because the difference between the wavenumbers of the two waves interacting with protons is 
much larger than that for electrons, the resonant interaction is more strongly dominated by 
one of the resonant waves for protons than it is for electrons.  The factor $1-R_2^2/R_1$ 
for protons can therefore be several orders of magnitude smaller than that for electrons at 
a given energy and pitch angle (Figure \ref{fig6.ps}). Consequently, in the intermediate 
energies where $R_1\sim R_2^2\sim 1$, the pitch angle averaged acceleration time for 
protons has a more prominent increase than that for electrons. At still higher energies, 
particles with four resonances dominate the acceleration rate because $R_2^2\ll R_1$ for 
the interactions. For both electrons and protons, the new resonant wave modes come from the 
EC branch. In the relativistic limit, the acceleration is dominated by resonances with the 
Alfv\'{e}n waves and the interaction rates for electrons and protons become comparable.

\begin{figure}[thb]
\begin{center}
\includegraphics[height=8.4cm]{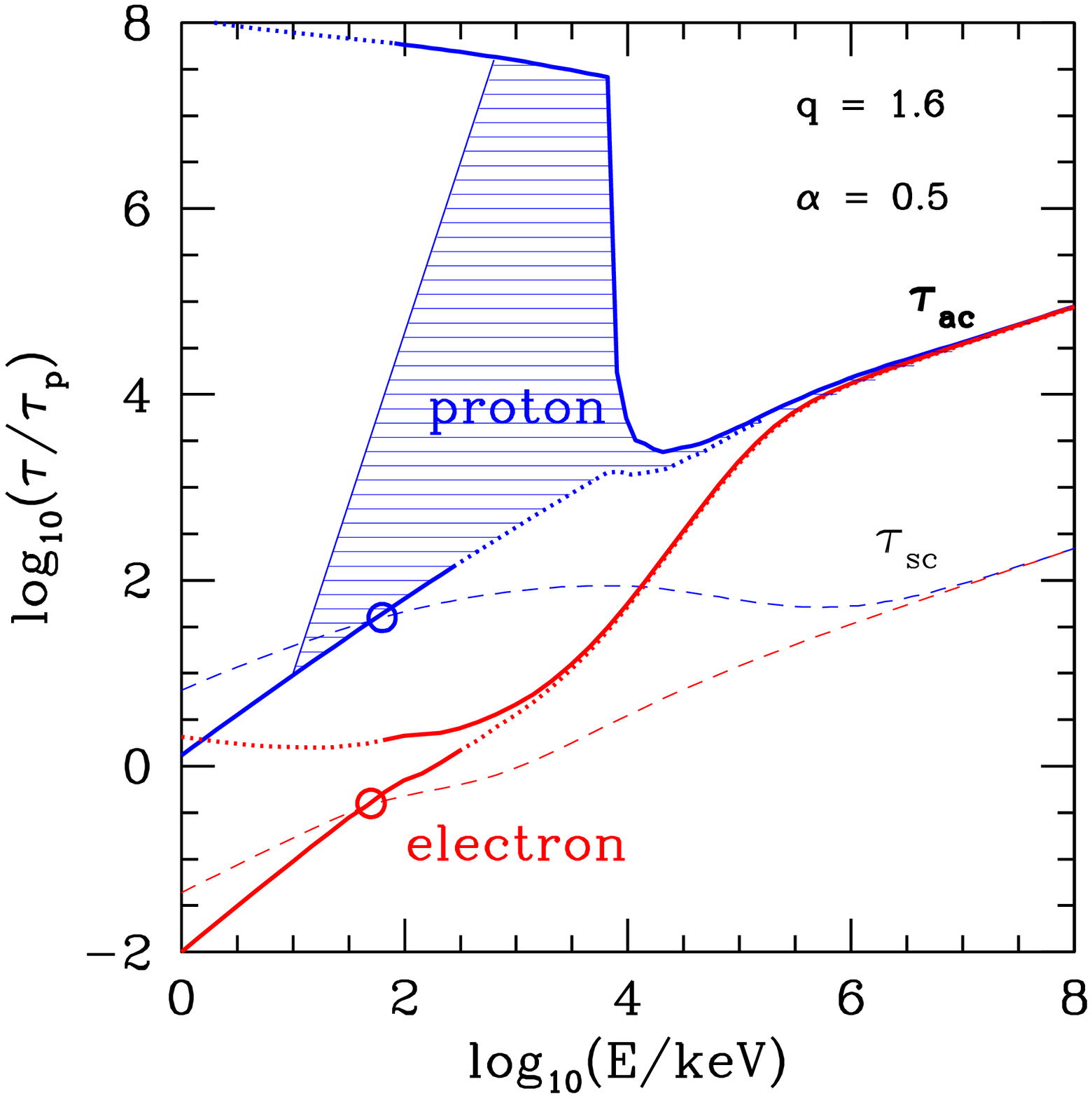}
\hspace{-0.6cm}
\includegraphics[height=8.4cm]{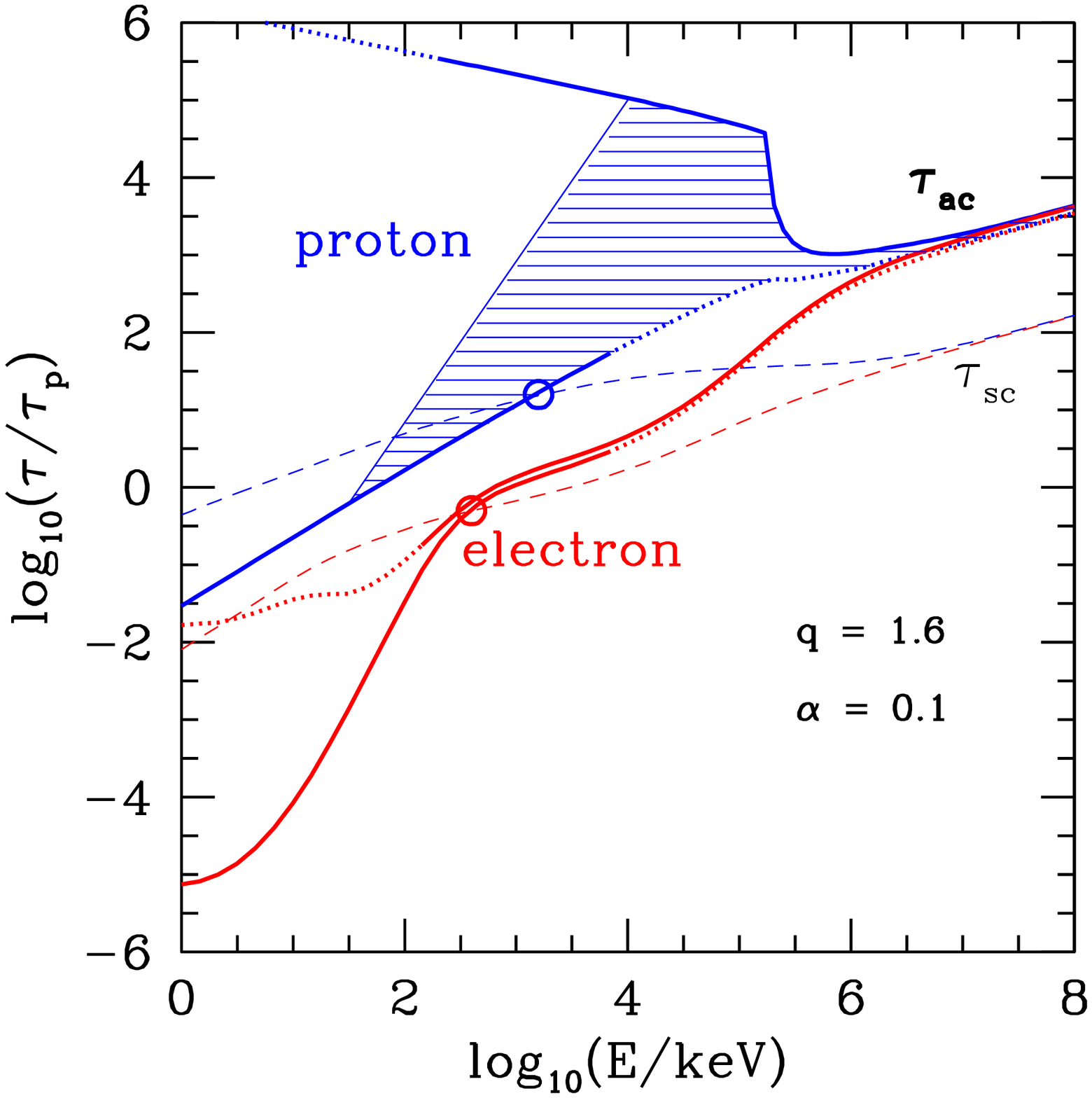}  
\end{center}
\caption{ \small
Pitch angle averaged acceleration (thick curves) and scattering (thin dashed curves) times 
in units of $\tau_{\rm p}$ in plasmas with $\alpha=0.5$ ({\bf a: left panel}) and 
$\alpha=0.1$ ({\bf b: right panel}). Here the turbulence spectral index $q=1.6$. The 
acceleration times defined for both cases of equation (\ref{tacclall}) are plotted with the 
corresponding invalid segments indicated by the dotted curves. The upper two curves are 
for protons and the lower two are for electrons. The circles indicate the points of 
transitions between low and high energies where $R_1\sim |R_2|\sim 1$. The transitions of 
the electron acceleration times are quite smooth. The thin solid lines show schematically 
the transitions of the acceleration time for protons. The 
acceleration barrier (as indicated by the hatched areas) in the proton acceleration times 
are prominent. The sharp drop of the electron acceleration time with the decrease of energy 
for $\alpha=0.1$ is due to interactions with the EM branch (PP97).
}
\label{fig8.ps}
\end{figure}

Figure \ref{fig8.ps} shows the pitch angle averaged acceleration (thick curves) and 
scattering (thin dashed curves) times in units of $\tau_{\rm p}$ for electrons (lower 
curves) and protons (upper curves) in plasmas with $\alpha=0.5$ (left panel) and 
$\alpha=0.1$ (right panel).  The turbulence spectral index $q=1.6$ here.  The acceleration 
times for both cases in equation (\ref{tacclall}) are plotted with the invalid segments 
plotted as dotted curves. We see that the pitch angle averaged acceleration times are much 
shorter than the corresponding scattering times for keV particles. The particle 
distributions can be anisotropic at these energies unless there are other scattering 
processes (e.g. Coulomb collisions). In the high energy range, the scattering time is always 
shorter than the corresponding acceleration time when $\beta_{\rm A}<1$. The transitions 
where $R_1\sim|R_2|\sim 1$ (as indicated by the circles in the figure) occur between 
$10^2$ to $10^3$ keV, increase with the decrease of $\alpha$ and depend on the turbulence 
spectral index $q$ as well.

\clearpage

\clearpage

There is clearly an acceleration barrier (as indicated by the shaded area) in the proton 
acceleration time. The thin solid line shows schematically the acceleration time of protons 
in the transition region.  The sharp increase of the proton acceleration time at lower 
energies is caused by their low acceleration efficiency when the scattering rate already 
overtakes the acceleration rate as discussed above. The sharp drop of the proton 
acceleration time at a higher energy is due to interactions with the Whistler waves. 
Because protons with small pitch angles ($\mu \simeq 1$) interact with the Whistler waves 
at the lowest energy and the interaction is very efficient, this energy corresponds to the 
critical energy $E_{\rm cr}$ identified in equation (\ref{betacr}). 

These characteristics are not true for electrons. In section \S\ \ref{cri}, we have shown 
that a small fraction of particles with $\mu<\mu_{\rm cr}$ can resonate with four waves and 
$|R_2|\ll R_1$ for the interactions (Figure \ref{fig5.ps}). Compared with protons, more 
electrons can be accelerated this way (eq. [\ref{mucr2}]). Because the acceleration of 
electrons with two resonances is very inefficient, the acceleration of this small fraction 
of electrons already dominates the electron acceleration processes where the scattering 
rate becomes comparable with the acceleration rate. At even higher energies, there is no 
extra wave modes which can enhance the electron acceleration processes. Consequently, 
electrons have a smooth acceleration time profile.

Expressions for estimating the difference between electron and proton acceleration time are 
derived in the appendix.  Briefly, because particles with $\mu<\mu_{\rm cr}$ have 
significant contributions to the acceleration in the low energy region ($\beta\ll 1$), one 
can estimate the pitch angle averaged acceleration times with the approximate expressions 
for the critical pitch angles (eq. [\ref{mucr2}]):
\begin{equation}
{\tau_{\rm ac}\over\tau_{\rm p}} = 2\left[\int_{-1}^1\d \mu 
D_{\mu\mu}(R_1-R_2^2)\tau_{\rm p}\right]^{-1}\simeq 7 
\alpha^{q+2}\left({E\over m_{\rm e}c^2}\right)^{(1-q)/2}\cases{1, & for electrons; \cr
\delta^{-5/2} & for protons, \cr} 
\end{equation}
which is consistent with the numerical result within a factor of two.  In the relativistic 
limit ($\gamma\gg 1$), particles interact with the Alfv\'{e}n waves and we find
\begin{equation}
{\tau_{\rm ac} \over \tau_{\rm p}}= {q(q+2)\alpha^2\over 4 \delta}\left({e\over 
q_i}\right)^{2-q}\left({E\over m_{\rm e}c^2}\right)^{2-q}\,.
\end{equation}
The difference between these two time scales at the critical energy (eq. 
[\ref{betacr}]) gives an estimate of the height of the acceleration barrier.

\subsection{Application to Solar Flare Conditions}
\label{application}

\begin{figure}[htb]
\begin{center}
\includegraphics[height=8.4cm]{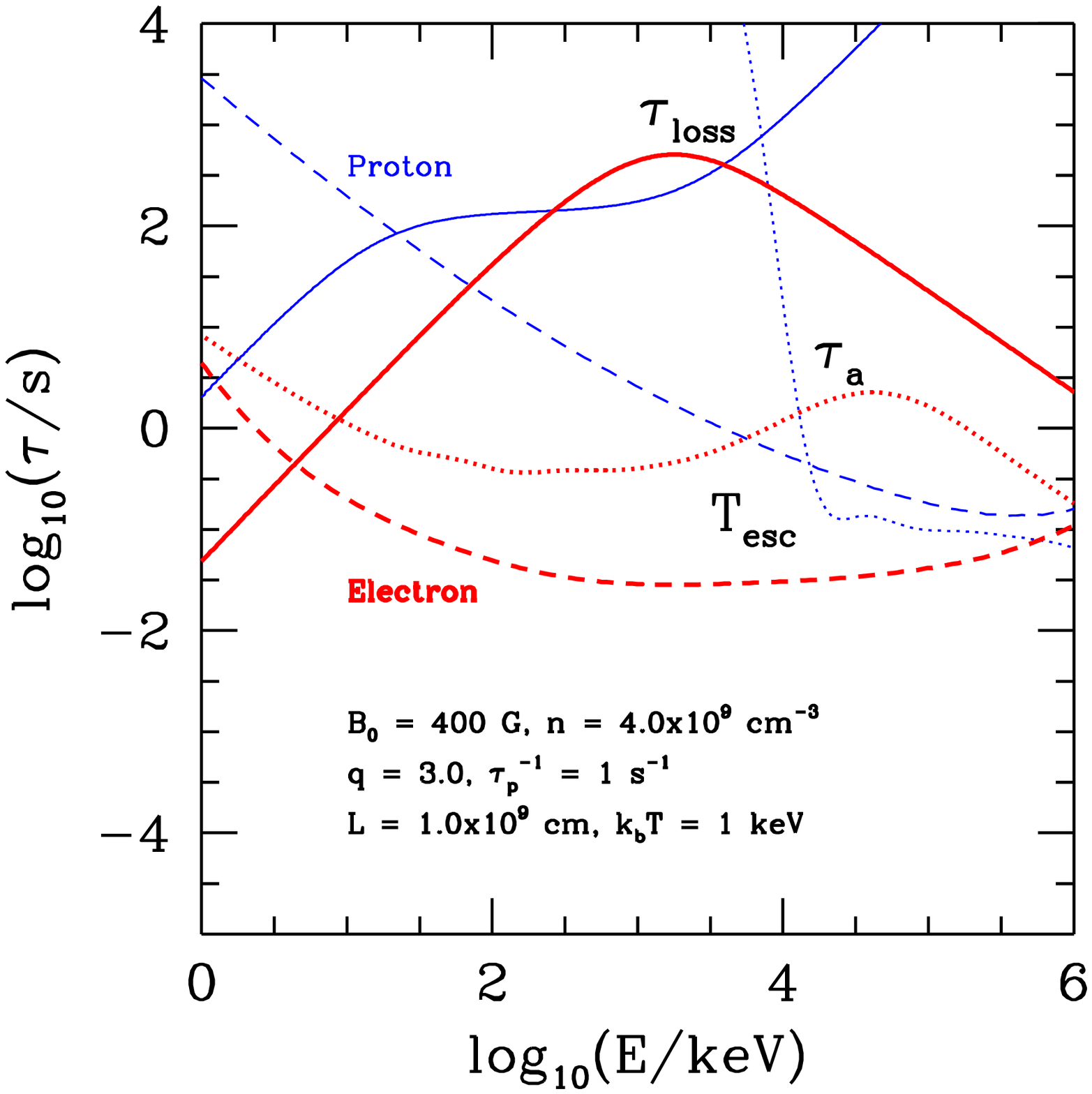}
\hspace{-0.60cm}
\includegraphics[height=8.4cm]{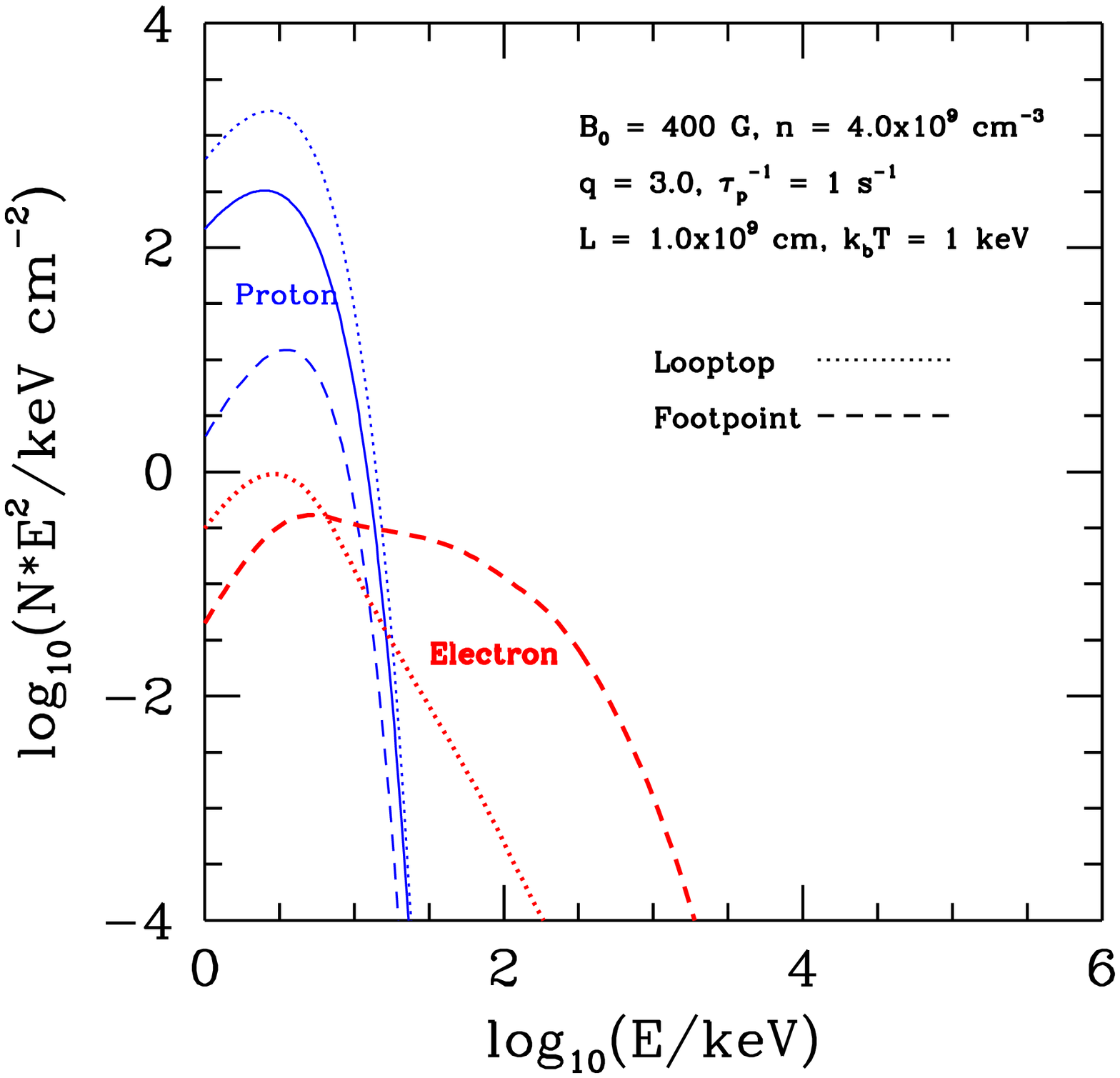}
\end{center}
\caption{ \small {\bf Left panel:} Time scales for protons (thin curves) and electrons 
(thick curves) in a strongly magnetized plasma ($\alpha=0.5$) with a steep turbulence 
spectrum ($q=3$). The direct acceleration times $\tau_{\rm a}= E/A$, which are related to 
the diffusion acceleration times $\tau_{\rm ac}$, are shown by the dotted curves. The solid 
curves are for $\tau_{\rm loss}$ and the dashed curves are for $T_{\rm esc}$. {\bf Right 
panel:} The corresponding distributions $N*E^2$ of the accelerated electrons (thick curves) 
and protons (thin curves). Here the dotted curves give the thin target LT particle 
distributions, the dashed curves indicate the effective thick target particle distributions 
at the FPs. The solid curve gives the injected thermal particle distribution with arbitrary 
normalization. One can see that because of the presence of the strong acceleration barrier 
(exceeding the range of the graph) protons are basically not accelerated.
} 
\label{fig10.ps}
\end{figure}

Figure \ref{fig10.ps} shows a model of electron (thick curves) and proton (thin curves) 
acceleration in a strongly magnetized plasma.  The LT size $L=10^9$ cm, the 
temperatures of the injected electrons and protons are the same $k_{\rm b}T=1$ keV.  The 
magnetic field and gas density are $400$ G and $4.0\times 10^9$ cm$^{-3}$, respectively, 
i.e. $\alpha=0.5$ (see equation [\ref{alpha}]).  The relevant time scales are shown in the 
left panel, where we have defined the direct acceleration time $\tau_{\rm a} = E/A$, which 
is related to $\tau_{\rm ac}$ (eqs. [\ref{accel}] and [\ref{ae}]). The corresponding 
accelerated particle distributions are shown in the right panel. The dotted and dashed 
curves show the LT and FP spectra, respectively. 

We note that for the above conditions the electrons can be accelerated to a few hundreds of
keV while the proton acceleration is suppressed due to the acceleration barrier. The
electron distribution steepens with the increase of energy because the escape time becomes
shorter than the acceleration time ($T_{\rm esc}<\tau_{\rm a}$). At low energies where
Coulomb collisions dominate, the LT electrons have a quasi-thermal distribution. The solid
curve gives the thermal distribution of the injected particles with arbitrary
normalization. Due to the absence of acceleration at low energies, the steady state proton
distributions are almost identical with the injected proton distribution.

To produce a near power law electron distribution, as suggested by solar flare 
observations, the escape time must be comparable with the acceleration time in the relevant 
energy band.  Because the escape time of the nonrelativistic electrons always decreases 
with the increase of energy, we adopt a turbulence spectral index of $q = 3$ in the model.  
The plasma time $\tau_{\rm p}=1$ s.  Then we have the ratio of the turbulent wave energy 
density to the magnetic field energy density $8\pi {\cal E}_{\rm tot}/B_0^2=4.4\times 
10^{-10}k_{\rm min}^{-2}$.  To ensure that this ratio is much less than one so that the 
quasilinear approximation for the F-P treatment stays valid, one needs $k_{\rm 
min}>10^{-5}$ or an injection length of the turbulent waves of less than $10^6$ cm (note 
that $k_{\rm min}$ is in units of $\Omega_{\rm e}/c\simeq0.24$ cm$^{-1}$ here).  Otherwise, 
the turbulence spectrum must flatten at low $k$ so that there is less energy content in 
long wavelength waves.

\clearpage

\clearpage

\begin{figure}[htb]
\begin{center}
\includegraphics[height=8.4cm]{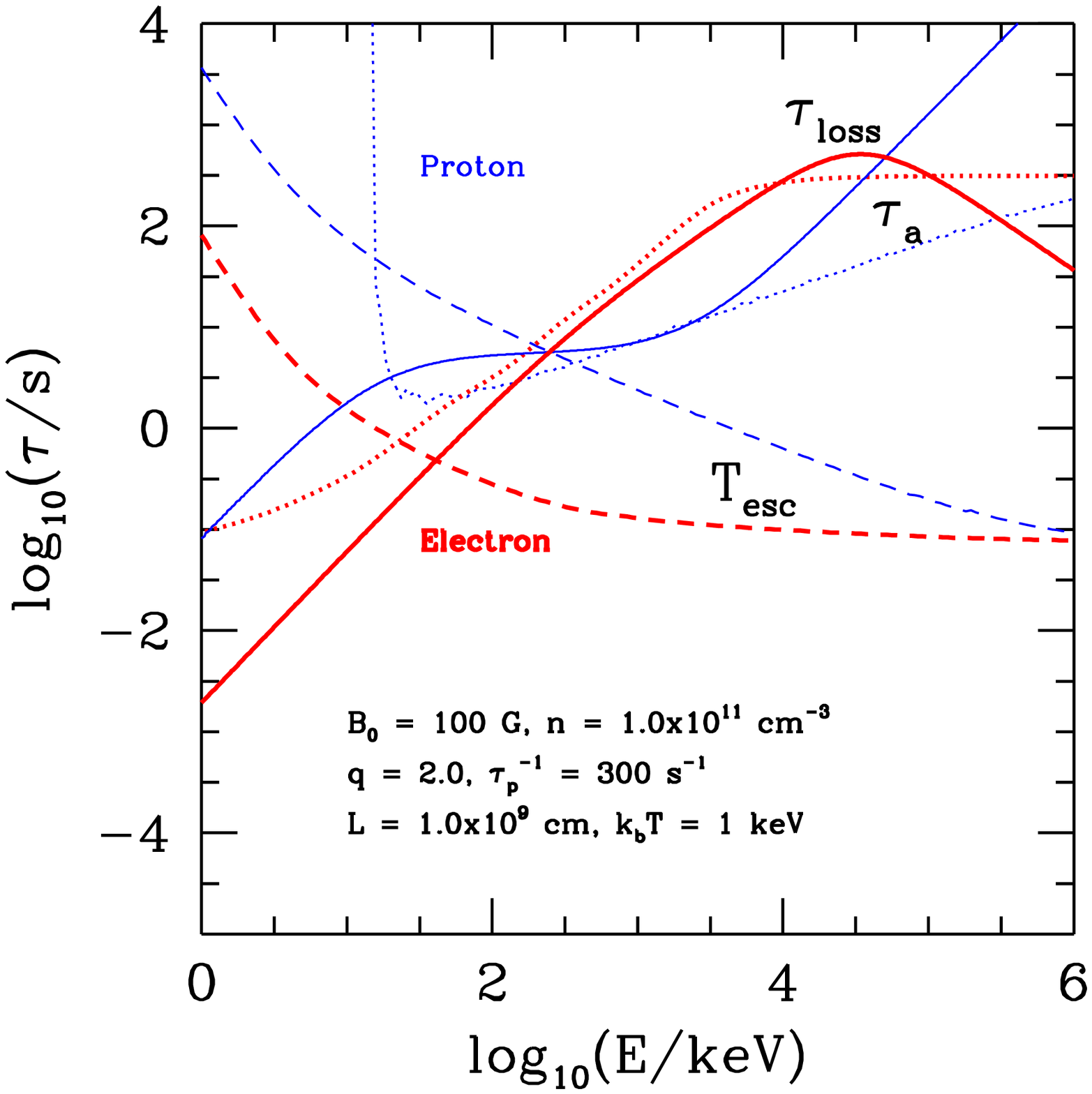}
\hspace{-0.6cm}
\includegraphics[height=8.4cm]{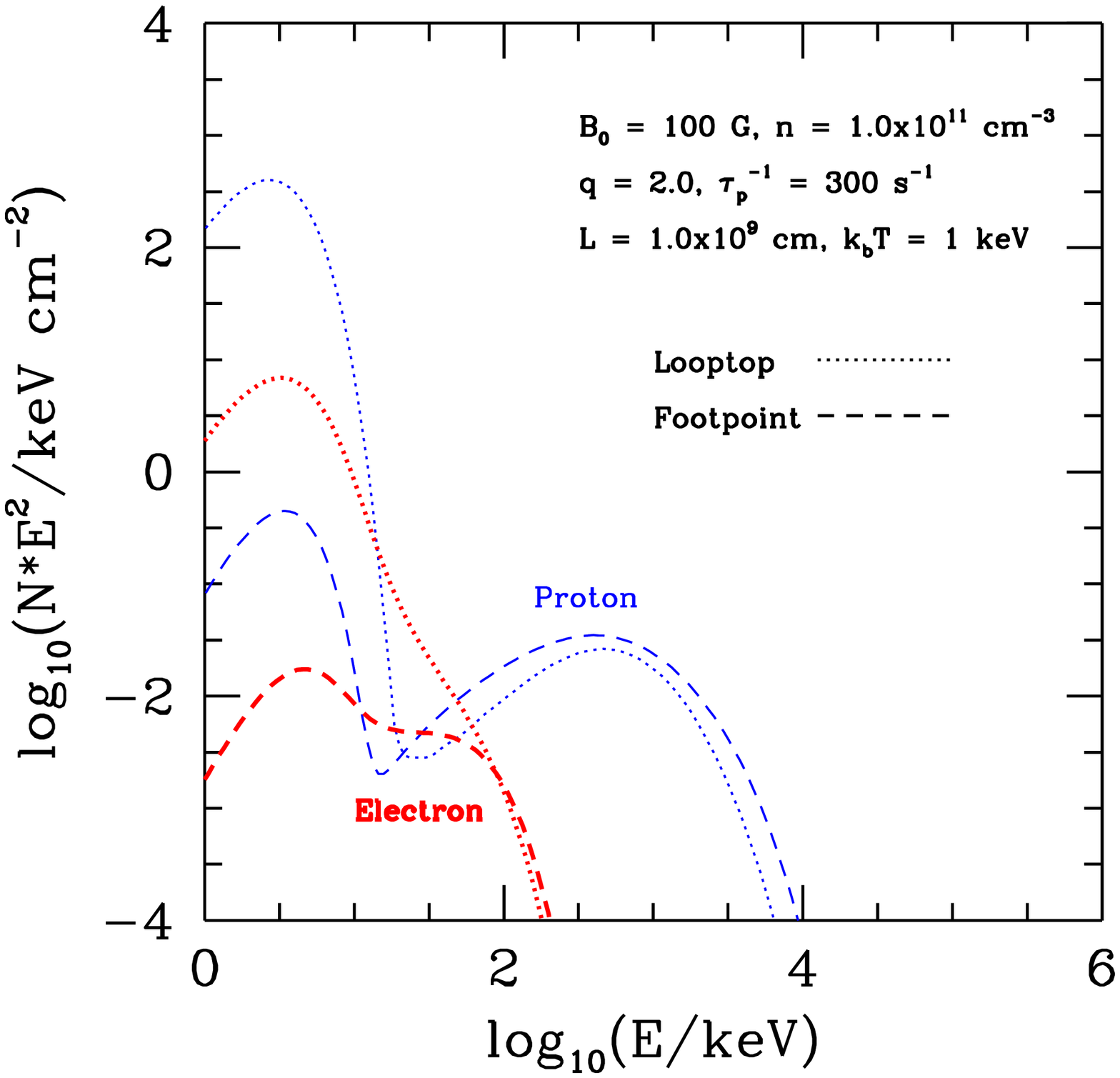}
\end{center}
\caption{ \small
Same as Figure \ref{fig10.ps} but for a model in a weakly magnetized plasma ($\alpha=10$). The 
model parameters are shown in the figure. Here the proton acceleration becomes more efficient than 
the electron acceleration because its barrier is close to the energy of the injected particles and 
does not affect the acceleration processes.
}
\label{fig11.ps}
\end{figure}

In \S\ \ref{cri} and \S\ \ref{barrier}, we showed that the proton acceleration barrier
moves toward lower energies with the increase of $\alpha$.  So in very weakly magnetized
plasmas, this barrier can be close to the thermal energy of the injected particles and thus
has little effect on the acceleration of protons. Protons can be accelerated efficiently in
the case because their loss time is long. Figure \ref{fig11.ps} shows such a model, where
$B = 100$ G and $n = 10^{11}$ cm$^{-3}$. The size of the LT and the injected particle
temperatures remain the same as those in the previous model. Because the turbulence
spectrum is flat ($q=2$), we have a pretty hard accelerated proton distribution below 1
MeV.  Above this energy, there is a cutoff due to the dominance of the escape term over the
acceleration terms. The accelerated electron distribution has a cutoff at less than 100 keV
which is also due to the quick escape of electrons with higher energies from the
acceleration site. At a few keV, both electron and proton distributions are quasi-thermal
because of the dominance of Coulomb collisions.

The above results show that electrons can be accelerated to very high energies by parallel 
propagating turbulent waves in pure hydrogen plasmas, but the presence of the 
acceleration barrier in the intermediate energy range makes the acceleration of protons
very inefficient. Only in very weakly magnetized plasmas where the barrier is close to the 
background particle energy does the acceleration of protons become efficient. The required 
value of the plasma parameter $\alpha$ is above 10 which is much larger than that believed 
to be the case for solar flares. However, most astrophysical plasmas including solar flares 
are not made of pure hydrogen. They contain significant numbers of $^4$He. These particles 
modify the dispersion relation used above. Abundances of elements heavier than He are too 
small to have a significant effect but $^4$He with an abundance (by 
number) of about $8\%$ can have important effects. 

To produce a near power law distribution of the accelerated particles, the index of the 
turbulence spectrum must be larger than 2 at high wavenumbers. If the turbulence is 
generated on a scale comparable to the size of the flaring loops, such a steep turbulence 
spectrum must flatten at low wavenumbers to ensure that the turbulence energy density is 
less than the energy density of the magnetic field, which is presumably the dominant source 
of energy for solar flares.  Such a steepening of the turbulence spectrum at high $k$ is 
expected if one includes the thermal damping effects of the waves with high wavenumbers. 
We will incorporate these effects in the following discussion.

\clearpage

\clearpage


\section{ACCELERATION IN HYDROGEN AND HELIUM PLASMAS}
\label{ephe4}

We now repeat the derivation of the previous section for more realistic plasmas 
containing e, p and $\alpha$-particles. We assume a fully ionized H and $^4$He plasma with 
the following relative abundances: ${\rm electron}/{\rm proton}/\alpha$-particle$ = 
1/0.84/0.08$.

\subsection{Dispersion Relation and Resonant Interactions}
\label{disphe}

The dispersion relation for such a plasma can be written as:
\begin{equation}
{k^2\over\omega^2}=1 -{\alpha^2\over \omega}\left[{1\over\omega-1}+{(1-2Y_{\rm 
He})\delta\over\omega+\delta}+{Y_{\rm He}\delta\over \omega+\delta/2}\right]\,,
\end{equation}
where the $^4$He abundance $Y_{\rm He}=0.08$.  The other symbols are the same as 
those defined in \S\ \ref{dis}.

\begin{figure}[h]
\begin{center}
\includegraphics[height=8.4cm]{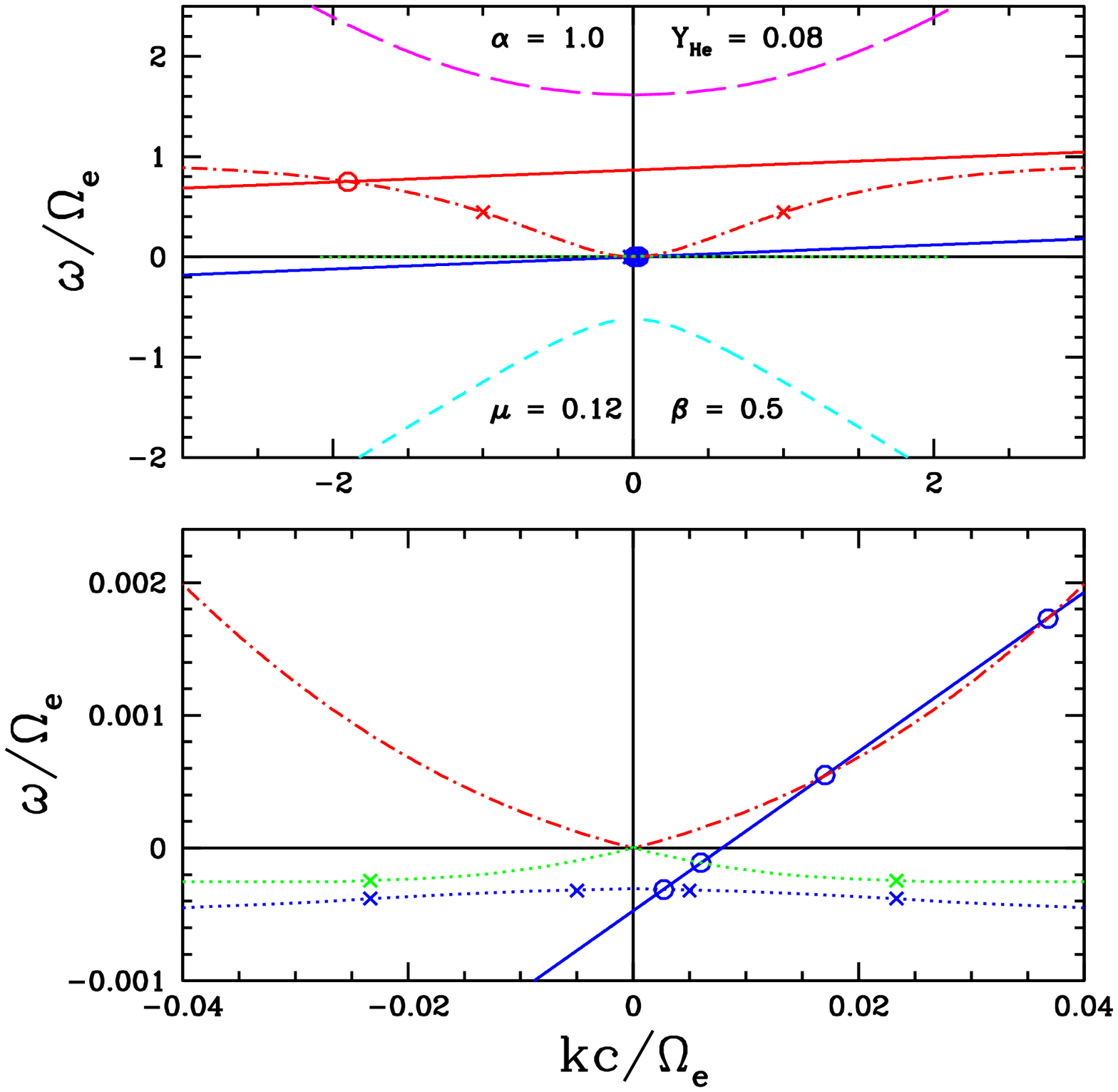}
\hspace{-0.6cm}
\includegraphics[height=8.4cm]{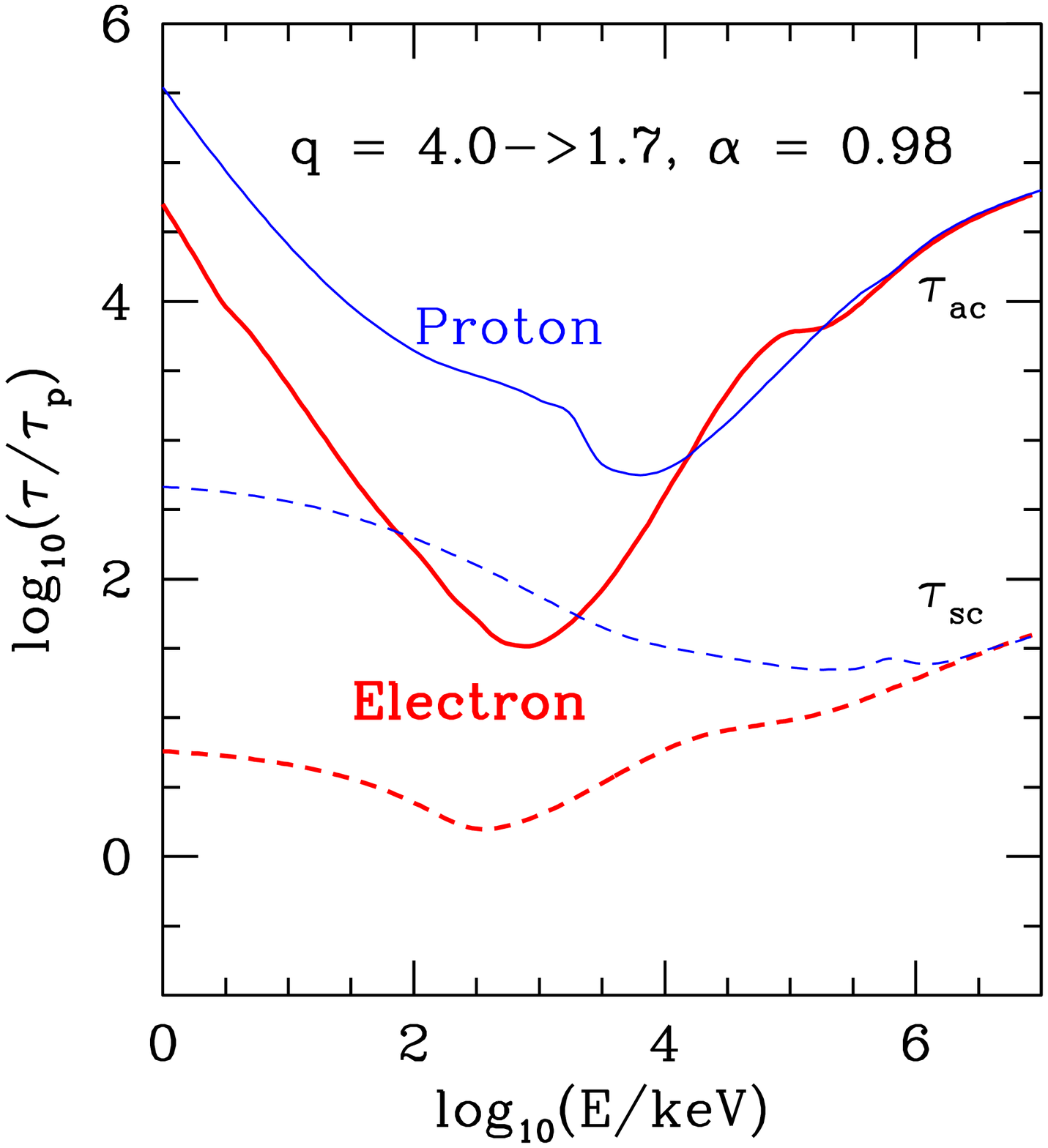}
\end{center}
\caption{ \small
{\bf Left panel}: Same as Figure \ref{fig1.ps} but in an electron-proton-$^4$He plasma 
with $\alpha = 1.0$ and $Y_{\rm He}=0.08$. The resonant interactions (circles) are for 
electrons (upper line) and protons (lower line) with $\beta = 0.5$ and $\mu=0.12$. Here we 
note that there are two ion cyclotron branches (both 
indicated by the dotted curves), the upper one approaches to the $^4$He cyclotron waves at 
high $k$ (HeC) while the lower one approaches to the modified proton cyclotron waves 
(PC'). The crosses indicate the expected breaks in the turbulence spectrum as 
discussed in the text. {\bf Right panel}: The pitch angle averaged acceleration (solid 
curves) and scattering (dashed curves) times in units of $\tau_{\rm p}$ for electrons 
(thick curves) and protons (thin 
curves).  The model parameters are indicated in the figure. The minimum of the electron 
acceleration time corresponds to the spectral break $k_{\rm max}$ of the turbulence in the 
EC branch. See text for details.
} 
\label{fig12.ps}
\end{figure}

The inclusion of $^4$He splits the PC branch into two: one covers the frequency
range of 0 to $-\delta/2$ and the other covers the frequency range of
$-\omega_{\rm PC}$ to $-\delta$, where $\omega_{\rm PC}\simeq (0.5+Y_{\rm He})\delta$ 
is the lowest frequency of the branch (remember that the minus sign only indicates that the 
waves are left-handed polarized). We refer the former as a $^4$He cyclotron branch (HeC) 
and the later as a modified proton cyclotron branch (PC') because at high wavenumbers they 
approach to the $\alpha$-particle and proton cyclotron frequencies, respectively. Figure 
\ref{fig12.ps} (left panel) shows the dispersion relation in such a plasma and the 
resonance conditions for electrons and protons with $\beta=0.5$ and $\mu = 0.12$.

With this additional branch, particles with $\mu\ne 0$ can interact at least with three
waves with one from each of the three cyclotron branches (particles with $\mu=0$ always
interact with two waves from one of the wave branches). Some particles can resonate with 
five waves three of which would be from one of the three cyclotron branches (e.g. the 
protons in the left panel of Figure \ref{fig12.ps} have three resonances with the EC branch 
with two of them shown in the lower panel of the figure. The third one is at high $k$ and 
$w\rightarrow\Omega_{\rm e}$). In a strongly magnetized plasma the two 
additional resonances can also come from the electromagnetic branches. The general 
results are quite similar to those in pure hydrogen plasmas; there are critical 
angles and critical energies, which now separate the $\mu-\beta$ and $\beta-\alpha$ spaces 
into regions with three and five resonant interactions. Given spectra for each of the wave 
branches one can proceed to calculate the F-P coefficients, which have sharp jumps across 
the critical angles or energies, and the times $\tau_{\rm ac}$ and $\tau_{\rm sc}$. In 
the next section we discuss the turbulence spectrum we shall use for this purpose.

\clearpage

\clearpage

\subsection{Spectrum of Turbulence}

In the discussion of the previous section we assumed a power law wave spectrum ${\cal 
E}\propto k^{-q}$, for $k>k_{\rm min}$, with $k_{\rm min}\ll k_{\rm A} = \alpha 
\delta^{1/2}$ so that there will be waves with the Alfv\'{e}nic dispersion relation for 
interactions with high energy electrons and protons. We now introduce an additional cutoff 
at high wavenumbers presumably caused by thermal damping. Thus we will have a broken 
power-law turbulence spectrum with three indexes $q$, $q_{\rm l}$, and $q_{\rm h}$ and two 
critical wavenumbers 
$k_{\min}$ and $k_{\max}$. 
\begin{equation}
{\cal E}(k) = (q-1){\cal E}_0/k_{\rm min} \cases{(k/k_{\rm min})^{q_{\rm l}}, & for 
$k<k_{\rm min}$; \cr
(k/k_{\rm min})^{-q}, & for $k_{\rm min}<k<k_{\rm max}$; \cr
(k_{\rm max}/k_{\rm min})^{-q}(k/k_{\rm max})^{-q_{\rm h}}, & for $k>k_{\rm max}$,}
\end{equation}
where $q_{\rm l}>0$ (we choose $q_{\rm l}=2$ because its value is almost irrelevant), $q = 
1.7$ is the Kolmogorov value, and $q_{\rm h}=4$, a typical value of the spectral index for 
waves subject to strong damping (Vestuto, Ostriker \& Stone 2003).  A self-consistent 
treatment of wave-particle interactions is required for an exact description of these 
spectral breaks.  This is beyond the scope of the current investigation.  In stead, we make 
some reasonable assumptions on these breaks:

{\it The low $k$ or large scale cutoff}  $k_{\rm min}\gtrsim c/(\Omega_{e} L_{\rm max})$ 
where $L_{\rm max}$ is the largest scale of the turbulence, which must be less than the 
size $L$ of the region, and is most likely much less than it.  To accelerate 
particles to high energies, we also need $k_{\rm min}\ll k_{\rm A} = \alpha/43$. For 
most waves we choose $k_{\rm min}$ as before, i.e. by the value of the highest energy we 
want the particles to achieve. However, such a choice for the PC' branch results in a 
sharp feature in the spectrum of the accelerated protons.  As can be seen in the lower left 
panel of Figure \ref{fig12.ps}, the PC' branch, unlike the EC, PC or HeC branches, crosses 
the frequency axis ($k = 0$) at a finite frequency $\omega_{\rm PC}\ne 0$.  Such 
waves can resonantly scatter protons with a Lorentz factor of $\simeq2/(1+2Y_{\rm He})\sim 
1.7$ or an energy of a few hundreds of MeV.  If the spectrum of PC' branch 
wave extends to a very low wavenumber, one would get very efficient acceleration at such 
energies and a sharp spectral feature.  We assume that such a feature is not present 
(although there is no definite observation to rule it out) and cutoff the spectrum of 
the PC' waves at a higher $k_{\rm min}\simeq k_{\rm A}/5$ or a scale of $L_{\rm max}\simeq 
400\pi c/\alpha\Omega_{\rm e}$ .

{\it The high $k$ or small scale} cutoff is determined via the damping of the waves by low 
energy particles. For example, the cyclotron waves with high wavenumbers are subject to 
thermal damping in plasmas with a finite temperature (Schlickeiser \& Achatz 1993; 
Steinacker et al. 1997; Pryadko \& Petrosian 1998).  One can introduce an imaginary 
$\omega$ to include this effect.  The real part of $\omega$ is not very sensitive to the 
plasma temperature (Miller \& Steinacker 1992).  So the cold plasma dispersion relation 
still gives a good description of the resonant interactions between waves and particles. 
However, above a wavenumber where the thermal damping time is comparable with the time 
scale of the wave cascade, the thermal damping steepens the spectrum of the turbulence.  In 
the absence of a full treatment of these processes, we shall assume that the cyclotron 
waves have steeper spectra than the Whistler and Alfv\'{e}n waves and set $k_{\rm 
max}=k_{\rm W}=\alpha$ for the EC branch.  For the PC' and HeC branches we set $k_{\rm 
max}=k_{\rm A}=\alpha\delta^{1/2}$.  These spectral breaks are indicated by the crosses in 
the lower left panel of Figure \ref{fig12.ps}.

Recent studies of the transport of high energy particles in the solar wind indicate that 
the resonance broadening due to the dissipation of turbulent plasma waves plays an 
essential role in explaining the observed long mean free paths of the particles (Bieber et 
al. 1994; Dr\"{o}ge 2003).  Such broadening is also expected in our case.  It will modify 
only equation (\ref{chi}) if the broadening width is less than the separation between the 
resonances.  In the opposite (and unlikely) case the idea of resonant interaction is 
invalid.  Because the resonance broadening mostly affects the scatterings of particles at 
relatively low energies, where Coulomb collisions are important under solar flare 
conditions, it is less important in understanding the particle acceleration processes 
studied here.

All the formulae developed in the previous sections are still valid except that now 
\begin{equation}
\tau_{\rm p}^{-1} = {\pi\over 2}\Omega_{\rm e}\left[{{\cal E}_{0}\over
B_0^2/8\pi}\right](q-1)k_{\rm min}^{q-1}\,,
\label{taup1}
\end{equation}
and the Alfv\'{e}n velocity is given by 
$\beta_{\rm A} = \delta^{1/2}/[\alpha(1+2Y_{\rm He})^{1/2}]$. Because $q_{\rm l}>0$ and 
$q_{\rm h}>q$ (and the PC' branch contains much less energy than the other branches) the 
total turbulence energy density ${\cal E}_{\rm tot}\simeq {\cal E}_0$.

The right panel of Figure \ref{fig12.ps} gives the electron (thick curves) and proton (thin
curves) acceleration (solid curves) and scattering (dashed curves) times for a model with
$\alpha=1$. We see that in the high energy range where particles are mostly interacting 
with the Alfv\'{e}n waves, the times are almost the same as those in a pure hydrogen 
plasma. At low energies, the times rise sharply with the decrease of energy due to the 
thermal damping of the waves with high wavenumbers.  As we will see in the following 
discussion, the thermal damping makes the particle acceleration times match their escape 
times, giving near power law accelerated particle distributions. More importantly, in the 
intermediate energy range, an acceleration barrier in the proton acceleration time 
still exists even though it is not as prominent as it is in a pure hydrogen plasma. 
This is mainly due to interactions with the HeC branch which makes the acceleration of low 
energy protons more efficient. The sharp decrease of the acceleration time with energy near 
the critical energy is still due to interaction with the Whistler waves. Comparing with 
Figure \ref{fig8.ps}, we note that the electron acceleration and scattering times are also 
affected as evident by the wiggles seen at a few tens of MeV. These wiggles are due to 
interactions with the HeC and PC' branches.

We emphasize here that $q=1.7$ corresponds to the Kolmogorov spectrum.
Our only assumption which may be ad hoc is the large scale size cutoff for the PC' branch 
waves. This assumption is not driven by observations but primarily introduced to obtain a 
smooth proton spectrum. We shall explore consequences of the assumption in the future.
In the following discussion, we will fix these parameters at the specified values and 
investigate how the particle acceleration processes are affected by the strength of the 
turbulence ${\cal E}_{\rm tot}$, the size of acceleration site $L$, the plasma parameter 
$\alpha$ and the temperature of the injected particles.  We will show that this model gives 
much more reasonable explanations to solar flare observations than the previous one.

\subsection{Relative Acceleration of Electrons and Protons}

We now present some results on the relative numbers of accelerated electrons and protons at
the acceleration site (LT) and escaping to the FPs. Here we explore its dependence on the 
model parameters. The normalization is as before (see Figure \ref{fig10.ps} and
discussion in eq. [\ref{Nfp}]), namely we assume that the escape fluxes for the total
numbers of electrons and protons are equal. 

\begin{figure}[h]
\begin{center}
\includegraphics[height=8.4cm]{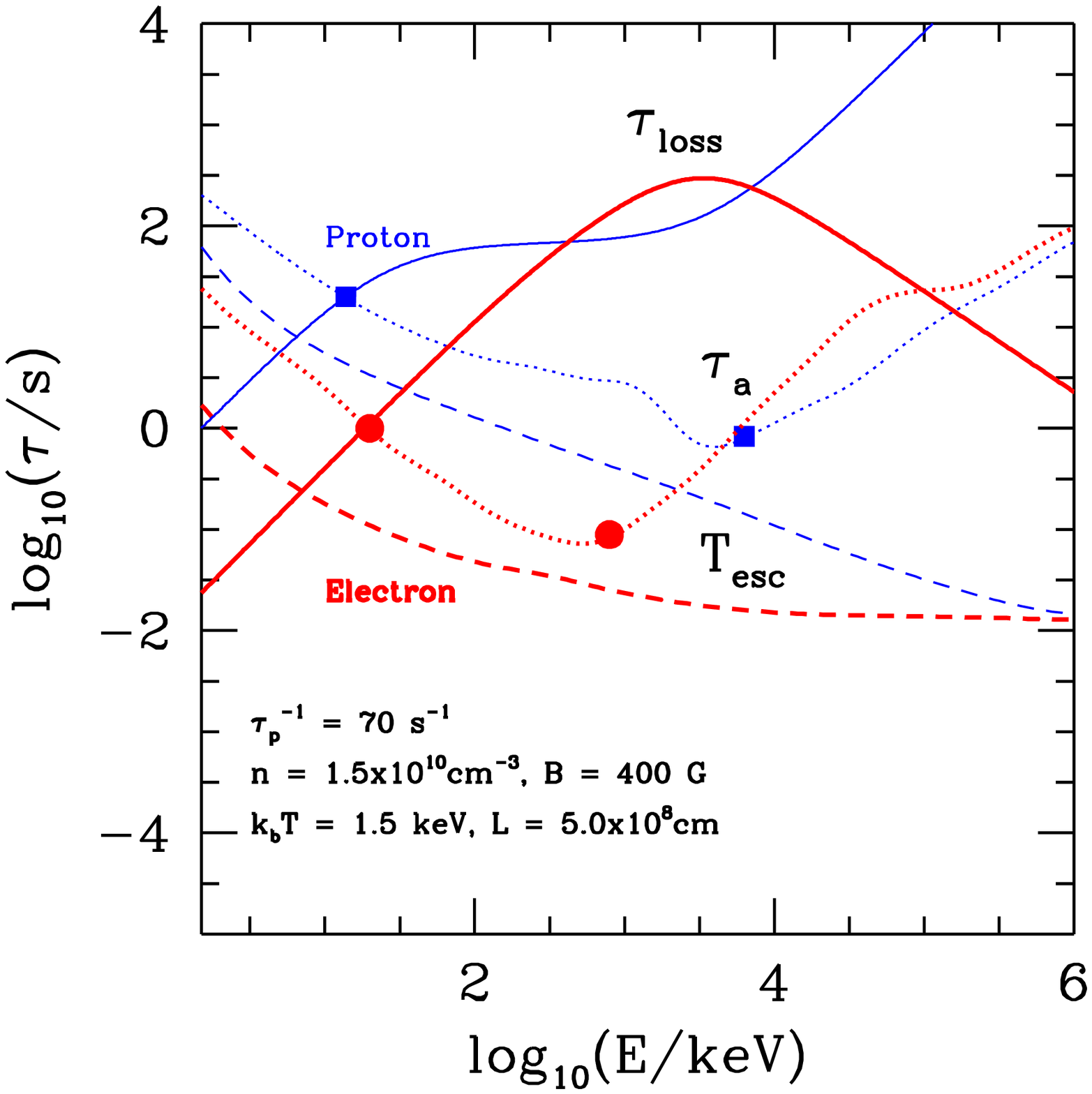}
\hspace{-0.6cm}
\includegraphics[height=8.4cm]{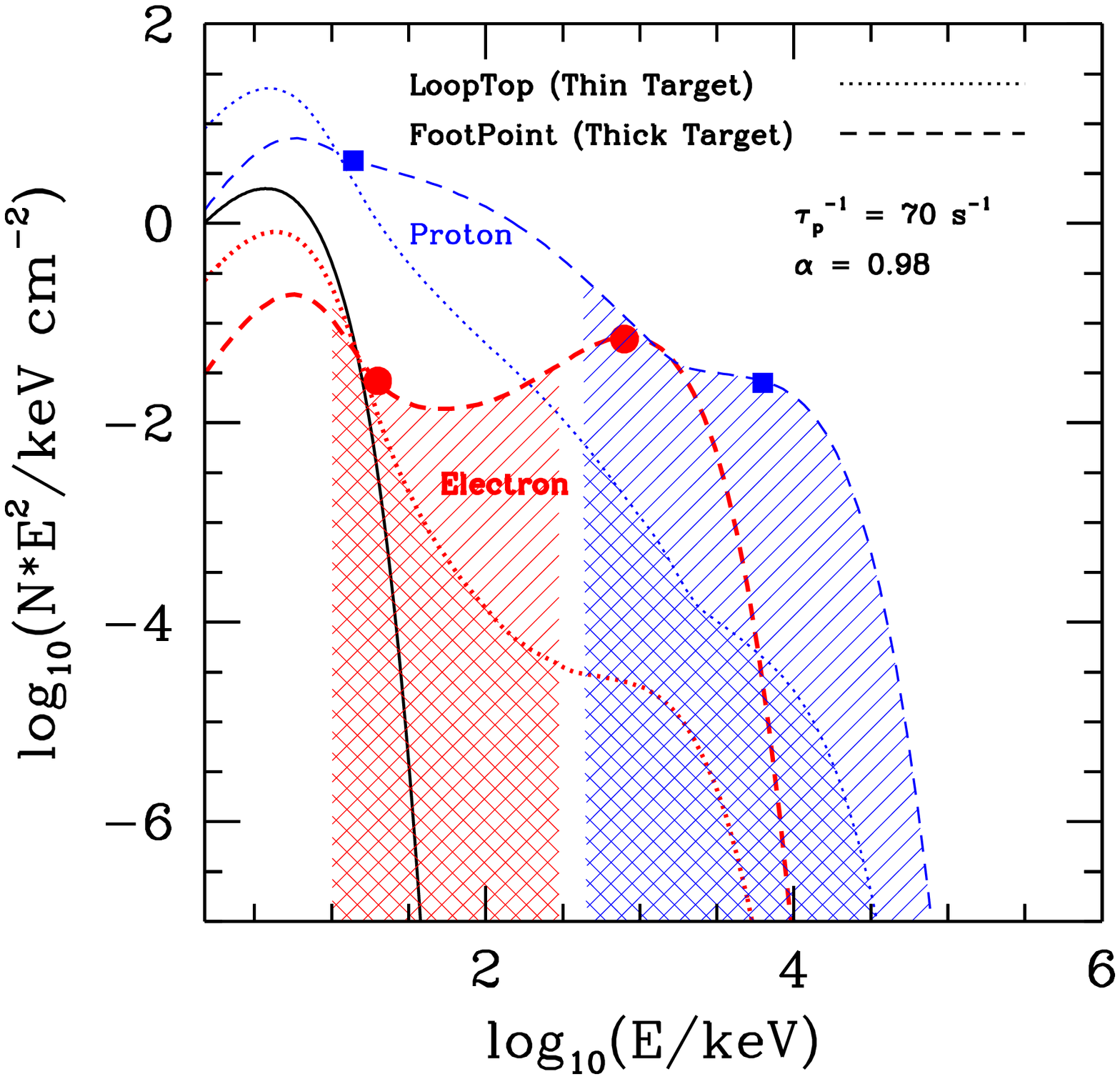}
\end{center}
\caption{ \small
Same as Figure \ref{fig10.ps} but for different model parameters (listed) and in a 
hydrogen and helium plasma. The energy content in high energy electrons and protons is 
comparable. The hatched regions in the right panel correspond to the energy bands related 
to observations of the hard X-ray and gamma-ray emissions during the impulsive phase of 
solar flares. The signs indicate the spectral breaks of the accelerated particle 
distributions.
}
\label{fig13.ps}
\end{figure}

Figure \ref{fig13.ps} gives our fiducial model for solar flares where the energy content 
in the accelerated electrons and protons in the relevant observational energy bands 
(indicated by the shaded regions) is comparable. The time scales are given in the left 
panel and the corresponding particle distributions are shown in the right panel. The line 
types are the same as those in Figure \ref{fig10.ps}. The temperatures of the injected 
electrons and protons are $T = 1.5$ keV. The other model parameters are: the size of the LT 
source $L=5\times 10^8$ cm, $n_{\rm e} = 1.5\times 10^{10}$ cm$^{-3}$, $B = 400$ Gauss 
($\alpha=0.98$) and $\tau_{\rm p}=1/70$ sec. These imply $8\pi{\cal E}_{\rm tot}/B_0^2 
\simeq 8.8\times 10^{-9}k_{\rm min}^{-0.7}$, which is $<10^{-3}$ for $k_{\rm min}>2\pi 
c/L\Omega_{e}= 5.2\times 10^{-8}$. Compared with models in pure hydrogen plasmas, the 
turbulence required to accelerate the particles is much weaker in the current model. This is 
mainly because of the adopted Kolmogorov (instead of $q=2$ or $3$) turbulence spectrum at 
long wavelengths.

\clearpage

\clearpage

\begin{figure}[h]
\begin{center}
\includegraphics[height=8.4cm]{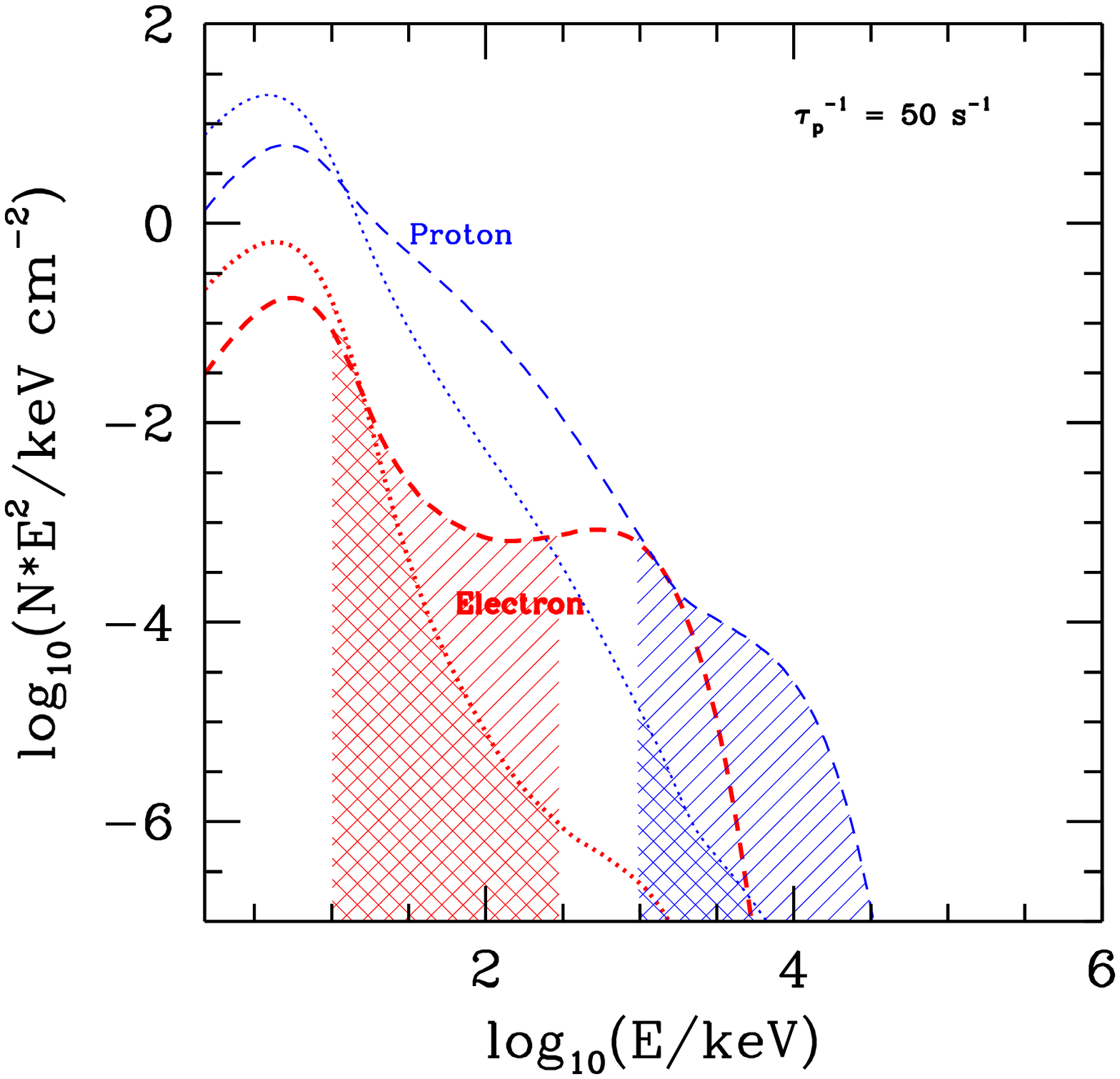}
\hspace{-0.6cm}
\includegraphics[height=8.4cm]{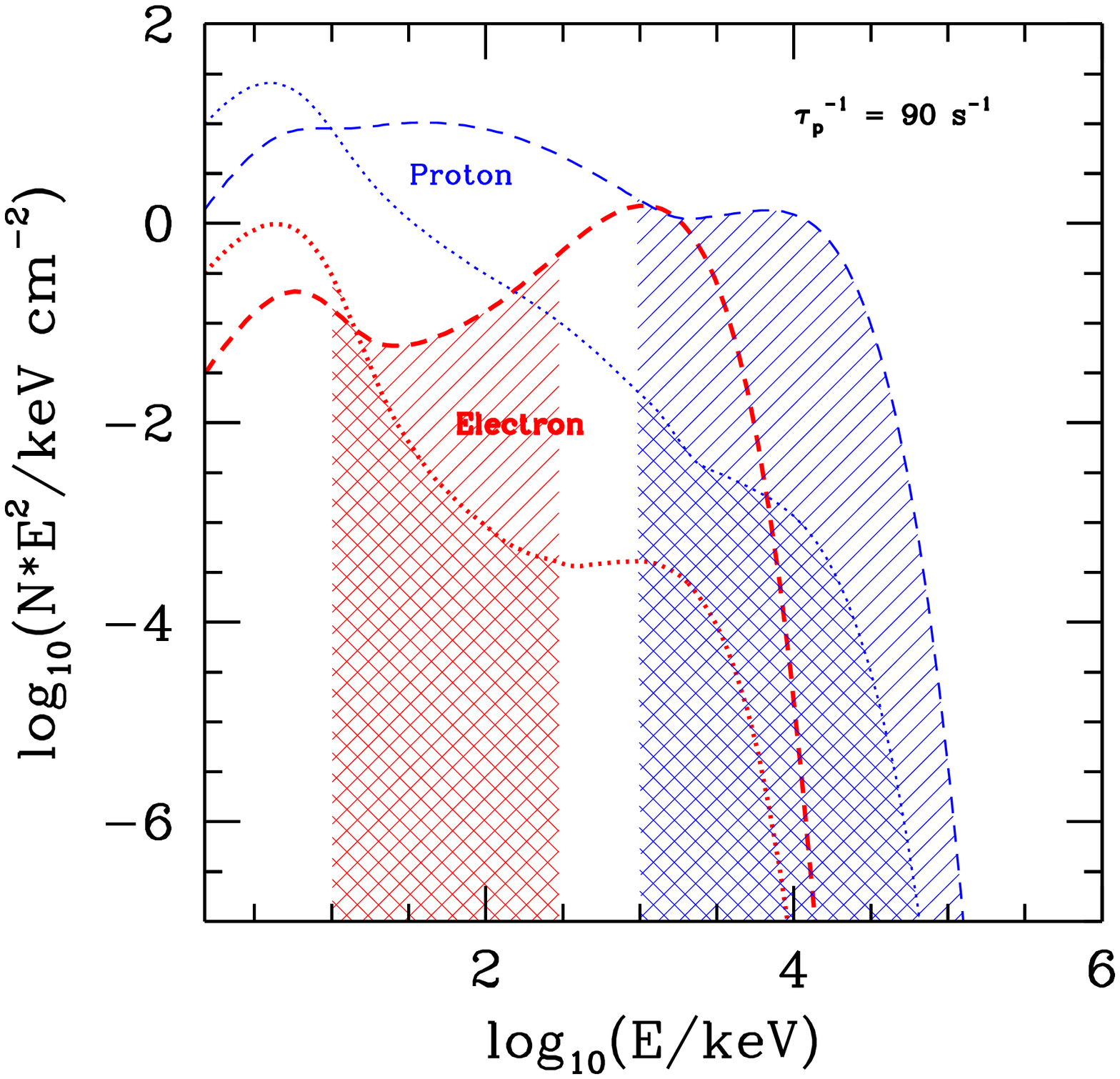}
\end{center}
\caption{ \small
Same as the right panel of Figure \ref{fig13.ps} but for two different values of $\tau_{\rm 
p}$, one larger (left panel) and one smaller (right panel).
}
\label{fig14.ps}
\end{figure}

\clearpage

In Figure \ref{fig14.ps}, we show the effects of the strength of turbulence on the particle 
acceleration. Because the acceleration time is proportional to $\tau_p$ but the escape time 
is inverse proportional to it, a small change in $\tau_p$ can make the acceleration and 
escape times off balance very quickly. The spectra of the accelerated particles then 
change dramatically with $\tau_p$. However, because this is true for both electrons and 
protons, changes in the energy partition between electrons and protons are much smaller. 

Another parameter that affects the spectra of the accelerated particles is the size L of 
the acceleration region. An increase of $L$ results in an increase of the escape time 
and harder spectra of the accelerated particles.  However, again because the escape times 
of electrons and protons increase by the same factor, the energy partition between them is 
not changed significantly.  For example, for a model with $L=10^9$ cm and $\tau_{\rm 
p}^{-1} = 50$ s$^{-1}$, we find that the accelerated particle spectra which are harder than 
those in the left panel of Figure \ref{fig14.ps} and more like those in the right panel of 
Figure \ref{fig13.ps}. 

\begin{figure}[h]
\begin{center}
\includegraphics[height=8.4cm]{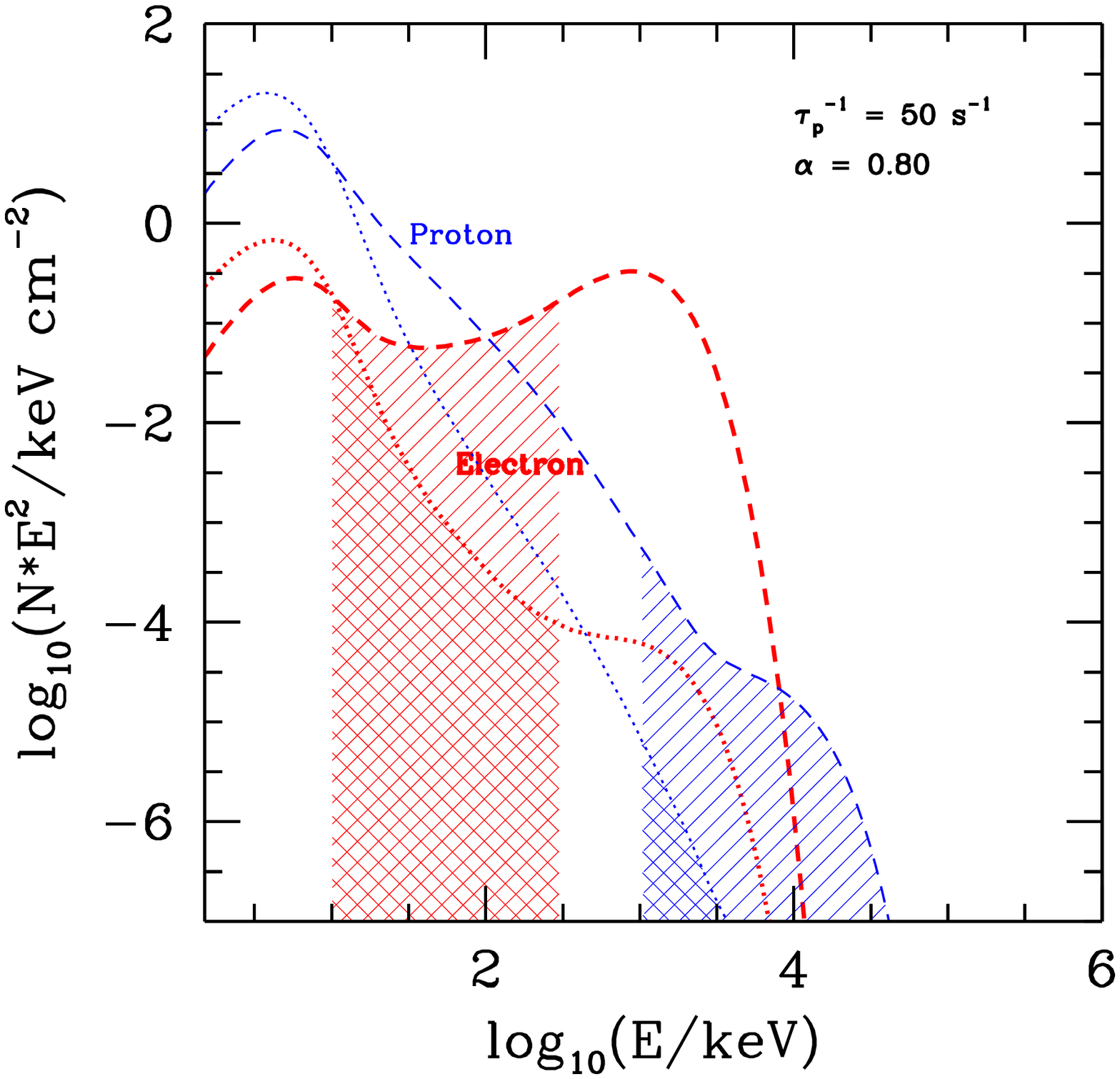}
\hspace{-0.6cm}
\includegraphics[height=8.4cm]{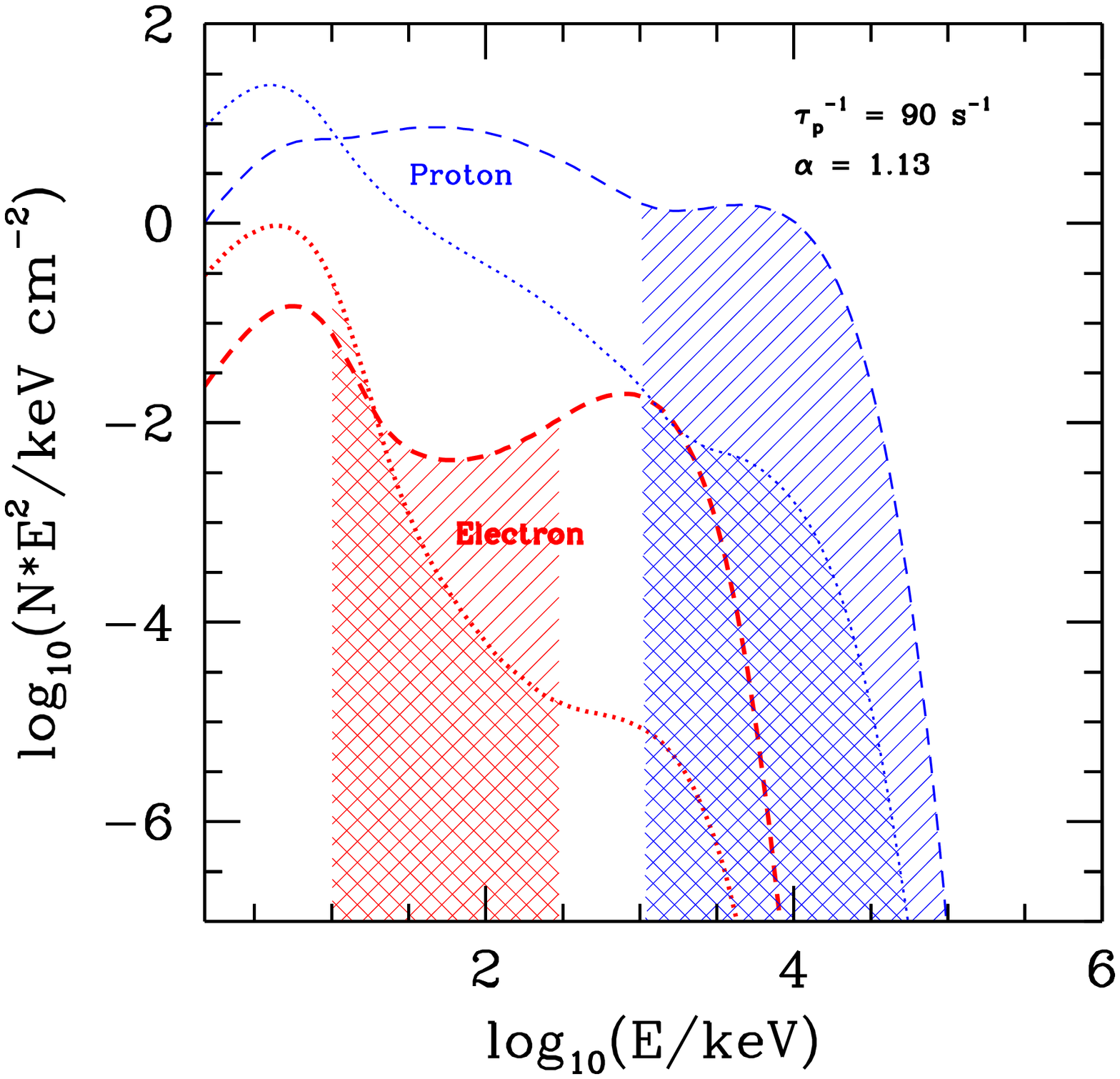}
\end{center}
\caption{ \small
Same as Figure \ref{fig13.ps} but for models with different plasma parameters $\alpha$, 
which are indicated in the figures. $\tau_{\rm p}$ is chosen such that the 
accelerated particle distributions are similar to that required to explain the impulsive 
hard X-ray emission from solar flares. The other model parameters are the same as those 
shown in Figure \ref{fig13.ps}. We see that the electron acceleration is favorable in 
strongly magnetized plasmas while the proton acceleration dominates in weakly magnetized 
plasmas. 
} 
\label{fig15.ps} \end{figure}

Next we examine the effects of the plasma parameter $\alpha\propto n^{1/2}/B$. Figure
\ref{fig15.ps} shows how the relative acceleration of electrons and protons changes with
the change in the value of $\alpha$. It turns out that for the range of parameters used
in the current study, it does not matter whether $\alpha$ is changed by changing the
value of the density $n$ or the magnetic field $B$. The difference between these two
possibilities will appear as relatively small changes in the spectra at the low and high 
energy ends where Coulomb collisions and synchrotron losses become important, respectively. 
To make the spectral shapes compatible with solar flare observations, the value of 
$\tau_{\rm p}$ also needs adjustment. But as we showed above, $\tau_{\rm p}$ affects 
primarily the spectral hardness but not the relative acceleration of electrons and protons. 
Thus the most relevant cause of the changes in the relative acceleration of protons and 
electrons is the variation of $\alpha$; proton (and consequently other ions) acceleration 
is more efficient in high density and/or low magnetic field plasmas. Given that the 
acceleration of electrons and protons are dominated by different wave branches, it is not 
surprising that their relative acceleration depends on the plasma parameter $\alpha$.

\clearpage

\clearpage

\begin{figure}[h]
\begin{center}
\includegraphics[height=8.4cm]{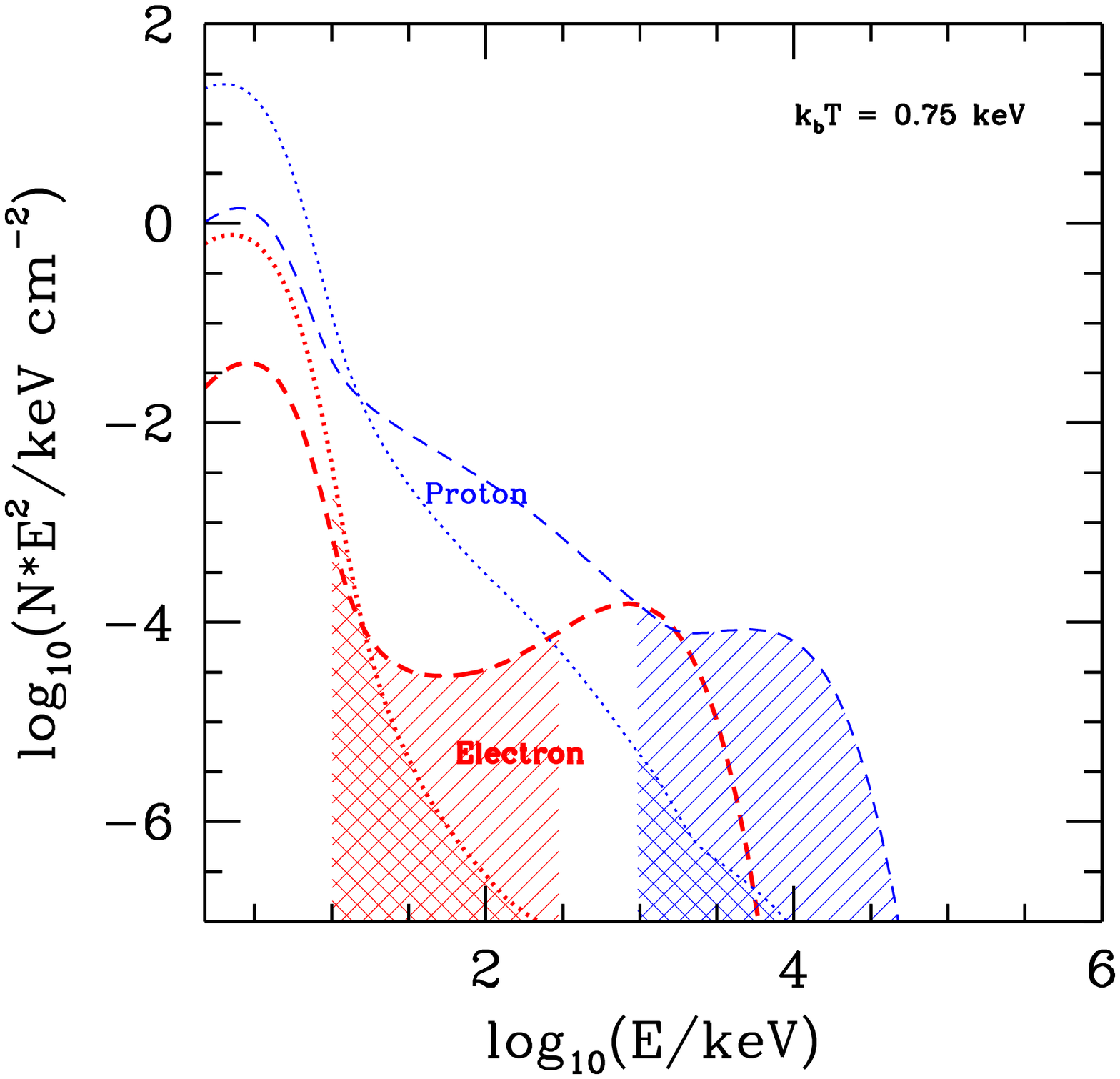}
\hspace{-0.6cm}
\includegraphics[height=8.4cm]{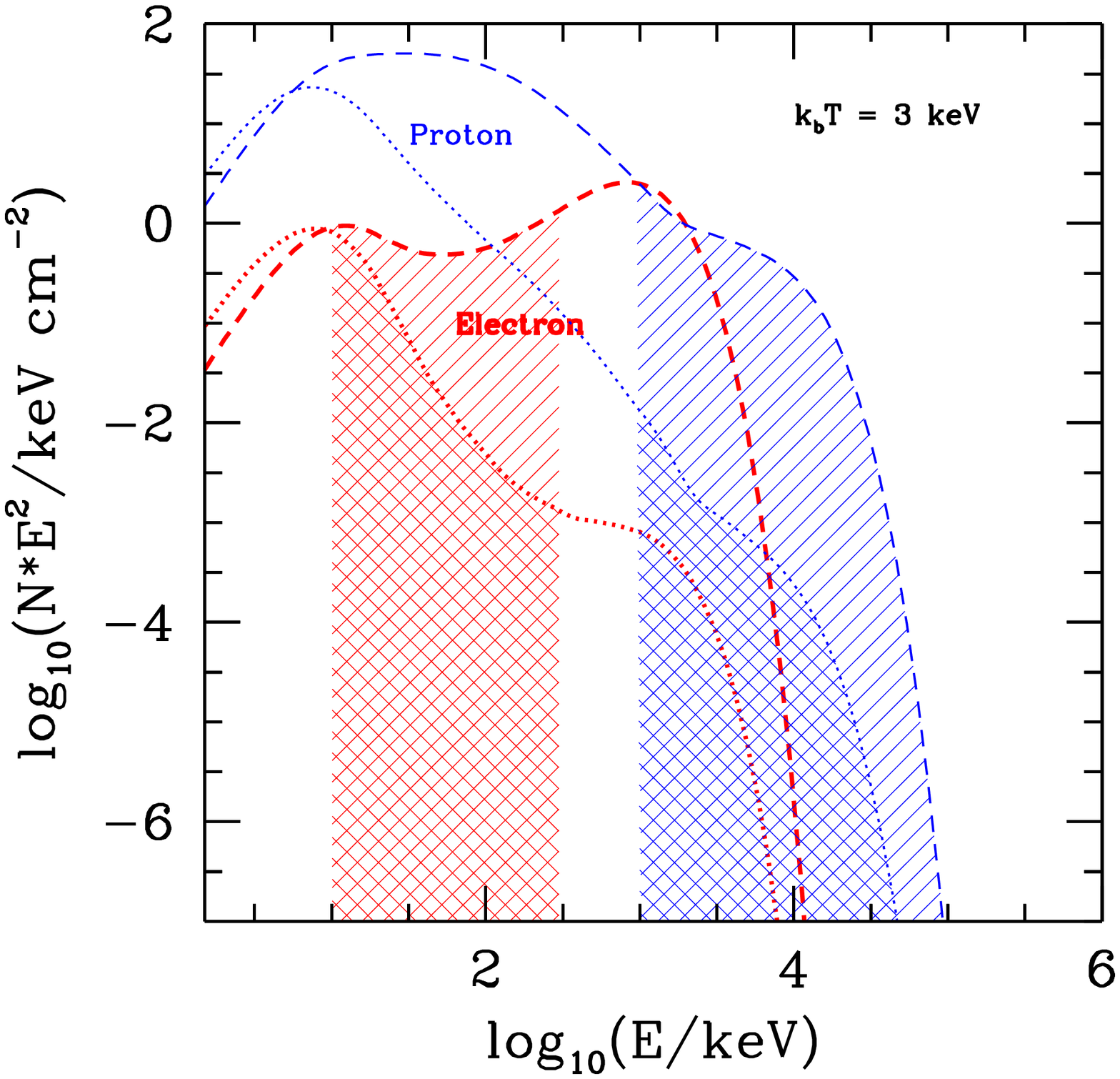}
\end{center}
\caption{ \small
Same as Figure \ref{fig13.ps} but for models with different temperatures 
for the injection plasma. The
temperatures are indicated in the figures. Other model parameters remain 
the same as the fiducial model. The
distributions of high energy particles are not affected by the 
injection process. However, at low energies
where Coulomb collisions dominate, much less particles are accelerated for 
an injection plasma with a lower
temperature.
}
\label{fig16.ps}
\end{figure}

\clearpage

Finally, we consider the effects of the temperature of the background or injected 
particles.  In the models discussed above, we use a high value of temperature of a few keV,
which requires a pre-heating of the flaring plasma to a temperature above the quiet coronal
value.  {\it GOES} and {\it RHESSI} observations do suggest such a characteristic energy 
for the particles before the impulsive phase of X-ray flares.  For example, {\it RHESSI}'s 
high resolution spectra indicate that the electrons in the soft X-ray emitting plasma 
always deviate from an isothermal distribution, implying a significant pre-heating of the 
flare plasma (Holman et al. 2003).  The effects of the injection temperature are 
demonstrated in Figure \ref{fig16.ps}, which shows particle spectra for models with 
temperatures different than that in the fiducial model (Figure \ref{fig13.ps}).  All other 
model parameters remain the same. We see that the shapes of the spectra in the high energy 
range do not change significantly.  However, with the increase of the
temperature, more particles reach the energy range where the acceleration rate is larger
than the Coulomb collisional loss rate and are eventually accelerated to higher energies.
At lower temperature, the quasi-thermal part of the spectra (similar to that of the
injected particles) is more prominent, while at higher T the spectra of the accelerated
particles are dominated by the nonthermal tails.

\section{SUMMARY AND DISCUSSION}
\label{discs}

The primary aim of this work is the determination of the relative acceleration of electrons
and protons by waves in a stochastic acceleration (SA) model. In this paper, we present the
results of the investigation of the resonant interaction of the particles with a broad
spectrum of waves propagating \underline{parallel} to the large scale magnetic field. We
calculate the acceleration and transport coefficients and determine the resulting spectra
for both particles for physical conditions appropriate for solar flares. We show that the
injection of such a turbulence in a magnetized hot plasma can accelerate both electrons and
protons of the thermal background plasma to high energies in the acceleration site. Some of
the accelerated particles escape the site and reach the FPs. The parameters that govern
these processes are the density, temperature, magnetic field, size of the acceleration
region and the intensity and spectrum of the turbulence.

We first describe two general features of our results. 

A. The first has to do with the general characteristics of the accelerated particle 
spectra. The outcome of the SA of a background thermal plasma is the presence of two 
distinct components. The first is a quasi-thermal component at low energies where Coulomb 
collisions play important roles. This can be considered as a simple heating process of the 
background plasma. The second is a nonthermal tail with a somewhat 
complex spectral shape. Technically, one can separate the two components at the energy 
where the Coulomb collisional loss rate $\dot{E}_{\rm Coul}$ is equal to the direct 
acceleration rate $A$. We can then calculate the fractions of the turbulence energy that 
go to ``heat'' and to``acceleration''. This explains the observation of both thermal and
nonthermal emissions during the impulsive phase of solar flares.

The spectra of particles reaching the FPs are in general harder than the 
corresponding LT particle spectra because high energy
particles in the nonthermal tails escape more readily. The relative size of the two
components in the acceleration site (LT) vs the FPs and for electrons vs protons depends
sensitively on the model parameters, which can explain the large variation of the observed
nonthermal emission among solar flares and other astrophysical sources.

B. The second feature has to do with the relative acceleration of electrons and protons. 
{\it To our knowledge, a new result of our investigation is that there is a significant 
difference in the acceleration of protons and electrons.} While the transport and 
acceleration coefficients for electrons are smooth functions of energy, this is not 
true for protons. There appears to be a barrier (lower rate) of acceleration for 
intermediate energy protons. This can have a dramatic effect on the relative production 
rate and spectra of the accelerated protons and electrons. 

To demonstrate some of this and other more subtle effects, we first investigated the 
acceleration of electrons and protons in pure hydrogen plasmas by turbulence with a simple 
power-law spectrum. We find that this simple model does not agree with some qualitative 
aspects of the observed accelerated particle distributions in solar flares. The barrier for 
protons is too strong for reasonable physical conditions. We then explore more realistic 
models, where we include the effects of the background $^4$He particles and the thermal 
damping of the waves. These more realistic models are in better concordance with solar 
flare observations. 

Specifically we find the following:

\begin{enumerate}
\item In general, electrons are preferentially accelerated in more strongly magnetized 
plasmas (small $\alpha\propto n^{1/2}/B$) while the proton acceleration is efficient in 
more weakly magnetized plasmas. The ratio of the energy that goes into the accelerated 
electrons to that into protons is very sensitive to $\alpha$, which can explain the wide 
range of the observed energy partition between these particles. The proton acceleration 
will be more efficient in larger loops where the magnetic field is presumably weaker and 
during the later phase of flares when the corona loops have been filled by plasmas 
evaporated from the chromosphere, giving a higher gas density. This can explain the offset 
of the centroid of the gamma-ray line emission (due to accelerated protons and ions) from 
that of the hard X-rays indicated by a recent {\it RHESSI} observation \citep{Hurford03}. 
It can also account for the observed delay of the nuclear line emission relative to the 
hard X-ray emission \citep{Chupp90}.

\item The acceleration rates and spectra of both electrons and protons are very 
sensitive to the intensity of the turbulence and the size of the acceleration site. Models 
with more intense turbulence and/or larger acceleration region give rise to harder spectra.
This result can explain the observed soft X-ray emission in advance of the impulsive phase 
hard X-ray and gamma-ray emissions (the so-called pre-heating) and the slower than expected 
decline of the temperature of the LT plasma in the gradual phase. When the turbulence is 
weak, as will be the case at the beginning and end of a flare, almost all the dissipated 
turbulence energy goes into the quasi-thermal component and there is no significant 
hard component, producing soft X-ray emission without obvious hard X-ray or gamma-ray 
emission.  When the strength of the turbulence exceeds a threshold, nonthermal tails and 
high energy radiations ensue.  On the other hand, for a turbulence energy much above this 
threshold, one would expect harder spectra than observed in solar flares. This may indicate 
that the sudden presence of a large amount of high energy particles also introduces 
significant dissipation of the turbulence over a broad frequency range such that the 
strength of the turbulence is limited to a level close to the threshold. Consequently, we 
do not see flares with very flat X-ray spectra. To address these processes in detail, one 
needs to treat the wave generation, cascade and damping by both low and high energy
particles properly. Such an investigation is clearly warranted now but is beyond the scope 
of this paper.

\item In general the spectra of both electrons and protons at the acceleration site (LT) 
are softer (stronger quasi-thermal component and steeper nonthermal spectrum) than the 
equivalent thick target spectra at the FPs. This is in excellent agreement with the 
results from the {\it YOHKOH} \citep{Petrosian02} and with the more convincing evidence 
from {\it RHESSI} observations \citep{Jiang03}. The most important parameter here is the 
energy dependence of the escape time (see equation [\ref{escall}]) which depends on the 
pitch angle diffusion coefficient and the size of the acceleration region. Unlike the 
acceleration time (see item B above), the scattering and escape times for protons and 
electrons have similar general behaviors. Consequently, the difference between the LT and 
FP spectra is similar for both electrons and protons.  

\item For injected plasmas with high temperatures, most of the particles can be 
accelerated to very high energies and the steady state particle distribution at low 
energies can be quite different from a thermal distribution. The presence of a 
quasi-thermal component is typical for low temperature plasmas.

\item There are high energy cutoffs (at around 1 MeV for electrons and 10 MeV for protons)  
in the accelerated particle spectra. Both cutoffs are due to the quick escape and the
relatively inefficient acceleration of higher energy particles. The location of these
cutoffs are directly related to the higher wavenumber spectral breaks in the turbulence
spectrum.  In plasmas with stronger thermal damping, the acceleration of high energy
particles becomes relatively more efficient than that of low energy particles.
Consequently, the cutoffs shift toward higher energies. We would then expect a positive
correlation between the cutoff energies and the heating rate of the background plasma.
Observations over a broad energy range will be able to test this prediction.

\end{enumerate}

Finally, we summarize several improvements that are required for direct comparisons with 
observations:
\begin{enumerate}

\item The results presented here show that wave-particle interactions play crucial roles in 
solar flares, especially during the impulsive phase. A self-consistent treatment of this 
problem requires the solution of the coupled kinetic equations for both particles and 
waves. Previous studies on this aspect focused on the Alfv\'{e}n waves alone and ignored 
the energy dependent escaping processes (Miller \& Roberts 1995; Miller et al. 1996). 
Incorporation of the current investigation will make the models more realistic.
 
Waves propagating obliquely with respect to the large scale magnetic field will introduce
new features to the wave-particle interaction (Pryadko \& Petrosian 1999).  Earlier studies
have shown that the fast-mode waves are very efficient in heating or accelerating
super-Alfv\'{e}nic particles via Landau damping or transit-time damping (Miller et al.
1996;  Quataert 1998; Schlickeiser \& Miller 1998). These waves are expected to enhance the
acceleration of high energy electrons and protons and sub-Alfv\'{e}nic particles may also
be accelerated when one adopts the exact dispersion relation for the waves. Results similar
to what we present here are expected. Moreover, if the turbulence is dominated by the lower
hybrid waves, electrons will not be scattered efficiently so that the electron distribution 
is not isotropic (Luo, Wei, \& Feng 2003). The acceleration barrier may not exist if this
is also true for protons. A comprehensive study including these waves is needed to address
the heating and acceleration processes more completely.

\item A time dependent model is needed to address the temporal characteristics of solar
flares and the injection processes. Here we assume that the system is in a steady state
and the injection fluxes of protons and electrons are equal. This may be the case if the
plasmas are brought into the acceleration site by the reconnecting magnetic fields.  
However, e.g. if electrons have a shorter escape time than protons, there could be a net 
charge flux from the acceleration site, which would induce reverse currents consisting 
mainly of electrons so that the injection fluxes of electrons and protons into the 
acceleration site will be different.

\item The application of the formalism developed here to the acceleration of other ions is 
straightforward.  We are in the process of evaluating the relative acceleration of 
different ion species and isotopes and the results are promising and will be published in 
future papers.  It is also straightforward to apply the formalism to accretion systems of 
black holes and neutron stars.  Besides the magnetic reconnection, turbulent plasma waves 
can also be produced by the magneto-rotational instability in accretion disks (Balbus \& 
Hawley 1991). 

\end{enumerate}

\acknowledgements

The work is supported by NASA grants NAG5-12111, NAG5 11918-1, and NSF grant 
ATM-0312344.

\appendix

\section{Dispersion Relations for the EC and PC Branches}

Waves in the EM' branch interact resonantly with protons and those in the EM branch 
interact with electrons only for low values of $\alpha$.  These interactions mostly
affect the acceleration of very low energy particles (see discussion in \S\
\ref{cri}).  Waves in the EC and PC branches are the dominant modes for the acceleration of
protons and electrons for intermediate and high values of $\alpha$ and energies and 
therefore play key roles in determining the relative acceleration of the two species.
In what follows we give some approximate analytic descriptions of these modes
which are considerably simpler than equation (\ref{disp}).

At frequencies $\omega\ll\delta$, or $|k|\ll k_{\rm A} \equiv \alpha \delta^{1/2}$, both 
branches reduce to the Alfv\'{e}n waves with the dispersion relation $\omega= |k|\beta_{\rm 
A}=\delta (|k|/k_{\rm A})$ for the EC branch.  The middle portion of the EC branch, $k_{\rm 
A}\ll |k| \ll k_{\rm W} \equiv \alpha$, corresponds to the Whistler waves with the
dispersion relation $\omega\simeq k^2/\alpha^2=(k/k_{\rm W})^2$.  At still higher 
wavenumbers, ($|k|\gg k_{\rm W})$, $\omega\rightarrow 1$ and the dispersion relation can be
approximated as $\omega\sim 1-\alpha^2/k^2$.  The transition between the
Whistler and this portion occurs at $k_{\rm W}=\alpha$.  This suggests that the
dispersion relation for the Whistler and electron-cyclotron portions can be
approximately described by $\omega=k^2/(k^2+\alpha^2)$. For the EC branch we can use the 
simple approximation
\begin{equation} 
\omega={\delta^{1/2}x+x^2\over 1+x^2}\,; \ \ \ \ {\rm with} \ \ \ \ x=|k|/k_{\rm 
W}=|k|/\alpha\,,
\label{ecapprox}
\end{equation} 
which agrees with the exact expression
within $40\%$ for $\alpha\ge 0.6$.  For highly magnetized plasmas with $\alpha
\ll 1$  the Whistler branch disappears. One then
has $\omega=|k|$ for $|k|<1$ and $\omega=1$ for $|k|>1$ for the EC branch (and 
the reverse is true for the EM branch).

Similarly for the PC branch one gets the Alfv\'{e}n waves with $\omega= - |k|\beta_{\rm A}$ 
at low wavenumbers ($|k|\ll k_{\rm A}$).  The proton-cyclotron waves 
whose dispersion relation doesn't have a simple form can be roughly approximated as 
$\omega_{\rm PC}\sim -\delta(1- 1/(2+(k/k_{\rm A})^2)$ for high $k$.  We can combine these 
two forms into one simple expression
\begin{equation}
\omega=-\delta {y+y^2\over 1+y+y^2}\,;\ \ \ \  {\rm with} \,\,\,\, y=|k|/k_{\rm 
A}=|k|/(\alpha\delta^{1/2}),
\label{pcapprox}
\end{equation}
which agrees with the exact expression to within $10\%$ for $\alpha\ge 0.6$.
For very highly magnetized plasmas, $\alpha \ll \delta^{1/2}$, one again has 
$\omega=-|k|$ for $|k|<\delta$ and $\omega=-\delta$ for $|k|>\delta$ for the PC branch 
(and the reverse for the EM' branch).

\section{
Critical Energies and Angles for Resonance with the EC and PC Branches} 

With the approximate analytical expressions (\ref{ecapprox}) and (\ref{pcapprox}) for the 
dispersion relation, one can derive the critical velocity (\ref{betacr}) and critical angle 
(\ref{mucr2}) for resonant interactions of low energy particles with the EC and PC 
branches. The critical velocity is for protons with $\mu = 1$ interacting with the EC 
branch. From the resonance condition (\ref{reson}) and the dispersion relation 
(\ref{ecapprox}) for the EC branch, we have
\begin{equation}
\alpha\beta x-{\delta\over\gamma} ={\delta^{1/2}|x|+x^2\over 1+x^2}\,,\ \ \ \ \ \ x
=k/k_W=k/\alpha.
\label{crb1}
\end{equation}
This equation has three roots with two of them being equal and $\ll 1$ at the 
critical velocities (see Figure \ref{fig1.ps}, and equation [\ref{betacr}]). We can 
therefore ignore the $x^2$ term in the denominator of the right hand side of equation 
(\ref{crb1}). One can show that 
\begin{equation}
\beta_{\rm cr} = 3\sqrt{\delta}/\alpha\simeq0.07/\alpha\,,
\label{crb}
\end{equation}
which agrees with the numerical result within $15\%$ [eq.(\ref{betacr})]. In general, we 
have
\begin{equation}
\beta_{\rm cr} = 3\sqrt{\delta}/\mu\alpha\simeq0.07/\mu\alpha\,.
\end{equation} 

For electron resonances with the EC branch, we have
\begin{equation}
\alpha\beta\mu x+{1\over\gamma} ={\delta^{1/2}|x|+x^2\over 1+x^2}\,.
\end{equation}
The equation has three roots with two of them being equal when
\begin{equation}
(\alpha\beta\gamma\mu_{\rm cr})^2 = {8(\gamma-1)^3\over 
8+20\gamma-\gamma^2+\sqrt{(8+20\gamma-\gamma^2)^2+64(\gamma-1)^3}}\,.
\end{equation}
We then have
\begin{equation}
\mu_{\rm cr} \simeq \beta^2/\sqrt{54}\alpha\,, \hspace{1cm} {\rm for}\ \ \ \beta\ll 1\,,
\end{equation}
which agrees with equation (\ref{mucr2}) within $5\%$.

For proton resonance with the PC branch, we have
\begin{equation}
\alpha\beta\mu\sqrt{\delta} y-{\delta\over\gamma} =-\delta{|y|+y^2\over 1+|y|+y^2}\,,\ \ \ 
\  \ \ \ y = k/\sqrt{\delta}\alpha. 
\end{equation}
For low energy protons at critical pitch angles, the two equal roots of the equation are 
much larger than unity. We can therefore approximate 
the right hand side of this equation as $-\delta(1-1/y^2)$. Then we have
\begin{equation}
{4\over 27}{\delta\over(\alpha\beta\gamma\mu_{\rm cr})^2} = \left({\gamma\over 
\gamma-1}\right)^3\,,
\end{equation}
which becomes
\begin{equation}
\mu_{\rm cr} = {\sqrt{\delta}\beta^2/\sqrt{54}\alpha}\, \hspace{1cm} {\rm for}\ \ 
\beta\ll 1\,.
\end{equation}
This agrees with equation (\ref{mucr2}) within $5\%$.

\section{
Approximate Analytic Expressions for the Acceleration and Scattering Times in 
the Relativistic and Low Energy Limits}

It is useful to have some approximate analytical expressions for the acceleration and 
scattering times under certain limits. In the relativistic region where $\gamma\gg1$, the 
results for electrons are relatively simple and have been studied under different 
context (Schlickeiser 1989; Pryadko \& Petrosian 1997). Here we discuss the results for 
ions. For relativistic particles, and in general for weakly magnetized plasmas 
($\alpha>\delta^{1/2}$), $R_1\gg R_2^2$. As a result, the acceleration time defined by 
equation (\ref{accel}) can be approximated as:
\begin{equation}
\tau_{\rm ac}^{-1} = \left<D_{pp}/p^2\right>\,,
\end{equation}
where ``$<>$" denotes average over pitch angle.  Relativistic particles with $\gamma\gg 
|\omega_i|/\delta$ resonate with the Alfv\'{e}n waves with $\omega = \pm \beta_{\rm A} k$. 
From the resonance condition (\ref{reson}), we have the wavenumbers of 
the resonant waves: $k_{\pm} = \omega_i/[\gamma(\beta\mu\mp\beta_{\rm A})]\simeq 
\omega_i\beta\mu/\gamma$. Then we have
\begin{eqnarray}
\left<{D_{pp}\tau_{{\rm p}i}\over p^2}\right>&=& {\gamma^{q-2}\beta_{\rm A}^2\over 
|\omega_i|^q 
\beta^2}\int_0^1\d\mu(1-\mu^2)[(|\beta\mu-\beta_{\rm A}|)^{q-1}+(|\beta\mu+\beta_{\rm 
A}|)^{q-1}] 
\nonumber \\
&\approx& {\gamma^{q-2}\beta_{\rm A}^2\over |\omega_i|^q 
\beta^2}\int_0^1\d\mu(1-\mu^2)2(\beta\mu)^{q-1}
={4\gamma^{q-2}\beta^{q-3}\delta\over q(q+2)\alpha^2|\omega_i|^q}
\,,
\label{tach}
\end{eqnarray} 
which is consistent with the numerical results within a factor of two.

Similarly, one can estimate the scattering time in the relativistic limit:
\begin{eqnarray}
{\tau_{\rm sc}\over\tau_{{\rm p}i}} &=& 2|\omega_i|^q\gamma^{2-q} 
\left<{(1-\mu^2)\over(\beta\mu-\beta_{\rm A})^{q-1}+(\beta\mu+\beta_{\rm A})^{q-1}}\right> 
\nonumber \\
&\approx& {|\omega_i|^q\over\gamma^{q-2}\beta^{q-1}}\left[ 
{1-(\beta_{\rm A}/\beta)^{2-q}\over 2-q}-{1-(\beta_{\rm A}/\beta)^{4-q}\over 4-q}\right]\,,
\end{eqnarray}
which is in agreement with the numerical results within $50\%$. Note that the integral over 
$\mu$ has been taken from $\beta_{\rm A}/\beta$ to 1 here. Like the acceleration 
time, protons and electrons with the same energy have the same scattering time. 
This is consistent with the result of Pryadko \& Petrosian (1997). We note that in the 
relativistic region, the acceleration and scattering times are identical for all charged 
particles except for the $\omega_i$ term.

In the nonrelativistic region where $\gamma\approx 1$, from the resonance condition $\omega 
= -\omega_i+\beta\mu k$, and the dispersion relation (\ref{disp}), one can 
show that 
\begin{equation}
k = -\omega\alpha(\beta\mu k)^{-1/2}\,.
\end{equation}
So $k =(\omega_i^2\alpha^2\beta^{-1}\mu^{-1})^{1/3}$. Using equation (\ref{coeff}), we 
have
\begin{eqnarray}
\left<{D_{pp}\tau_{pi}\over p^2}\right> &\approx& \left<{1-\mu^2\over 
3\alpha^2\beta^2}\left({|\omega_i|^2\alpha^2\over\beta\mu}\right)^{(1-q)/3}\right> 
\nonumber \\
&=& 
{6\over(q+2)(q+8)}\left[\alpha^{-2(q+2)}|\omega_i|^{2(1-q)}\beta^{q-7}\right]^{1/3}\,.
\end{eqnarray}
We note that $\beta_{\rm g} \approx -2\beta\mu$ under the resonance condition. In a 
previous study (Pryadko \& Petrosian 1997), the minus sign was missed, which causes their 
acceleration time three times shorter than ours. Equation (\ref{coeff}) then gives
\begin{eqnarray}
<D_{\mu\mu}\tau_{\rm pp}> &\approx& 
\left<\left[1+{\mu(|\omega_i|\alpha\beta\mu)^{1/3}\over\beta\alpha}\right]^2
{1-\mu^2\over3\beta\mu}\left({\omega_i^2\alpha^2\over \beta\mu}\right)^{-q/3}\right> 
\nonumber \\ &\approx& 6\beta^{q/3-1}(|\omega_i|\alpha)^{-2q/3}\left[{1\over 
q(q+6)}+{2|\omega_i|^{1/3}\over (q+4)(q+10)(\alpha\beta)^{2/3}}\right. \nonumber \\
&&\left.+{|\omega_i|^{2/3}\over(q+8)(q+14)(\alpha\beta)^{4/3}}\right]\,,
\end{eqnarray}
which agrees with the numerical results within $20\%$.

\section{
Approximate Analytic Expression for the Acceleration Barrier}

To estimate the acceleration time at the barrier in pure hydrogen plasmas, we notice that 
resonant interactions of protons with nearly 90$^\circ$ pitch angle ($\mu<\mu_{\rm cr}$) 
with proton-cyclotron waves moving in both directions have significant contributions to the 
proton acceleration below the critical energy.  Because $D_{\mu\mu}(R_1-R_2^2)$ is a smooth 
function near $\mu=0$ (Figure \ref{fig5.ps}) and $\mu_{\rm cr}\ll 1$ (eq. [\ref{mucr2}]), 
we have 
\begin{equation}
\tau_{\rm ac}(p)\approx {p^2\over D_{pp}(\mu=0,p)\mu_{\rm cr}(p)}\,.
\end{equation}
Beyond the critical energy, some protons start to resonate with Whistler waves (see 
Figure \ref{fig2.ps}) and the proton acceleration rate increases sharply with energy.

From equation (\ref{coeff}), we have
\begin{equation}
{D_{pp}(\mu = 0, p)\over p^2} = {2\beta_{\rm ph}^2|k|^{-q}\over \tau_{\rm pp} 
\gamma^2\beta^2|\beta_{\rm g}|}\,,
\end{equation}
where $k$ can be obtained from the dispersion relation (\ref{disp}) and 
$\omega=-\delta/\gamma$ is given by the resonance condition (\ref{reson}). 
Then we have
\begin{eqnarray}
\tau_{\rm ac}&\approx &{\tau_{\rm p}\gamma^{2-q}\beta^2\delta^{q-2}\over 2 \mu_{\rm cr}}
\left[1+{\alpha^2\gamma^2(1+\delta)\over 
\delta(\delta+\gamma)(\gamma-1)}\right]^{(3+q)/2}
\left[1+{\alpha^2(1+\delta)\{\gamma-0.5(1-\delta)\}\gamma^3\over 
(\delta+\gamma)^2(\gamma-1)^2\delta}\right]^{-1} \nonumber \\
&\approx& {2^{(q-1)/2}\tau_{\rm p}\delta^{(q-5)/2}\alpha^{q+1}\over \mu_{\rm 
cr}}\beta^{3-q} 
\hspace{2.3cm} \textrm{for }\beta\ll 1 \nonumber \\ 
&=& 7 \alpha^{q+2}(2\delta)^{(q-1)/2}\beta^{1-q}\delta^{-5/2} \tau_{\rm p} \nonumber \\
&=& 7 \alpha^{q+2}\delta^{-5/2}\left({E\over m_{\rm e} c^2}\right)^{(1-q)/2}\tau_{\rm p}\,. 
\label{tappro} 
\end{eqnarray}
Combining this with equations (\ref{crb}) and (\ref{tach}), one can estimate the height of 
the acceleration barrier in logarithmic scale at the critical velocity
\begin{equation}
\delta \log{(\tau_{\rm ac})} \simeq 
\log{\{3.1\times
2^{(q-1)/2}\alpha^{q+2}\delta^{-1/2-q}/[q(q+2)]\}}\,.
\end{equation}
Similarly, one can estimate the acceleration time for low energy electrons:
\begin{eqnarray}
\tau_{\rm ac} &=& 2\left[\int_{-1}^1\d \mu
D_{\mu\mu}(R_1-R_2^2)\tau_{\rm p}\right]^{-1} \nonumber \\
&\approx &{\tau_{\rm p}\gamma^{2-q}\beta^2\over 2 \mu_{\rm cr}}
\left[1+{\alpha^2\gamma^2(1+\delta)\over
(\gamma-1)(1+\gamma\delta)}\right]^{(3+q)/2}
\left[1+{\alpha^2(1+\delta)\{\gamma\delta+0.5(1-\delta)\}\gamma^3\over
(\gamma-1)^2(1+\gamma\delta)^2}\right]^{-1} \nonumber \\
&=& 7 \alpha^{q+2}\left({E\over m_{\rm e} c^2}\right)^{(1-q)/2}\tau_{\rm p}
\hspace{2.3cm} \textrm{for }\beta\ll 1\,. \nonumber
\label{tapproe}
\end{eqnarray}
These expressions are consistent with the numerical results within a factor of two. The 
discrepancy is large for turbulence with a flat spectrum. This is mainly due to 
contributions from electron-cyclotron and proton-cyclotron waves to the acceleration of 
particles with two resonances. When the turbulence spectrum becomes flatter, their 
contributions to the pitch angle averaged acceleration time becomes more important. 
However, as we discussed in \S\ \ref{ephe4}, this effect is not important in real 
astrophysical situation where the cyclotron waves are damped. So the analytical expressions 
give a good estimate of acceleration time in the intermediate energy range.

{}

\end{document}